\documentclass[fleqn,usenatbib]{mnras}

\usepackage{newtxtext,newtxmath}
\usepackage{xcolor}
\usepackage{threeparttable}
\usepackage{multirow}

\usepackage[T1]{fontenc}

\DeclareRobustCommand{\VAN}[3]{#2}
\let\VANthebibliography\thebibliography
\def\thebibliography{\DeclareRobustCommand{\VAN}[3]{##3}\VANthebibliography}

\usepackage{graphicx}	% Including figure files
\usepackage{amsmath}	% Advanced maths commands
\title[Phase dependent metallicity effect on Leavitt Law]
{The effect of metallicity on the Leavitt Law using phase-dependent properties of classical Cepheids}
\author[Bhuyan et al.]{Gautam Bhuyan$^{1}$\thanks{E-mail: gautam.bhuyan2825@gmail.com},
Shashi Kanbur$^{2}$\thanks{E-mail: shashi.kanbur@oswego.edu},
Sukanta Deb$^{1}$\thanks{E-mail: sukanta.deb@cottonuniversity.ac.in},
Louise Breuval$^{3}$,
Anupam Bhardwaj$^{4}$, 
Mami Deka$^{5}$, \and
Earl P.\ Bellinger$^{6}$, 
Kerdaris Kurbah$^{1}$\\ 
$^{1}$Department of Physics, Cotton University, Guwahati 781001, Assam, India \\
$^{2}$Department of Physics, State University of New York Oswego, Oswego, NY 13126, USA\\
$^{3}$European Space Agency (ESA), ESA Office, Space Telescope Science Institute, 3700 San Martin Drive, Baltimore, MD 21218, USA \\
$^{4}$Inter University Centre for Astronomy \& Astrophysics, Pune, Maharashtra, India \\
$^{5}$INAF-Osservatorio astronomico di Capodimonte, Via Moiariello 16, I-80131 Napoli, Italy\\
$^{6}$Department of Astronomy, Yale University, CT, USA\\
}

\begin{document}
\label{firstpage}
\pagerange{\pageref{firstpage}--\pageref{lastpage}}
\maketitle

\begin{abstract}
The absolute calibration of period-luminosity (PL) relations of Cepheids in the Milky Way (MW) and its nearby galaxies has been a cornerstone in determining extragalactic distances and the current local expansion rate of the Universe. However, the universality of PL relations is still debated; in particular, the effect of metallicity on the Cepheid PL relation is not well understood. Due to the hydrogen ionization front (HIF) and the stellar photosphere interactions in Cepheids, different period-color (PC) relations at different phases can influence the corresponding period-luminosity relations at those phases. In this work, we have considered the PL relations at multiple pulsation phases as they capture the ensemble radiation hydrodynamic properties at those phases. We investigate the effect of metallicity on PL relations based on multiphase analysis of classical Cepheid light curves in the MW and its nearby satellite galaxies: the Large Magellanic Cloud (LMC) and the Small Magellanic Cloud (SMC). Multiphase metallicity coefficients $(\gamma)$ are derived in five different photometric bands ($V$, $I$, $G$, $G_{\rm BP}$, $G_{\rm RP}$) and two Wesenheit indices ($W_{VI}$, $W_{G}$). We show that the coefficients of multiphase period-luminosity-metallicity (PLZ) relations vary dynamically as functions of Cepheid pulsation phases over a complete pulsation cycle. Significant differences in the metallicity effect are found between the short- $(0.4 \leq \log{P} < 1)$ and long-period $(1 \leq \log{P} < 2)$ Cepheids at multiple pulsation phases, in two bands, $G_{\rm RP}$ and $W_{G}$. The weighted averages of the multiphase $\gamma_{\lambda}$ values are found to be in good agreement with the latest results published in the literature. Our methods and results provide new insights into understanding the metallicity effect on the Leavitt law, which can be useful in constraining pulsation models as well as show that the metallicity effect on mean-light PL relations can be recovered from the phase-dependent nature of the metallicity effect found in this study.    
\end{abstract}


%%%%%%%%%%%%%%%%%%%%%%%%%%%%%%%%%%%%%%%%%%%%%%%%%%
%%%%%%%%%%%%%%%%% BODY OF PAPER %%%%%%%%%%%%%%%%%%
%%%%%%%%%%%%%%%%%%%%%%%%%%%%%%%%%%%%%%%%%%%%%%%%%%

\section{Introduction} \label{sec:intro}
Classical Cepheids (hereafter, Cepheids) belong to a class of intrinsic variable stars which undergo pulsations due to perturbations in hydrostatic equilibrium driven by the $\kappa$-$\gamma$ mechanisms in the stellar envelope \citep{cox80, cate15, bhar20}. They are Population I stars in the core post-hydrogen and helium-burning phase, located in the instability strip, with luminosities $L \sim 10^{5}~{\rm L_{\odot}}$ and typical masses $M \sim 2-20 ~{\rm M_{\odot}}$ \citep{muse22, deka24}. The luminosity of Cepheids is strongly correlated to their pulsation periods. This correlation is known as the period-luminosity (PL) relation or the Leavitt law, in which the brighter stars have the longer periods \citep{leavitt12}. 

The absolute calibration of the Cepheid PL relation remains central to the determination of the current expansion rate of the Universe or the Hubble constant $(H_{0})$ using the ``cosmic distance ladder'' \citep{free01, ries16, ries22}. The measured value of $H_{0}$ based on extragalactic distances calibrated using Cepheids and type Ia Supernovae (SNe Ia) by the SH0ES collaboration favours a higher value of $(73.17 \pm  0.86) {\rm ~km s^{-1} Mpc^{-1}}$ \citep{ries22, breu24}. This value differs by more than $ 5\sigma$ from the value of $H_{0} = (67.4 \pm 0.5) \rm{~km s^{-1} Mpc^{-1}}$ obtained from the Planck mission cosmic microwave background (CMB) data \citep{plan20} assuming the $\Lambda {\rm CDM}$ cosmological model. The Tip of the Red Giant Branch (TRGB) method yields intermediate values for $H_0$ \citep{free20, yuan2019, anand2022, anderson2024, scol23}. This discrepancy between the late universe measurement of $H_0$ and the early universe inference based on $\Lambda {\rm CDM}$ is known as the \textit{Hubble tension} \citep[see the review by][]{verd24}. The source of this tension could be attributed either to the existence of unidentified systematic effects in the calibration of the cosmic distance ladder or in the CMB data, or to new physics beyond the standard $\Lambda$CDM cosmological model \citep{ries22}. 

One of the major and least understood systematic effects on the calibration of the Cepheid PL relation is the influence of chemical abundance on Cepheid brightness, known as the metallicity effect. Several studies have investigated the metallicity effect on Cepheid PL relations by incorporating the $\gamma$ term \citep{ries16, ries22, breu22, bhar24} as:
\begin{align}
M & = \alpha (\log{P} - \log{P_{0}}) + \beta + \gamma {\rm [Fe/H]}. 
\label{eq:1}
\end{align}
Empirical studies based on Cepheids in different galactic environments are found to have $\gamma$ values ranging from $\gamma = 0$, \citep{udal01, weil17,mado25} to $\gamma < 0$ in different wavelengths \citep{ripe21, breu22, bhar24, tren24}. Studies reporting $\gamma=0$ include all LMC and SMC Cepheids without applying geometry corrections, which is known to increase the PL scatter significantly and reduce the metallicity effect, see \citet{breu22, breu24}, while \citet{mado25} applies an offset of 0.26 mag to the MW Cepheid distance moduli to match the scatter of the LMC PL relation when combined with earlier HST FGS parallaxes \citep[see review by][]{breu2025}. A negative value of $\gamma$ indicates that a Cepheid with higher metal content is brighter for a given pulsation period. On the other hand, earlier theoretical studies have predicted a value of $\gamma > 0$, indicating that a metal-rich Cepheid should be fainter for a given pulsation period \citep{bono99, bono08, capu00, marc05}. However, recent theoretical studies based on Cepheid evolutionary and pulsation models considering rotational mixing \citep{ande16} and convective pulsation physics \citep{deso22}, respectively, have found $\gamma < 0$ in the period-Wesenheit plane. The results from these two theoretical studies are consistent with the recent observational studies. Metallicity differences among Cepheids in the SNe Ia host galaxies and anchor galaxies of the distance ladder: the MW, NGC4258, the LMC, and the SMC are known to contribute $\sim 0.5\%$ to the total error budget in $H_{0}$ measurement \citep{ries22, breu24}. The inclusion of the $\gamma$ term has been shown to improve the accuracy of PL calibration in the cosmic distance ladder, resulting in consistent distances for different anchor galaxies in recent empirical studies \citep{ries22, bhar23, breu24}. Therefore, a careful investigation of metallicity effects is crucial to quantify the systematic uncertainties.  

Cepheids in anchor galaxies, the Milky Way (MW), LMC, and the SMC differ in their mean chemical abundances $(\rm [Fe/H])$. The MW Cepheids are significantly metal-rich as compared to their counterparts in the LMC and SMC \citep{roma22, breu22, bhar24, breu24}. The combination of these three galaxies provides a unique opportunity for investigating the effect of metallicity on the Cepheid PL relation. Most studies in the recent literature investigated the metallicity effect by relying on the mean-light PL relations of Cepheids \citep{gier18, breu21, breu22, bhar23, bhar24, tren24}. 

The classical PL relation connects the two global properties of Cepheids namely the pulsation period, and mean luminosity. The pulsation period of Cepheids is determined by their period-mean density relation \citep{sand58,kanb96}, while the mean luminosity reflects the star's energy output averaged over the pulsation period. These time-averaged properties are not affected by short-term physical effects due to pulsation. Normally, PL relations are formulated in terms of these global properties: the pulsation period and mean magnitudes. These mean magnitude PL relations have a metallicity dependence which is of interest. However, the relative location of the hydrogen ionization front (HIF) and stellar photosphere plays a key role in determining the magnitudes at a particular pulsation phase. The HIF and stellar photosphere are not always comoving during a pulsation period. Sometimes they are engaged, and at other times they are quite far apart in the mass distribution. Hence the mean magnitudes at a given period are an average over the phase of these detailed radiation hydrodynamic considerations.

In the earliest study establishing the PL relation, \citet{leavitt12} examined the SMC Cepheids at phases of maximum and minimum light to demonstrate the correlation between luminosity and pulsation period. More recent observational studies have also shown the existence of distinct PL relations at different pulsation phases of Cepheids, particularly near the phases of maximum and minimum brightness \citep[and references therein]{ngeo06, ngeo12, kanb09, bhuy24}. It is also supported by the results from the theoretical studies such as \citet{kurb23} based on the \textsc{mesa-rsp}\footnote{\url{https://docs.mesastar.org/en/latest/test_suite/rsp_Cepheid.html}} \citep{paxt19} pulsation models of Cepheids. The period-color (PC) relations of long-period (P>10 days) Cepheids at the phase of maximum light have flat slopes, first observed by \citet{code47}. Later \citet*{simon93} and subsequent studies \citep{kanb04, kanb04ii,kanb06, kanb07, kanb09} have shown how such properties result from the HIF-stellar photosphere interactions at the phase of maximum-light. Their separation is small at low temperatures, and the photosphere lies at the base of the HIF (hence they are engaged). At this phase, their temperature properties are similar to the temperature at which hydrogen ionizes. It leads to the flat PC slopes at the phase of maximum light. On the contrary, the HIF and photosphere are well separated at the phase of minimum light, and the photosphere temperature depends only on the global stellar properties. It leads to a sloped PC relation at minimum light. A similar explanation was provided for the flat PC 
relation of RR Lyraes \citep{kanb95, kanb96} and later extended to type II Cepheids 
\citep{das20}. Their results demonstrate that changes in physical conditions within the stellar envelope lead to distinct observational implications. This can affect the Cepheid PC relations as a function of phase and hence also the PL relations as a function of phase because these relations are just manifestations of the period-luminosity-color (PLC) relation \citep{kanb96}. However, these effects at phases near maximum light are diluted at mean light due to averaging over a full pulsation cycle. 

Cepheids with the same period may also show large differences in amplitude and color. This scatter in the PL relation is well documented and arises from several factors, including metallicity, evolutionary stage, and temperature differences within the instability strip \citep{bono99, sand04, roma08, free12, ande16, bhar15}. Metallicity, in particular, affects the envelope opacity and modifies the efficiency of energy transport, leading to variations in pulsation amplitude, light-curve shape, and intrinsic color at a fixed period \citep{bono99, marc05, roma08}. These effects manifest differently across the pulsation cycle, which is lost when averaging over a pulsation cycle. A phase-resolved study enables us to trace these metallicity-driven effects across the cycle, disentangle them from other sources of variation, and better quantify where the metallicity effects are strongest.

The availability of multi-wavelength evenly sampled light curves from various ground and space-based missions such as OGLE-IV \citep{sosz15} and Gaia DR3 \citep{ripe19}, therefore, provides the opportunity to investigate the metallicity effect on Cepheid PL relations and its dependence on Cepheid pulsation phases.

In this study, we determine the multiphase PL relations of Cepheids in the MW, LMC, and SMC. We then investigate the metallicity coefficients of period-luminosity-metallicity (PLZ) relations at multiple phases over a complete pulsation cycle of Cepheids. The paper is structured in the following manner. In Section \ref{sec:dm} we describe the data and methodology used in this study. Section \ref{sec:result} deals with the results obtained. Finally, in Section \ref{sec:conc} we provide the summary and conclusions of the present study. 

\section{Data \& Methodology} \label{sec:dm}
\subsection{Light curve data and sampling} \label{sec:dm1}
\begin{figure*}
    \begin{tabular}{c|c}
    \resizebox{0.48\linewidth}{!}{
    \includegraphics[width=0.44\linewidth, keepaspectratio]{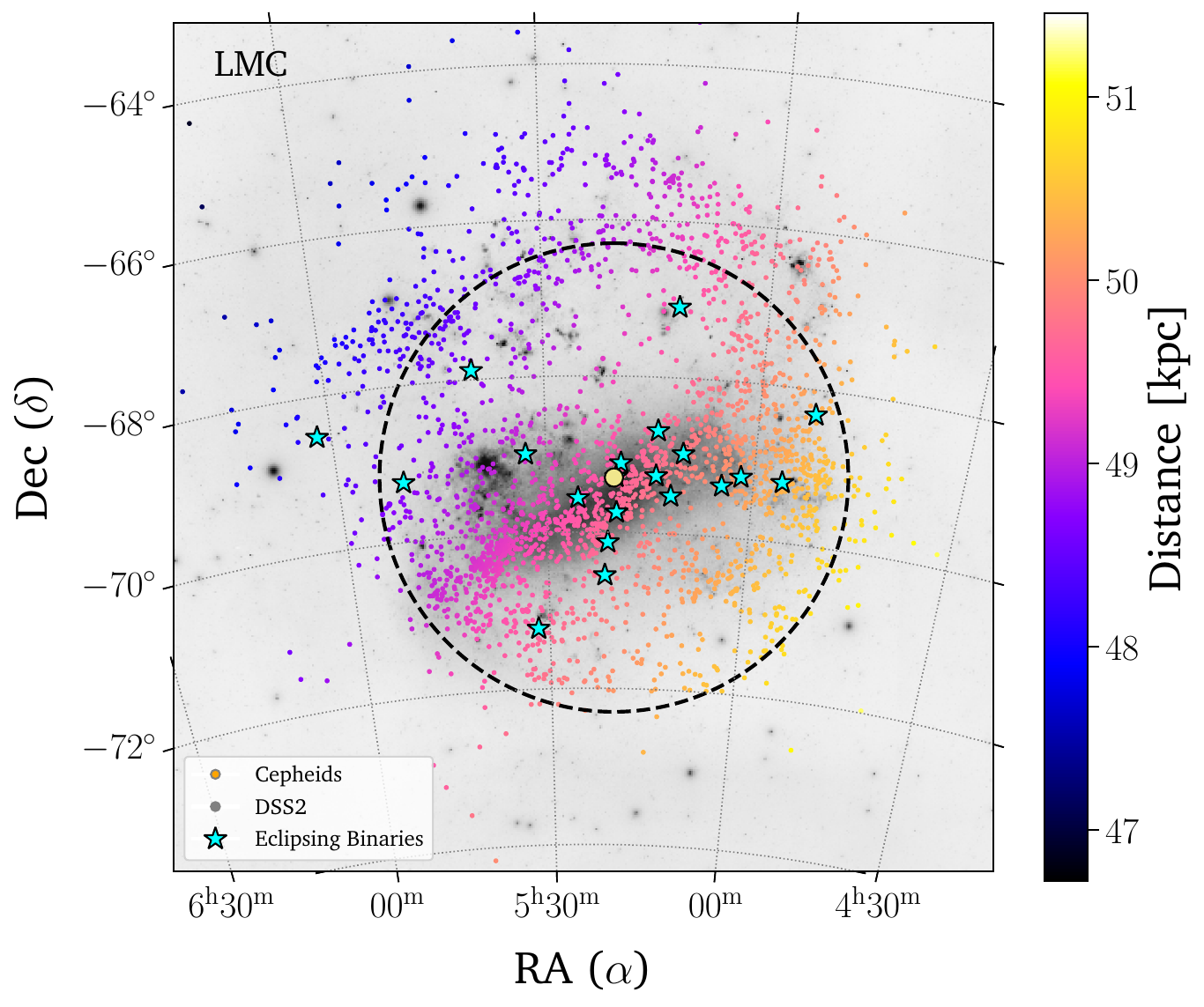}
    } 
    \resizebox{0.48\linewidth}{!}{
    \includegraphics[width=0.44\linewidth, keepaspectratio]{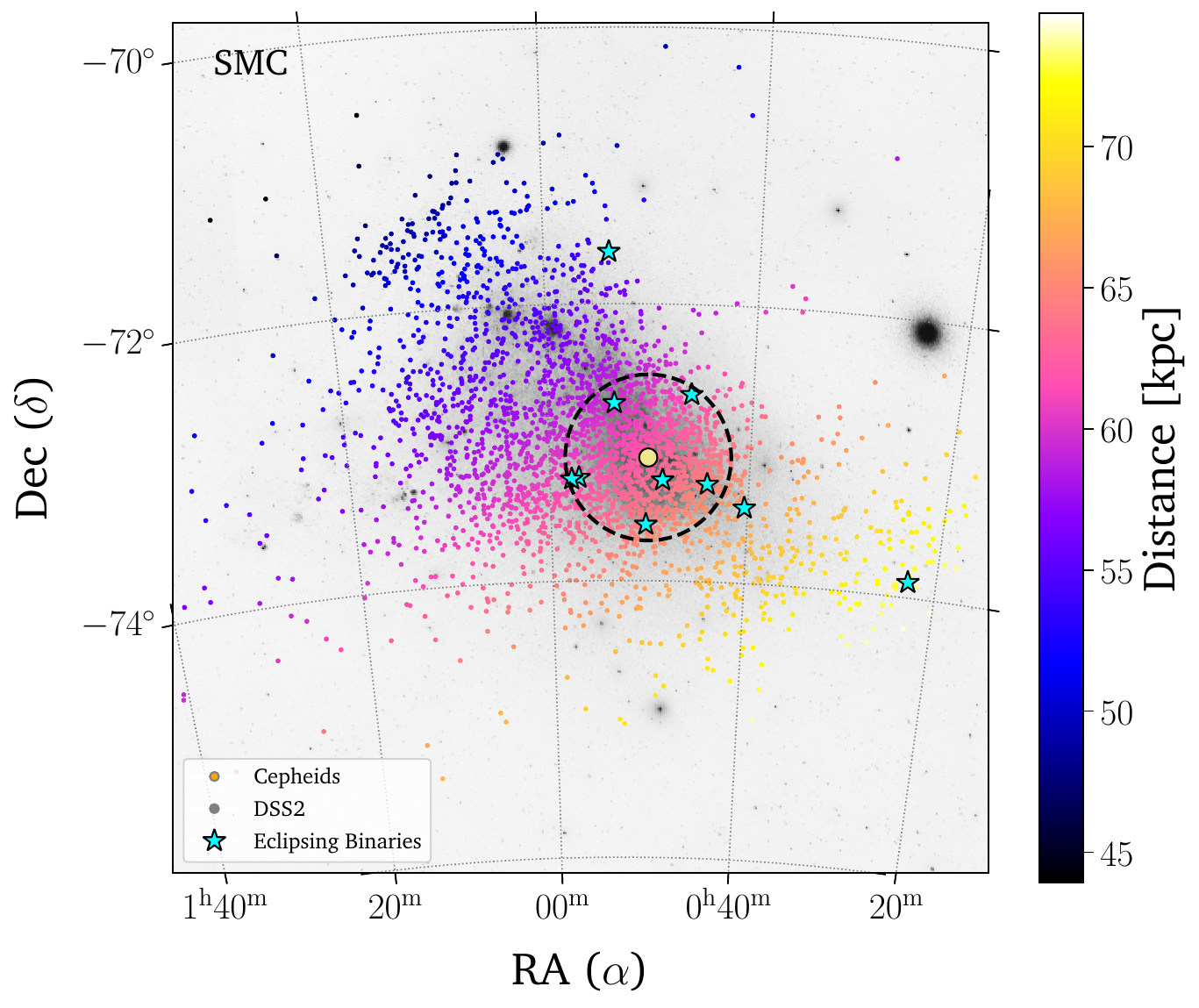}
    }
    \end{tabular}   
    \caption{Distribution of the sample of LMC (\textit{left panel}) and SMC (\textit{right panel}) Cepheids overplotted on gray-scale Digitized Sky Survey 2 (DSS2) blue-filter images of the LMC and SMC, respectively. These images are obtained using the Aladin virtual observatory tool. Cepheids plotted as colour-coded dots represent the sample before applying any selection criterion. The colour of the dots represents distances of individual Cepheids in both galaxies. Dashed circular boundaries in both panels denote the regions within $3^{\circ}$ and $0.6^{\circ}$ angular radii from the adopted LMC and SMC centres, respectively. The LMC and SMC centres are denoted by yellow coloured circles with RA-Dec coordinates: $(\alpha_{0, \rm LMC}, \delta_{0, \rm LMC}) = (80.05^{\circ}, -69.30^{\circ})$, and $(\alpha_{0, \rm SMC}, \delta_{0, \rm SMC}) = (12.54^{\circ}, -73.11^{\circ})$, respectively. Cyan coloured star symbols represent detached eclipsing binaries from \citet{pie19} and \citet{grac20}, respectively.}
    \label{fig:1}
\end{figure*}

\begin{figure*}
    \centering
    \resizebox{0.49\linewidth}{!}{
    \includegraphics[width=0.47\linewidth, keepaspectratio]{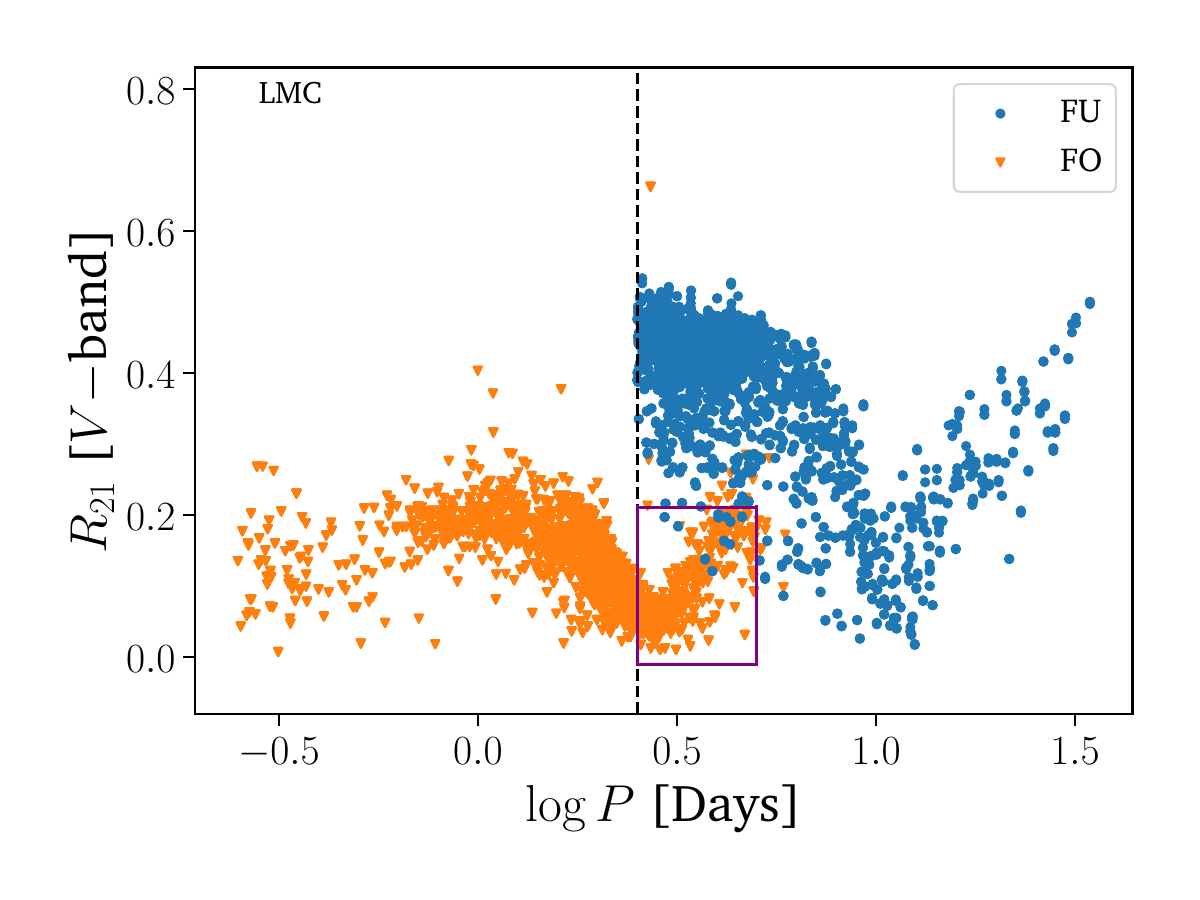 }
    } 
    \resizebox{0.49\linewidth}{!}{
    \includegraphics[width=0.47\linewidth, keepaspectratio]{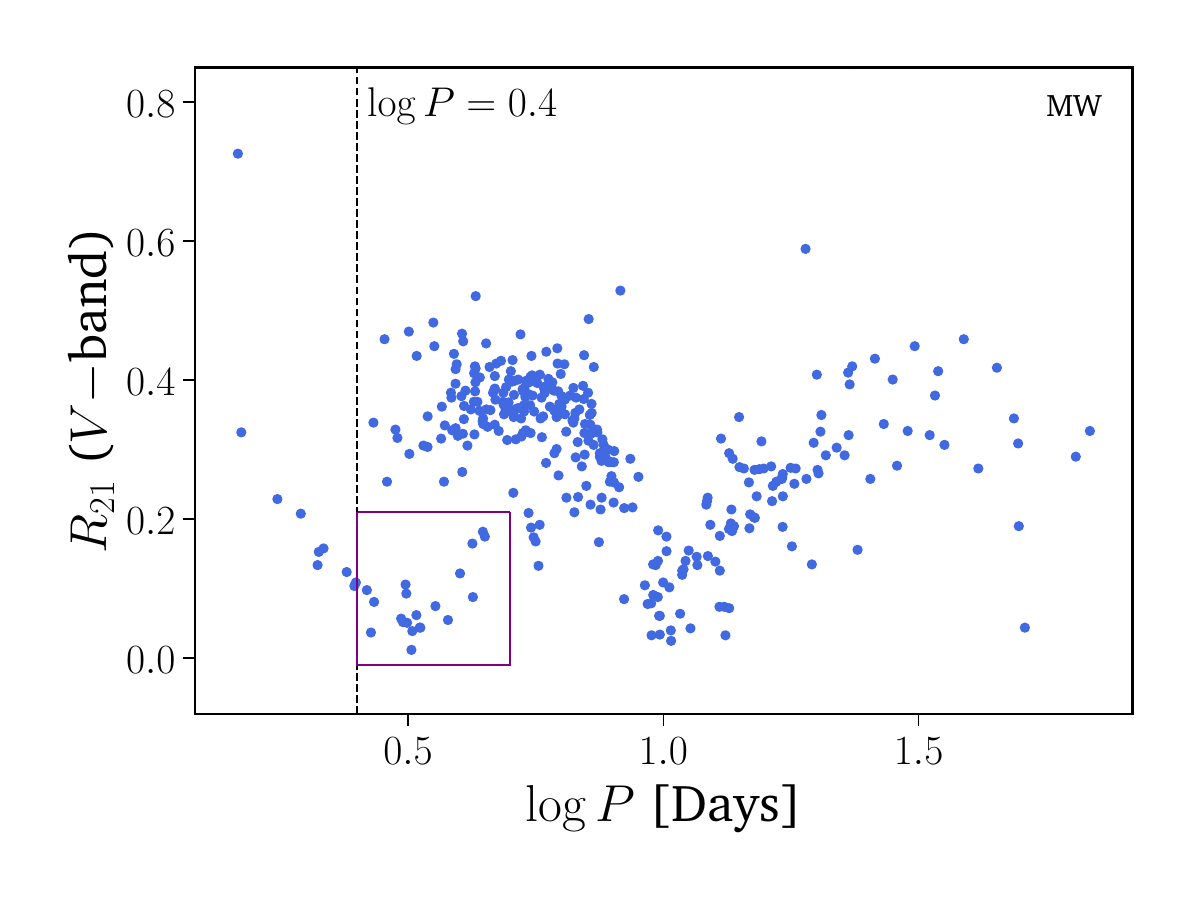}
    }
    \caption{Distribution of the Fourier parameter $R_{21}$ as a function of the logarithm of the pulsation period $(\log{P})$ for LMC and MW Cepheids in the $V$-band, respectively. LMC Cepheids are already classified by the OGLE survey, which are shown in the \textit{left panel}. The purple colored box in the \textit{both panels} represents the region in which the FO mode outliers are located in the period range $0.4 \leq \log{P} < 0.7$. MW Cepheids located inside this region in the $R_{21}-\log{P}$ plane in the \textit{right panel} are rejected from the sample. The dashed vertical line represents the $\log{P} =0.4$ period cut-off taken in sampling the Cepheids.}
    \label{fig:fou1}
\end{figure*}

\begin{figure*}
    \centering
    \includegraphics[width=1\linewidth]{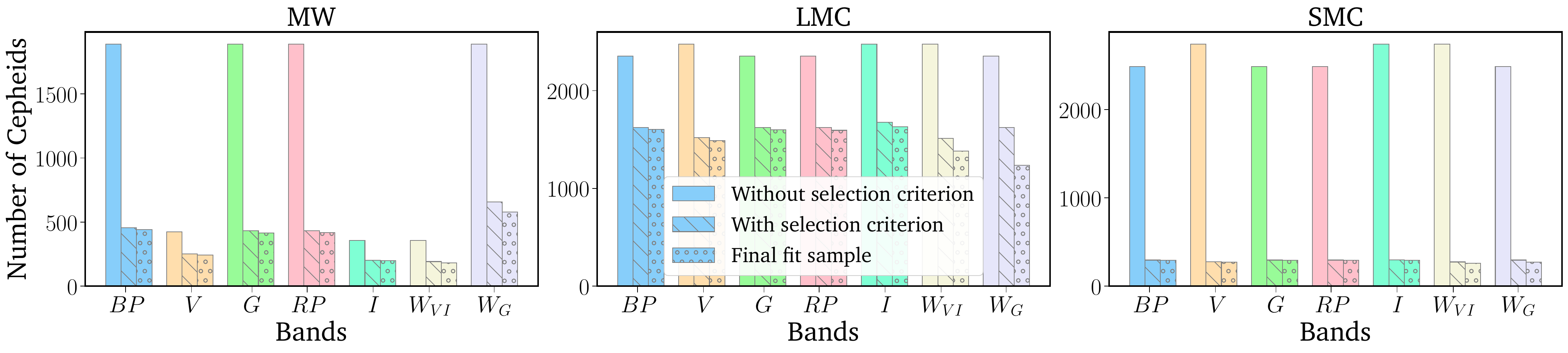}
    \caption{Number of Cepheids in the MW, LMC and SMC before (clear bar charts) and after (striped bar charts) the application of selection criterion in all five photometric bands: $G_{\rm BP}, V, G, G_{\rm RP}, I$ and two Wesenheit indices: $W_{VI}$ and $W_{G}$. The final numbers of Cepheids used for fitting the PL relations after removing the $3\sigma$ outliers are shown as dotted bar charts.}
    \label{fig:3}
\end{figure*}

We leverage the high-quality archival light curve data available for fundamental mode (FU) Cepheids in optical $V$, $I$, and Gaia $G$, $G_{\rm BP}$, and $G_{\rm RP}$ bands for this study. Nearly $15000$ stars recently available from the Gaia DR3 database have been classified as Cepheids by \citet{ripe23}. Based on the sky subregions as provided in \citet{ripe23}, light curve data for 1972, 2219, and 2398 FU mode Cepheids belonging to the MW, LMC, and SMC, respectively, are considered from the Gaia DR3 archival data. The photometric $V$- and $I$-band light curve data for 2477 and 2754 FU mode Cepheids in the LMC and SMC, respectively, are obtained from the OGLE-IV database \citep{sosz15}. The positions of Cepheids located in the LMC and SMC in equatorial coordinates are shown in Fig. \ref{fig:1}. The corresponding data for 422 Cepheids in the MW are available from \citet{bud08}. Furthermore, we utilize the extinction-free Wesenheit indices defined in these bands as: $W_{VI} = I - 1.55 (V-I)$ \citep{sosz15} and $W_{G} = G - 1.90 (G_{\rm BP} - G_{\rm RP})$ \citep{ripe19}, respectively. We employ various selection criteria for the appropriate identification of the sample of Cepheids suitable for this study. The selection criteria are as follows: 
\begin{enumerate}
\item The first selection criterion utilizes those light curves with at least 15 epochs of observation available in all five photometric bands for each galaxy. This is done to ensure evenly sampled light curves. 

\item We clean the MW Cepheid sample based on the renormalized unit weight error (RUWE) parameter available from Gaia DR3 and remove MW Cepheids having $\rm RUWE > 1.4$, as these objects are potentially astrometric binaries \citep{lind21a}.

\item Following \citet{breu22}, we have considered a minimum of $12\%$ uncertainty on the apparent magnitudes of MW Cepheids as a precision limit, applied uniformly to all photometric bands. It results in reduced scatter in the MW Cepheid PL relations. 

\item The LMC Cepheids located outside a radius of $3^{\circ}$ from the LMC centre as adopted from \citet{jacy16} with RA-Dec coordinates: $(\alpha_{0}, \delta_{0}) = (80.05^{\circ}, -69.30^{\circ})$, are removed from our sample. This spatial selection excludes Cepheids located far away from the detached eclipsing binaries (DEB) used by \citet{pie19} and from the LMC centre, so that their distances are less scattered from the mean LMC distance and reject those which are strongly affected by geometrical effects \citep{breu21, breu22}. The \textit{left panel} of Fig. \ref{fig:1} shows the distribution of LMC Cepheids available from OGLE-IV and Gaia DR3 databases, which are overplotted as colour-coded points on a gray-scale Digitized Sky Survey 2 (DSS2) blue-filter image of the LMC obtained from Aladin\footnote{\url{https://aladin.cds.unistra.fr/AladinLite/}} virtual observatory tool. The points are colour-coded based on individual distances obtained by applying geometrical corrections following \citet{breu21} as described in Section \ref{sec:dm3}. 

\item Similarly, we remove SMC Cepheids which are located beyond $0.6^{\circ}$ around the SMC centre with RA-Dec coordinates: $(\alpha_{0}, \delta_{0}) = (12.54^{\circ}, -73.11^{\circ})$ \citep{jacy16}. Furthermore, \citet{breu22} have shown that these sample cuts based on geometry result in a lower scatter in PL relations and more accurate PL intercepts and metallicity coefficients $(\gamma)$. The \textit{right panel} in Fig. \ref{fig:1} shows the distribution of SMC Cepheids available from OGLE-IV and Gaia DR3 databases as colour-coded points overplotted on a gray-scale DSS2 blue-filter image of the SMC obtained from Aladin.  

\item We further select Cepheids based on their pulsation periods. To avoid contamination from different mode pulsators which follow a different PL relation, we remove the Cepheids with $\log{P}< 0.4$. Furthermore, there may still exist some contamination of Cepheids from other pulsation modes, especially the first overtone (FO) mode at $\log{P}\geq 0.4$. We rely on the already available OGLE-IV \citep{sosz15} and Gaia DR3 \citep{ripe19} classifications to remove such stars from the LMC and SMC sample. On the other hand, such classifications are not available from \citet{bud08} for the MW sample. Therefore, any FO mode contamination in the MW Cepheid sample is removed based on the Fourier parameter, $R_{21}$. Here $R_{21} = \frac{A_2}{A_1}$ and $A_1$, $A_2$ are Fourier coefficients (Section \ref{sec:dm2}).  Fig. \ref{fig:fou1} shows the distribution of $R_{21}$ against $\log{P}$ of the LMC and MW Cepheids, respectively, in $V$-band. The LMC Cepheids, classified by the OGLE survey into FU- and FO-mode pulsators, follow two distinct trends in the $R_{21}$-$\log{P}$ plane as seen in the \textit{left panel} of Fig. \ref{fig:fou1}. The FO-mode outliers in the $R_{21}$-$\log{P}$ plane of the MW Cepheids \textit{(right panel)} can be visually identified by direct comparison with the corresponding figure for LMC Cepheids. This comparison indicates that the contamination from FO-mode Cepheids are confined to the region defined by $R_{21} \leq 0.22$ and $0.4 \leq \log{P} \leq 0.7$. This region is denoted by the purple colored rectangles on both panels. Therefore, we remove any MW Cepheids with $R_{21} \leq 0.22$ with pulsation periods in the range: $0.4 \leq \log{P} \leq 0.7$, which defines the boundary between the two different trends, vertically separated in the $R_{21}$ plane as seen in the figure.

Cepheids in the higher pulsating period range can contain ultra long-period pulsators (ULP), which can introduce non-linearity at $\log{P} > 2$ \citep[and references therein]{muse21}. Therefore, we also reject these Cepheids by applying a period cut at $\log{P} > 2$. 

\item  We also remove the $3\sigma$ outliers from the fitted PL relations to the sample of Cepheids in the MW, LMC and SMC. The number of Cepheids for different photometric bands in all three galaxies before and after applying the selection criteria, and finally those used in fitting the PL relations are shown as bar charts in Fig. \ref{fig:3}. The median of the number of Cepheids in the MW, LMC, and SMC available in each photometric band after applying the selection criteria are presented in Table \ref{tab:1}, in different period ranges: $0.4 \leq \log{P} < 2$, $0.4 \leq \log{P} < 1$, and $1 \leq \log{P} < 2$, respectively.    
\end{enumerate}

\begin{table*}
    \centering
    \caption{Median number of Cepheids in the MW, LMC, and SMC available in each photometric band after the application of sampling criteria and $3\sigma$ clipping performed across all 50 pulsation phases. The number of Cepheids in the period ranges: $0.4 \leq \log{P} < 2$, $0.4 \leq \log{P} < 1$, and $1 \leq \log{P} < 2$, respectively, are also presented separately.}
    \begin{tabular}{ccccccccc}
    \hline 
    \hline 
    Period Range & Cepheid & \multicolumn{7}{c}{Number of Cepheids} \\
    & sample & & & & & & \\
    \hline 
    & & $G_{\rm BP}$ & $V$ & $G$ & $G_{\rm RP}$ & $I$ & $W_{VI}$ & $W_G$ \\
    \hline
    \multirow{3}{*}{$0.4 \leq \log{P} < 2$} & MW & 381 & 236 & 368 & 378 & 186 & 186 & 592 \\
     & LMC & 1579 & 1449 & 1545 & 1563 & 1604 & 1397 & 1369 \\
    & SMC & 290 & 270 & 291 & 289 & 290 & 256 & 273 \\
    \hline
    \multirow{3}{*}{$0.4 \leq \log{P} < 1$} & MW & 267 & 158 & 252 & 256 & 120 & 114 & 414 \\
     & LMC & 1432 & 1330 & 1409 & 1425 & 1485 & 1314 & 1253 \\
    & SMC & 244 & 230 & 245 & 244 & 249 & 217 & 229 \\
    \hline
    \multirow{3}{*}{$1 \leq \log{P} < 2$} & MW & 113 & 81 & 118 & 125 & 67 & 69 & 179 \\
     & LMC & 139 & 111 & 139 & 138 & 116 & 98 & 125 \\
    & SMC & 46 & 40 & 46 & 46 & 41 & 39 & 44 \\
    \hline
    \hline
    \end{tabular}
    \label{tab:1}
\end{table*}

\subsection{Light curve analysis} \label{sec:dm2}
The phased Cepheid light curves are obtained using the following relation \citep{deb09}:
\begin{align}
\Phi & = \frac{t-t_{0}}{P} - \textrm{Int} \left(\frac{t-t_{0}}{P}\right),
\label{eq:2}
\end{align}
where $P$ is the Cepheid pulsation period in days, $t$ the time of observation, $t_{0}$ the epoch of maximum brightness, and $\textrm{Int}$ the integer part. 

\begin{figure}
    \centering
    \includegraphics[width=1.\linewidth, keepaspectratio]{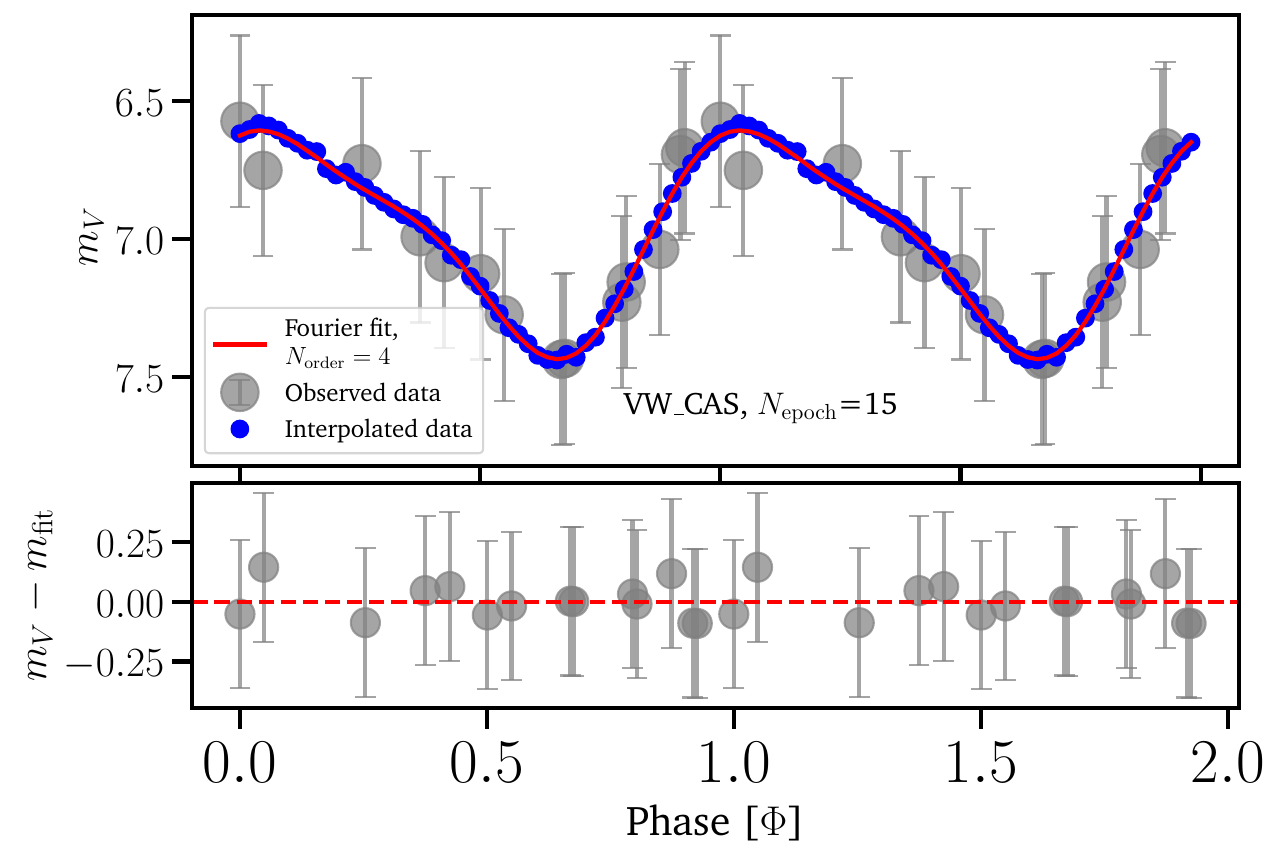}
    \caption{ \textit{Top panel} shows the Fourier fitted $V$-band light curve of the MW Cepheid VW-CAS and the \textit{bottom panel} shows residuals of the Fourier fit. The light curve has 15 epoch of observed data and is fitted well with a Fourier series of order $N=4$, represented by the red coloured solid line. The blue coloured points represent the interpolated points.}
    \label{fig:fourier_fit}
\end{figure}

The Fourier decomposition technique was put forward by \citet{simon81} to study light curve properties of Cepheids. It involves fitting the phased light curves with a Fourier series function of the form :
\begin{align}
m(t) & = A_{0} + \sum_{i=1}^{N} A_{i} \cos{(i\omega(t-t_{0}) + \phi_{i})},
\label{eq:3}
\end{align}
where $m(t)$ is the observed magnitude at the observation time $t$ and $\omega = 2\pi/P$. The quantities $A_{0}, A_{1}, \dots A_{N}$ and $\phi_{1}, \dots \phi_{N}$ are the Fourier coefficients that describe the light curve structure and can be used for comparison between theoretically computed light curves using pulsation models and observed light curves \citep[ and references therein]{kurb23}. Here $N$ represents the optimal order of the Fourier series function fitted to the observed light curve and is obtained using Baart's criterion \citep{peter86}. Fig. \ref{fig:fourier_fit} shows a Fourier fitted light-curve of a MW Cepheid, VW-CAS, with the residual plot in the bottom panel. VW-CAS has 15 measurements of observed data across one full pulsation cycle, and the light curve is well fitted with a Fourier series function of order $N=4$. Apparent magnitudes of Cepheids at 50 different phase points are then obtained by interpolating the observed light curve using the Fourier fit coefficients. 

\subsection{Cepheid Distances} \label{sec:dm3}
We have adopted the photogeometric distances of MW Cepheids available from \citet{bjon21}. The photogeometric distances have been found to be more accurate and precise than those based only on parallaxes, specifically for stars with poor parallax determinations \citep{bjon21}. These distances also include the parallax zero-point offset correction by \citet{lind21b}. However, \citet{ries21} has found the zero-point offset in Gaia DR3 parallax measurements of MW Cepheids to be overestimated by $\sim14~\mu{\rm as}$, and thus has to be taken into account in determining the individual distances of the MW Cepheids. To a good approximation, this can be achieved by adding a small additional correction term: $dr = -r^{2} d\varpi$, where $r$ denotes original \citet{bjon21} distances and $d\varpi = -0.014 $ milli-arcseconds \citep[see Section 2.4 of ][]{breu22}. 

The geometric distance to the LMC is available from \citet{pie19} based on detached eclipsing binaries (DEB): $49.59 \pm 0.09 {\rm (stat.)} \pm 0.54 {\rm (syst.)}$ kpc. We adopt this to be the mean LMC distance in our analysis. Earlier studies have revealed the planar geometry of the LMC with an inclined and rotating star-forming disk \citep{mare01, mare02, subrs10}. Cepheids in the LMC are distributed over its disk, which also traces its geometry as well as its orientation \citep{jacy16, deb18, bhuy24}. Due to the inclination of the LMC disk, some Cepheids are apparently closer or farther than the average LMC distance, depending upon their projected line of nodes \citep{ries19}. Therefore, precise distances to individual Cepheids in the LMC disk are determined by applying geometric corrections \citep[and references therein]{breu22}.       

Conversely, the SMC has an irregular morphology, with an extended bar and a wing region on the eastern side \citep{jacy16, deb19}. Cepheids in the SMC have been found to trace an elongated structure along the line of sight \citep[and references therein]{subrs15, ripe17}. Therefore, it induces significant scatter in distances to individual Cepheids with respect to the mean SMC distance: $62.44 \pm 0.44 ({\rm stat.}) \pm 0.81 {\rm (syst.)}$ kpc obtained by \citet{grac20} using DEBs. Hence geometrical corrections are also needed to apply to the SMC Cepheids. For a detailed description of geometrical corrections applied to Cepheids in the SMC and the LMC, interested readers are redirected to \citet{breu24} and references therein. 

\subsection{Extinction corrections} \label{sec:extinction}
Multiple reddening maps are available in the literature for the Milky Way, e.g., \citet[][SFD map]{schl98} or the 3D Dust maps constructed by \citet{lall19} and \citet[\texttt{\large Bayestar19}\footnote{\url{http://argonaut.skymaps.info/usage}}, available through \texttt{\large dustmaps}\footnote{\url{https://dustmaps.readthedocs.io/en/latest/}}]{green19}. There are also direct methods available to estimate the reddening values of Cepheids in the MW based on period-colour (PC) relations \citep{sand04, ripe21} or from the global fit of Cepheid pulsation using the spectro-photo-interferometry of pulsating stars (SPIPS) algorithm \citep{trah21}. Recently, \cite{bhar24} showed that using the \citet{ripe21} PC relation to derive the reddening of MW Cepheids results in lower PL scatter in comparison to using other available methods and estimates. Therefore, for the MW Cepheid sample, we determine reddening values by deriving their intrinsic colours $(V-I)_{0}$ based on the \citet{ripe21} PC relations. For the $V$- and $I$-band samples, observed colours $(V-I)$ are determined from the apparent magnitudes. We use the photometric transformations from \citet{panc22} to transform the magnitudes of the \textit{Gaia} band sample and derive the $(V-I)$ observed colours for them. 

Extinction corrections to the apparent magnitudes of Cepheids in the LMC and SMC are applied by utilizing the \citet{skow21} reddening maps. Making use of the relation $E(V-I) = 1.237 \times E(B-V)$ from \citet{skow21}, the $E(V-I)$ values are converted into $E(B-V)$ values. The extinction amplitudes are calculated using the following relation:
\begin{align}
A_{\lambda} & = R_{\lambda} E(B-V). \label{eq:4}
\end{align}
Here $A_{\lambda}$ and $R_{\lambda}$ denote extinction amplitude and ratios of total to selective absorption, respectively. The $R_{\lambda}$ values are determined using \citet{fitz99} reddening law with $R_{V}$ fixed to $3.1$. We utilize the \texttt{\large dust\_extinction}\footnote{\url{https://dust-extinction.readthedocs.io/en/latest/}} \textsc{Python} package to determine the extinction amplitude ratios: $A_{\lambda}/A_{V}$ and the $R_{\lambda}$ values. We have adopted the effective central wavelengths corresponding to the different bands from the Spanish Virtual Observatory (SVO) filter profile service\footnote{\url{http://svo2.cab.inta-csic.es/theory/fps/}}. The $A_{\lambda}/A_{V}$ and $R_{\lambda}$ values corresponding to each photometric band are listed in Table \ref{tab:2}.
\begin{table*}
    \centering
    \caption{The extinction amplitude ratios $(A_{\lambda}/A_{V})$, ratio of total to selective absorption $(R_{\lambda})$ in different photometric bands using \citet{fitz99} reddening law assuming $R_{V} = 3.1$ and the mean metallicity values of the MW Cepheid sample as well as their standard deviations within different period ranges calculated based on the individual [Fe/H] values available from \citet{geno14, geno15}. The effective central wavelengths of the photometric bands are adopted from the SVO filter profile service. $R_{\lambda}$ values for the Wesenheit indices ($W_{VI}$ and $W_{G}$) are adopted from the literature \citep{sosz15, ripe19}.}    
    \begin{tabular}{ccccccc}
    \hline
    \hline
    Bands & $\lambda_{\rm eff}$ & $A_{\lambda}/A_{V}$ & $R_{\lambda}$ & \multicolumn{3}{c}{${\rm \langle[Fe/H]\rangle_{MW}}$ (dex)} \\                        % & $\sigma_{\rm [Fe/H]}$ \\
    \hline
    & $(\micron)$ & & & $0.4 \leq \log{P} < 2$ & $0.4 \leq \log{P} < 1$ & $1 \leq \log{P} < 2$ \\
    \hline
    $BP$ & 0.5036 & 1.110 & 3.442 & $0.074 \pm 0.107$ & $0.068 \pm 0.097$ & $0.130 \pm 0.158$\\
    $V$ & 0.5468 & 0.986 & 3.056 & $0.090 \pm 0.114$ & $0.080 \pm 0.088$ & $0.168 \pm 0.138$\\
    $G$ & 0.5822 & 0.904 & 2.802 & $0.078 \pm 0.097$ & $0.070 \pm 0.092$ & $0.125 \pm 0.132$\\
    $RP$ & 0.762 & 0.600 & 1.859 & $0.075 \pm 0.102$ & $0.072 \pm 0.090$ & $0.133 \pm 0.158$\\
    $I$ & 0.7829 & 0.573 & 1.777 & $0.090 \pm 0.119$ & $0.075 \pm 0.089$ & $0.187 \pm 0.142$\\
    $W_{VI}$ & \dots & \dots & 1.550 & $0.090 \pm 0.116$ & $0.077 \pm 0.077$ & $0.169 \pm 0.169$\\
    $W_{G}$ & \dots & \dots & 1.900 & $0.075 \pm 0.105$ & $0.071 \pm 0.071$ & $0.131 \pm 0.131$\\ 
    \hline
    \hline
    \end{tabular}
    \label{tab:2}
\end{table*}

\subsection{Multiphase Period-Luminosity-Metallicity (PLZ) Relations}
In general, the Cepheid PL relation or the Leavitt law \citep{leavitt12} in any given photometric band can be mathematically represented as:
\begin{align}
M_{\lambda} & =  \alpha_{\lambda} (\log{P} - \log{P_{0}}) + \beta_{\lambda}. \label{eq:5}
\end{align}
Here $\alpha_{\lambda}$ and $\beta_{\lambda}$ denote the slope and intercept, respectively, $P_{0}$ represents the pivot period (in general, the mean period of the Cepheid sample), $M_{\lambda}$ is the absolute magnitude of Cepheids, which is determined from the apparent magnitude $(m_{\lambda})$, distance $(d)$, and reddening $E(B-V)$ using the following relation:
\begin{align}
M_{\lambda} & =  m_{\lambda} - R_{\lambda} E(B-V) - 5 \log{d} - 10.
\label{eq:6}
\end{align}
Here the distances $d$ of individual Cepheids are in kiloparsecs. The observed magnitudes of Cepheids change over a complete pulsation cycle. Therefore, the PL relation can be considered as a function of Cepheid pulsation phases, although it is usually calibrated with intensity-averaged mean magnitudes. Hence, the Leavitt law for the $i^{\rm th}$ Cepheid in its $j^{\rm th}$ pulsation phase in a photometric band $\lambda$ can be defined as:
\begin{align}
M_{\lambda, i, j} = \alpha_{\lambda, j} & (\log{P_{i}} - \log{P_{0}}) + \beta_{\lambda, j}
% + \nonumber \\ 
                        % & R_{\lambda} E(B-V)_{i} + 5 \log{d_{i}} + 10. 
\label{eq:7}
\end{align}
The PL relations defined by Equation \eqref{eq:7} are fitted to the absolute magnitudes of Cepheids in the MW, LMC and SMC at 50 different phases over a complete pulsation cycle. In our analysis, we adopt a pivot period of $\log{P_{0}} = 0.7$ from \citet{breu22}, which is very close to the $\log{P_0}$ found for our sample. We employ a Monte Carlo algorithm with 10,000 iterations to determine the optimized values of the PL parameters $\alpha_{\lambda, j}$ and $\beta_{\lambda, j}$, along with their corresponding uncertainties. In each iteration, the absolute magnitudes, distances, and $E(B-V)$ values are allowed to vary within their uncertainties to obtain a robustly fitted set of the PL parameters.  

Determination of multiphase period-luminosity-metallicity (PLZ) relations of Cepheids involves a two-step process. The PL slope parameters $(\alpha_{\lambda,j})$ are fixed to those of the LMC PL relations at each phase because of the large sample size and low PL dispersion in the LMC. Multiphase PL relations are then fitted to the absolute magnitudes of Cepheids in all three galaxies to obtain the corresponding intercept parameters $(\beta_{\lambda, j})$. We employ the Markov-chain Monte Carlo (MCMC) based maximum-likelihood estimation technique using the \texttt{emcee}\footnote{\url{https://emcee.readthedocs.io/en/stable/}} \citep{fore13} \textsc{Python} package. To determine the metallicity effect on PL relation at each pulsation phase, the PL intercept parameters of the three galaxies are fitted with the following linear function:
\begin{align}
\beta_{\lambda, j} & = \gamma_{\lambda, j} {\rm \langle[Fe/H]\rangle} + \delta_{\lambda,j}. \label{eq:8}
\end{align}
Here $\gamma_{\lambda, j}$ denotes the coefficient of metallicity effect, which is considered as a function of pulsation phase as well as the photometric band, $\delta_{\lambda,j}$ is the corresponding intercept of the fit, and ${\rm \langle[Fe/H]\rangle}$ are the mean metallicity of each galaxy. 

The metal abundances $({\rm [Fe/H]})$ of the MW Cepheids adopted in our analysis are taken from \citet{geno14} and \citet{geno15}. For Cepheids common to both these catalogues, we use the updated ${\rm [Fe/H]}$ values available from \citet{geno15}. The mean metallicity values $({\rm \langle[Fe/H]\rangle})$ of MW Cepheids are calculated by fitting a Gaussian to the distribution of individual [Fe/H] values in different bands and in different period ranges, and are provided in Table \ref{tab:2}. The ${\rm \langle[Fe/H]\rangle}$ values for different photometric bands and period ranges slightly differ from each other based on the number of available Cepheids. Furthermore, the ${\rm \langle[Fe/H]\rangle}$ values for Cepheids in the LMC and SMC, ($-0.409 \pm 0.076$) dex and ($ -0.750 \pm 0.050$) dex, respectively, are taken from \citet{roma22} and \citet{gier18}. 

\subsection{Error propagation}
The Cepheid PL relations have a finite width due to their period-luminosity-color relation. It also appears as the finite width of the instability strip (WIS). It is a source of additional scatter in the PL relation. We adopt the instability strip widths $0.21$, $0.15$ and $0.077$ mag, respectively for $V$-, $I$- and $W_{VI}$-band from \citet{sosz15}. Furthermore, in the Gaia $G_{\rm BP}$, $G$, $G_{\rm RP}$ bands and $W_{G}$ Wesenheit index the WIS are adopted as: $0.23$, $0.19$, $0.16$, and $0.10$ mag, respectively based on the PL dispersion of the LMC FU mode Cepheids obtained by \citet{ripe19}. They are quadratically added to the uncertainty in apparent magnitudes in the respective photometric bands and Wesenheit indices. 

The photometric data in $V$- and $I$-band of the MW Cepheids obtained by \citet{bud08} and those in the Magellanic Clouds were obtained by \citet{sosz15} using two different observational surveys based on different instruments. Therefore, we adopt a systematic uncertainty of $0.02$ mag between the two observations. However, photometric data in the Gaia bands ($G_{\rm BP}$, $G$, $G_{\rm RP}$) of all the Cepheids are available from the same instrument and hence no systematic uncertainty is adopted. We propagate the systematic uncertainties to the PL intercept errors in our analysis.    

\section{Results} \label{sec:result}
In this section, we present the results of the metallicity effect on Cepheid PL relations based on the pulsation phase-dependent properties of Cepheids, leveraging publicly available multi-band $(V, G_{\rm BP}, G, I$ and $G_{\rm RP})$ optical Cepheid light curves. The observed magnitudes of Cepheids at 50 pulsation phases over a complete pulsation cycle are fitted with Equation \eqref{eq:7} by employing the Monte Carlo technique with 10,000 iterations.

\subsection{Multiphase PL relations} \label{section:pl_result}
\begin{figure}
    \centering
    \includegraphics[width=\linewidth, keepaspectratio]{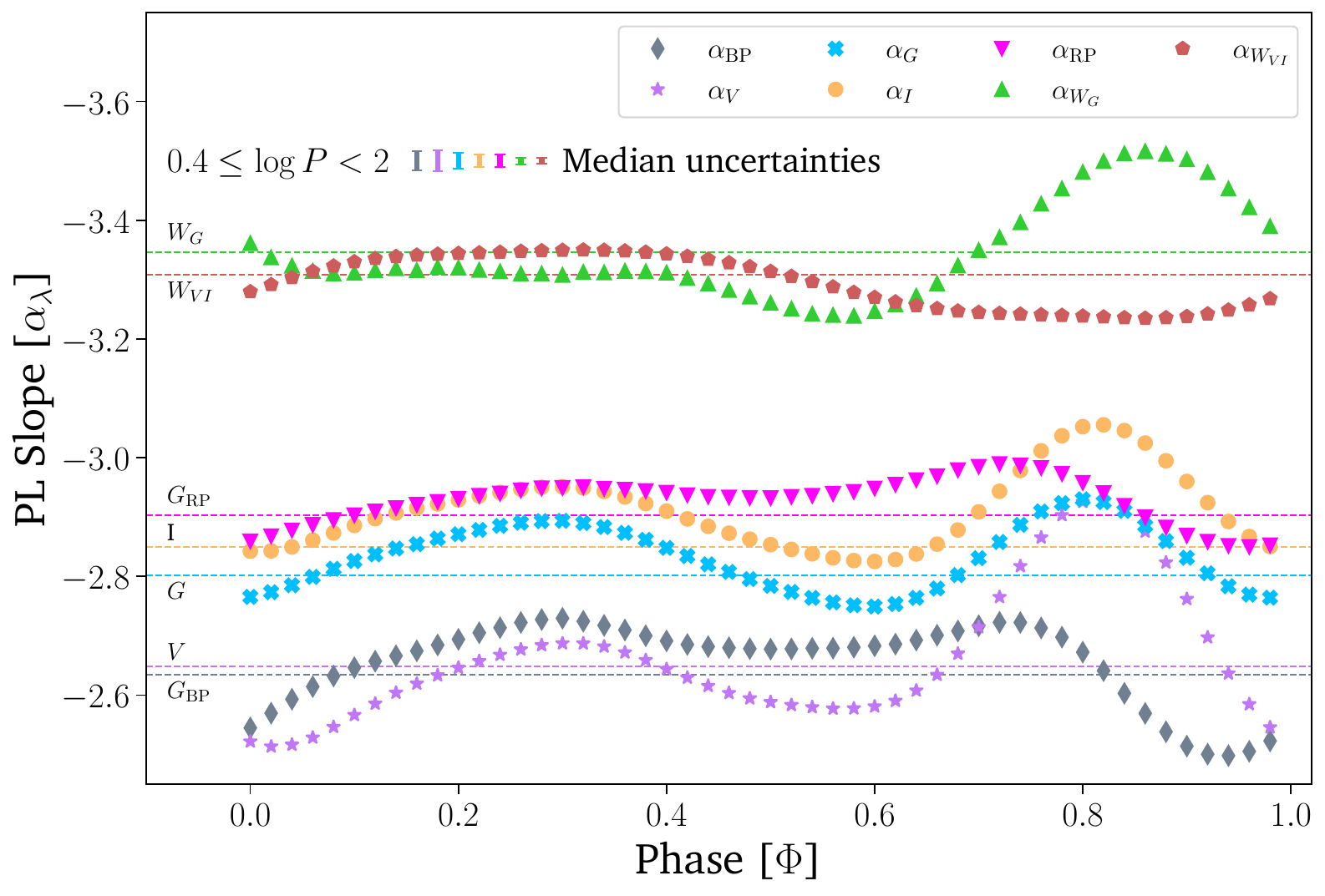}
    \caption{PL slopes of Cepheids in the LMC as a function of pulsation phases in five photometric bands: $G_{\rm BP}$, $V$, $G$, $G_{\rm RP}$, $I$, and two Wesenheit indices: $W_{VI}$ and $W_{G}$, respectively. Pulsation periods of the Cepheid sample used to determine the multiphase PL slopes are within the range: $0.4 \leq \log{P} < 2$. The dashed horizontal lines overplotted on the figure represent the PL slopes obtained using mean magnitudes in respective bands. The median uncertainties denote the typical uncertainties of the multiphase PL slopes.}
    \label{fig:4}
\end{figure}

\begin{figure}
    \centering
    \includegraphics[width=\linewidth, keepaspectratio]{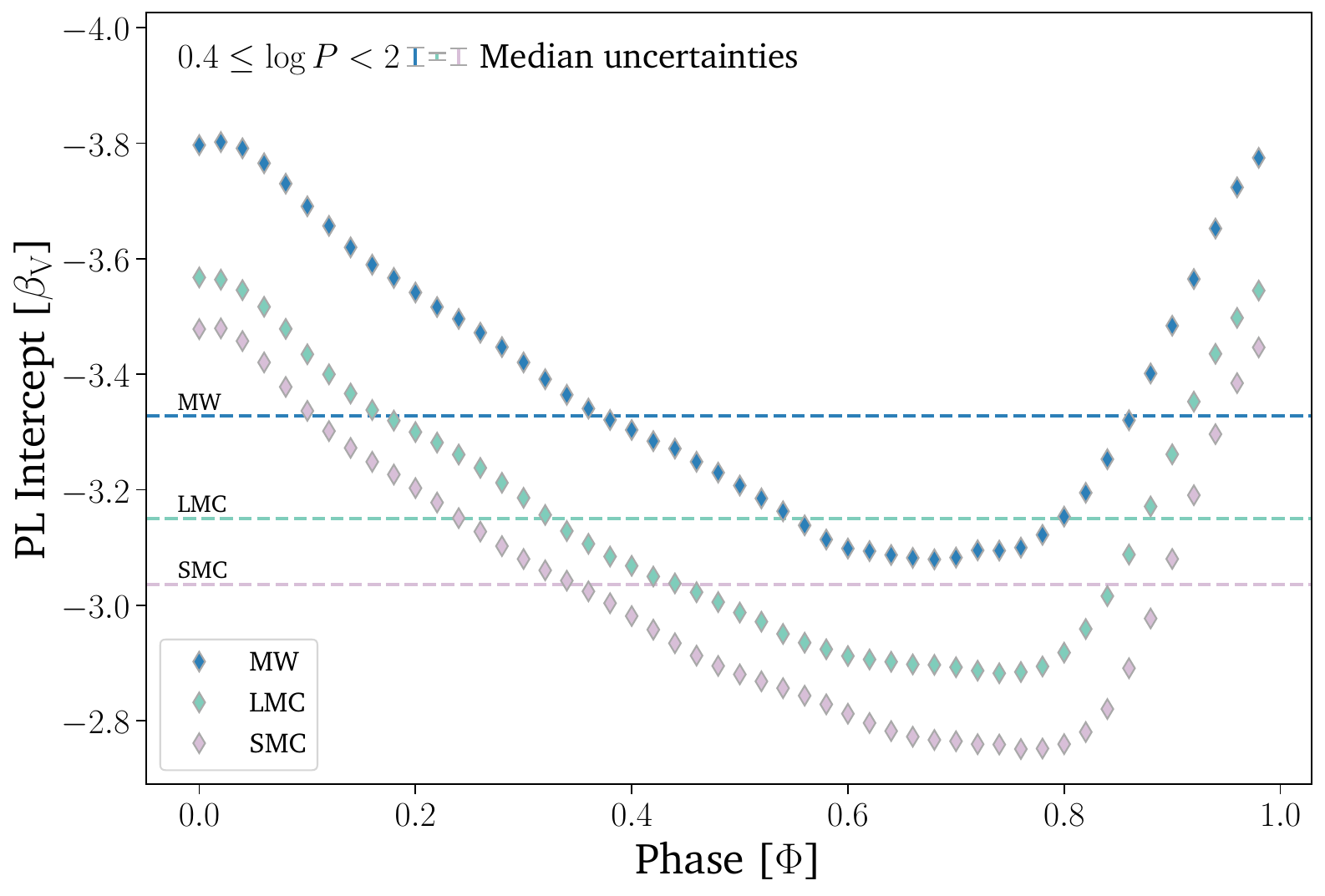}
    \caption{PL intercepts of Cepheids in the MW, LMC, and SMC as a function of pulsation phases in the photometric band $V$. The PL intercepts are determined assuming common PL slopes for Cepheids in the MW, LMC, and the SMC at the corresponding pulsation phases, and equal to those of LMC Cepheids. Pulsation periods of the Cepheid sample are in the range: $0.4 \leq \log{P} < 2$. The dashed horizontal lines in the figure represent the PL intercepts obtained using mean magnitudes of Cepheids for different galaxies. Median uncertainties here represent the typical uncertainties of multiphase PL intercepts.}
    \label{fig:5}
\end{figure}

The multiphase PL intercepts of Cepheids in the MW, LMC, and SMC are determined by keeping the slopes at corresponding phases fixed to those of the LMC Cepheids. Fig. \ref{fig:4} shows the multiphase PL slopes of the LMC Cepheids in the photometric bands: $V, I, G, G_{\rm BP}$ and $G_{\rm RP}$ and two extinction free Wesenheit indices: $W_{VI}$ and $W_{G}$, as functions of pulsation phases. Pulsation periods of the Cepheids used to determine the multiphase PL relations are in the range $0.4 \leq \log{P} < 2$ (all periods). The resulting PL intercept values for the MW, LMC, and SMC Cepheids in the $V$-band are shown in Fig. \ref{fig:5}. For other photometric bands, the results are shown in Fig. \ref{fig:a1}. We find that the PL slope and intercept values vary dynamically over a complete pulsation cycle, consistent with the results of earlier studies based on phase-dependent properties of Cepheids \citep{ngeo12, kurb23, bhuy24}. However, the PL slope values in the Wesenheit $W_{VI}$ band exhibit smaller amplitudes of phase dependence in comparison to all the other bands. The median uncertainties denote the typical uncertainties of multiphase PL slope and intercept values. We find that the median uncertainties in PL intercept values of MW Cepheids are higher than those in the LMC and SMC in the $G_{\rm BP}$, $V$, $G$, $G_{\rm RP}$, and $I$ band. However, in the $W_{VI}$ and $W_{G}$ Wesenheit indices, the MW intercept values have the lowest median uncertainties. It can be attributed to the statistically smaller size of the Milky Way sample as well as to the higher amount of reddening and larger reddening uncertainty of the Milky Way Cepheids compared to those in the LMC and SMC. The amplitudes of variation of PL slopes in all photometric bands and different period ranges are provided in Table \ref{tab:3}. We find that the PL slopes display most of the variations in the $V$ band in comparison to other bands, and the variation is least in $G_{\rm RP}$- and $W_{VI}$-band.

\begin{table*}
\centering
\caption{Amplitudes of variation and median uncertainties of the multiphase PL slopes in five photometric bands: $G_{\rm BP}$, $V$, $G$, $G_{\rm RP}$, $I$, and two Wesenheit indices: $W_{VI}$ and $W_{G}$, respectively. These values are determined with Cepheids in different period ranges, as shown in the table.}
\begin{threeparttable}
\begin{tabular}{ccccccccc}
\hline
\hline
Bands & $\lambda_{\rm eff}$ $(\micron)$ & \multicolumn{3}{c}{$^{\rm a}\Delta\alpha_{\lambda}$ (mag/dex) } & \multicolumn{3}{c}{ Median uncertainties $\left(\sigma_{\alpha_{\lambda}}^{\rm median} \right)$ (mag/dex)} \\
\hline
&  & $0.4 \leq \log{P} < 2$ & $0.4 \leq \log{P} < 1$ & $1 \leq \log{P} < 2$ & $0.4 \leq \log{P} < 2$ & $0.4 \leq \log{P} < 1$ & $1 \leq \log{P} < 2$ \\
\hline
 $BP$ & 0.5036 & 0.231 & 0.346 & 0.461 & 0.016 & 0.029 & 0.065 \\
 $V$ & 0.5468 & 0.415 & 0.852 & 1.496 & 0.018 & 0.028 & 0.097 \\
 $G$ & 0.5822 & 0.181 & 0.593 & 0.815 & 0.014 & 0.024 & 0.054 \\
 $RP$ & 0.7620 & 0.140 & 0.280 & 0.359 & 0.010 & 0.018 & 0.042 \\
 $I$ & 0.7829 & 0.231 & 0.485 & 1.221 & 0.011 & 0.016 & 0.067 \\
 $W_{VI}$ & \dots & 0.115 & 0.126 & 0.846 & 0.005 & 0.008 & 0.031 \\
 $W_{G}$ & \dots & 0.277 & 0.530 & 0.664 & 0.006 & 0.009 & 0.023 \\
\hline 
\hline
\end{tabular}
\begin{tablenotes}
    \item $^{\rm a}\Delta\alpha_{\lambda} = \left|{\rm min(\alpha_{\lambda}(\Phi))} - {\rm max(\alpha_{\lambda}(\Phi))}\right|$
\end{tablenotes}
\end{threeparttable}
\label{tab:3}
\end{table*}

\begin{figure*}
    \begin{tabular}{c|c}
        \resizebox{0.48\linewidth}{!}{
        \includegraphics[width=0.44\linewidth, keepaspectratio]{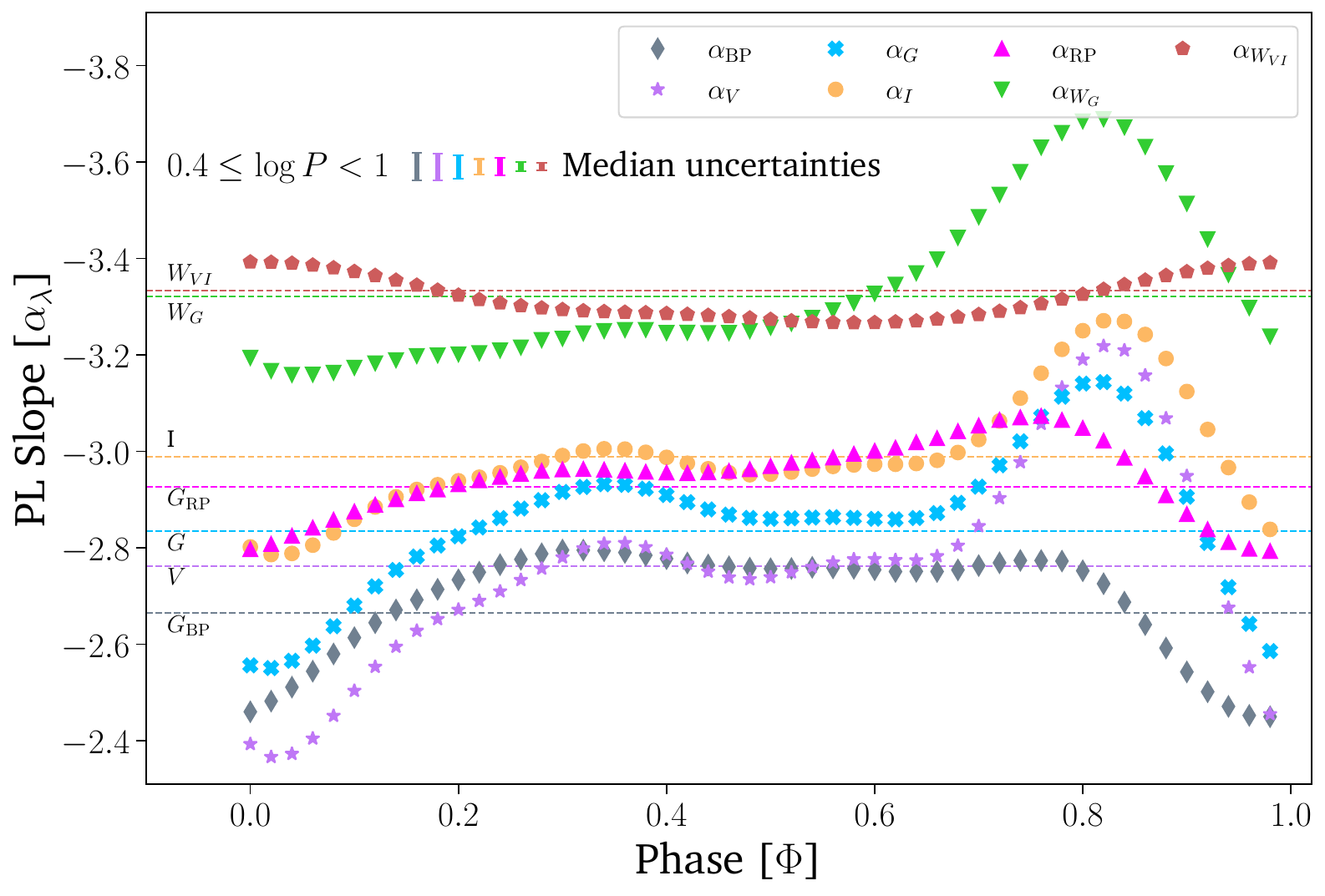}
        } 
        \resizebox{0.48\linewidth}{!}{
        \includegraphics[width=0.44\linewidth, keepaspectratio]{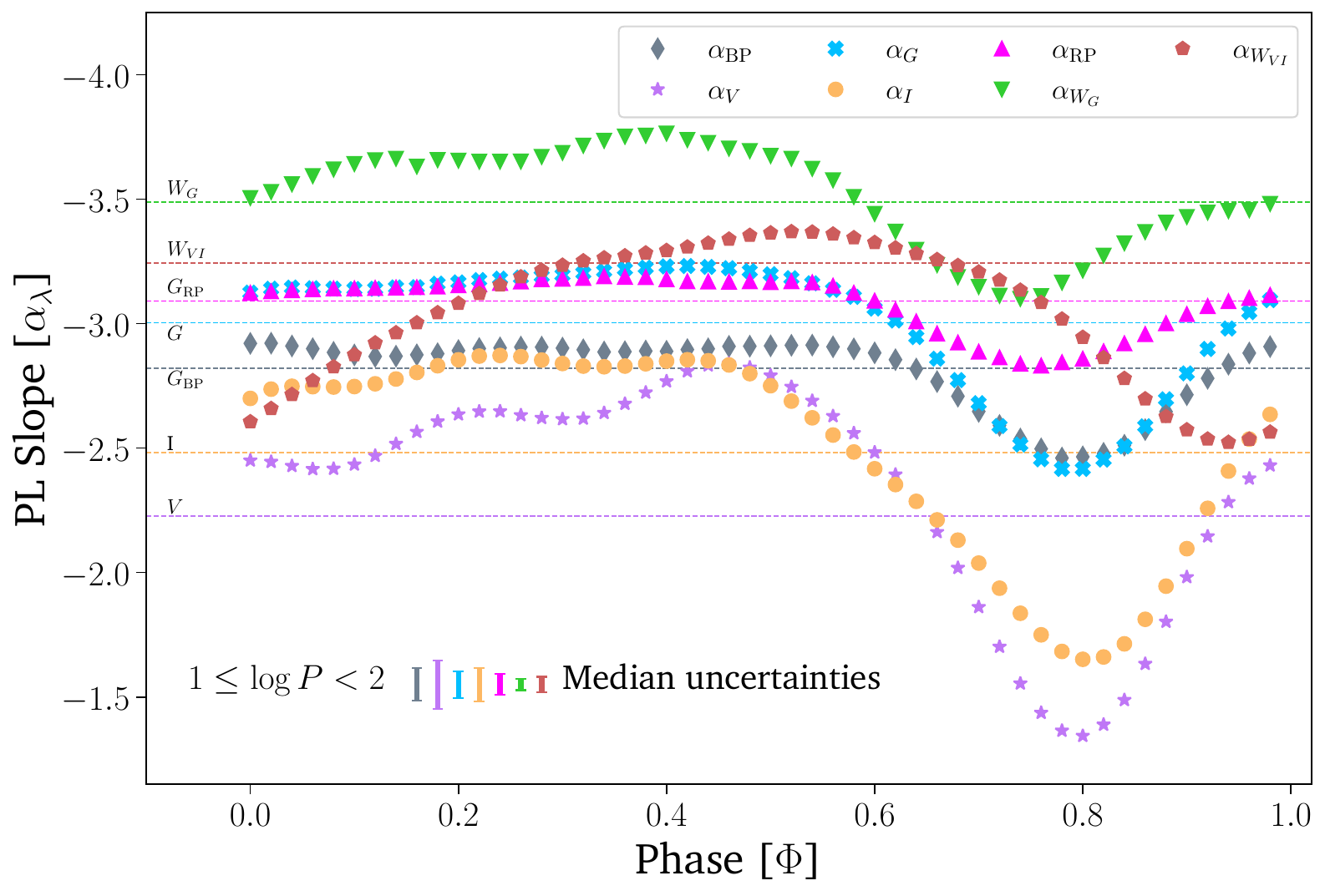}
        }
    \end{tabular}
    \caption{Same as Fig. \ref{fig:3} for short-period ($0.4 \leq \log{P} < 1$, \textit{left panel}) and long-period ($1 \leq \log{P} < 2$, \textit{right panel}) FU Cepheids in the LMC in five photometric bands: $G_{\rm BP}$, $V$, $G$, $G_{\rm RP}$, $I$, and two Wesenheit indices: $W_{VI}$ and $W_{G}$, respectively. The dynamical variations observed in the PL slopes of short- and long-period LMC FU Cepheids are opposite in nature within the phase range: $0.6 \leq \Phi < 1$. The dashed horizontal lines overplotted on the figure represent the PL slopes obtained using mean magnitudes in respective bands for both short- and long-period Cepheids. The median uncertainties denote the typical uncertainties of the multiphase PL slopes.}
    \label{fig:6}
\end{figure*}

We also determine the multiphase PL relations for Cepheids with pulsation periods in the range: $0.4 \leq \log{P} < 1$ (short-periods) and $1 \leq \log{P} < 2$ (long-periods) separately. Comparison of the multiphase LMC PL slopes for short- and long-period Cepheids is shown in Fig. \ref{fig:6}. We find that the PL slopes of LMC short-period FU Cepheids increase (in the absolute sense) to steeper values in the phase range: $0.6 \leq \Phi < 1 $. Whereas for the longer periods, it decreases to flatter values within the same phase range. This contrasting nature of multiphase PL slopes of short- and long-period Cepheids is consistent with the results from earlier studies \citep{ngeo12, kurb23, bhuy24}. Moreover, the overall dynamical variations of the PL slopes using Cepheids of all periods are similar to those obtained using only the short-period Cepheids. It might be due to the statistically larger sample of short-period Cepheids in the LMC as well as in the MW and SMC (Table \ref{tab:1}). 

\begin{figure*}
    \begin{tabular}{c|c}
     \resizebox{0.5\linewidth}{!}{
     \includegraphics[width=0.5\linewidth, keepaspectratio]{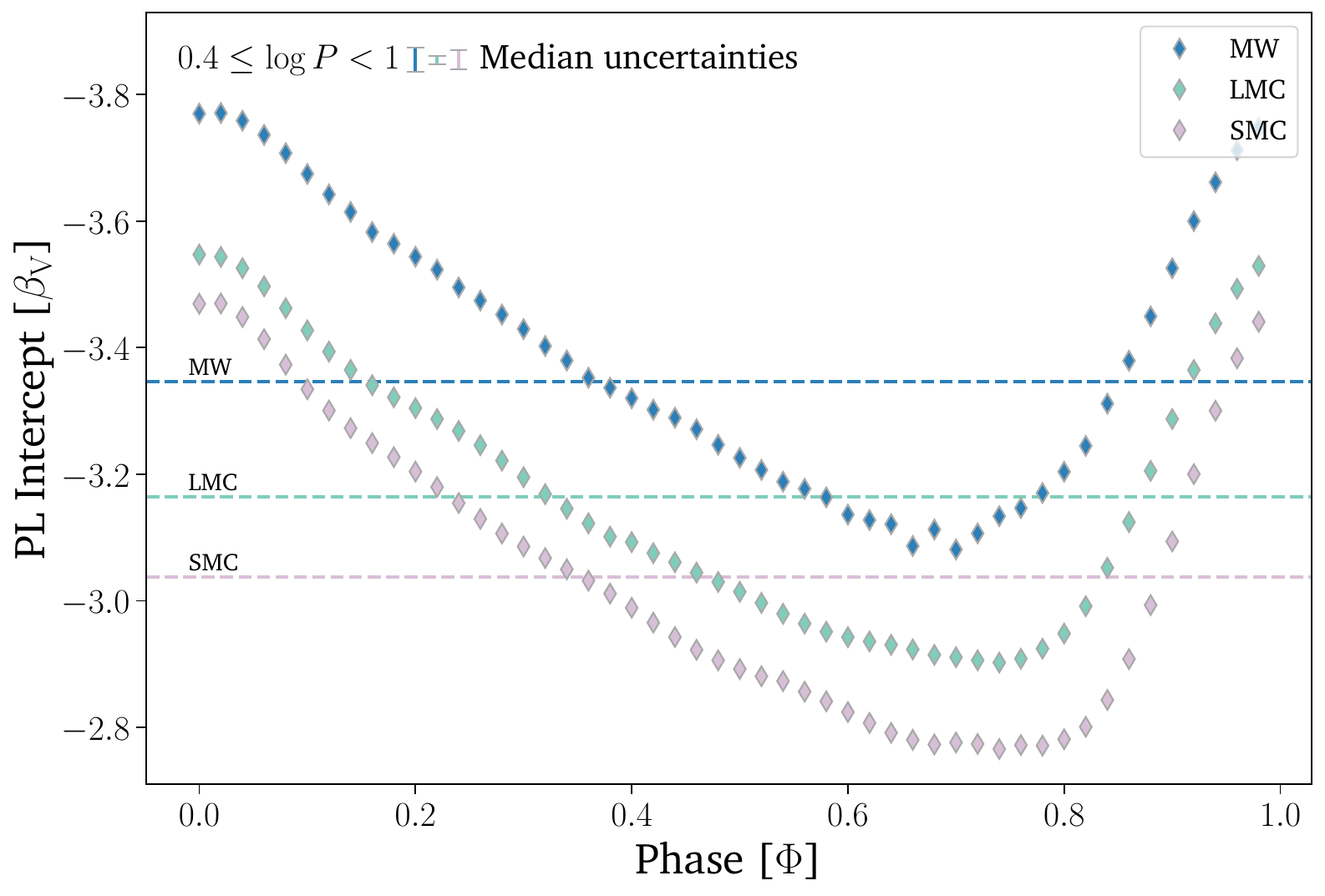}
     }
     \resizebox{0.5\linewidth}{!}{
     \includegraphics[width=0.5\linewidth, keepaspectratio]{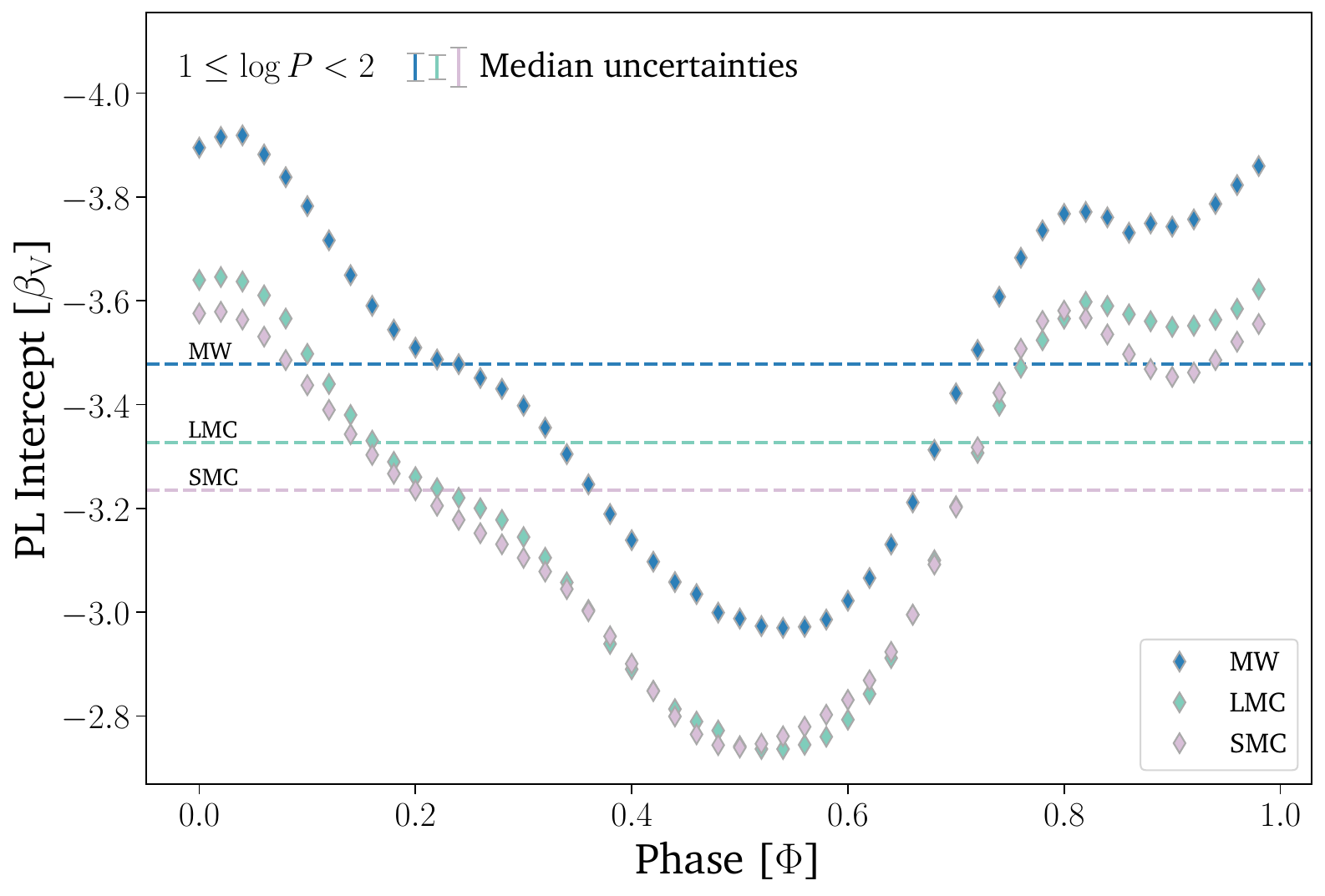}
     }
    \end{tabular}
    \caption{Comparison of the intercepts of multiphase PLZ relations for short-period $(0.4 \leq \log{P} < 1)$ and long-period $(1 \leq \log{P} < 2)$ Cepheids as functions of pulsation phases in the photometric band $V$. The \textit{left panel} compares PL intercepts determined using only the short-period Cepheids. The \textit{right panel} compares PL intercepts determined using only the long-period Cepheids. The dashed horizontal lines in the figures represent PL intercepts obtained using mean magnitudes of Cepheids for different galaxies. Median uncertainties here represent the typical uncertainties of multiphase PL intercepts.}
    \label{fig:7}
\end{figure*}

Similarly, the PL intercept ($\beta_{\lambda}$) values for short- and long-period Cepheids show opposite dynamical variations in the phase range: $0.6 \leq \Phi < 1$ for all three galaxies and all the photometric bands. The amplitudes of variation in the $\beta_{\lambda}$ values along with their median uncertainties are presented in Table \ref{tab:4}. 
\begin{table*}
    \centering
    \caption{Amplitudes of variation and median uncertainties of the multiphase PL intercepts, respectively, in five photometric bands: $G_{\rm BP}$, $V$, $G$, $G_{\rm RP}$, $I$, and two Wesenheit indices: $W_{VI}$ and $W_{G}$. These values are determined with Cepheids in different period ranges, as shown in the table.}
    \begin{tabular}{cccccccc}
    \hline
    \hline
      Bands  & Cepheid & \multicolumn{3}{c}{$^{\rm a}\Delta\beta_{\lambda}$} & \multicolumn{3}{c}{Median uncertainties $\left(\sigma_{\beta_{\lambda}}^{\rm median} \right)$} \\
      & sample & & & & & & \\
      \hline
      &  & $0.4 \leq \log{P} < 2$ & $0.4 \leq \log{P} < 1$ & $1 \leq \log{P} < 2$ & $0.4 \leq \log{P} < 2$ & $0.4 \leq \log{P} < 1$ & $1 \leq \log{P} < 2$ \\
    \hline
     \multirow{3}{*}{\shortstack{$BP$}} & MW &0.649 & 0.590 & 0.781 & 0.013 & 0.015 & 0.024   \\
     & LMC &0.501 & 0.484 & 0.443 & 0.007 & 0.006 & 0.021  \\
     & SMC &0.688 & 0.680 & 0.677 & 0.014 & 0.016 & 0.037 \\
     \hline
     \multirow{3}{*}{\shortstack{$V$}} & MW &0.723 & 0.690 & 0.949 & 0.016 & 0.019 & 0.027 \\
     & LMC &0.685 & 0.645 & 0.910 & 0.006 & 0.007 & 0.023 \\
     & SMC &0.728 & 0.704 & 0.842 & 0.015 & 0.016 & 0.038 \\
     \hline
     \multirow{3}{*}{\shortstack{$G$}} & MW &0.519 & 0.485 & 0.619 & 0.011 & 0.014 & 0.018 \\
     & LMC &0.578 & 0.536 & 0.629 & 0.005 & 0.006 & 0.018 \\
     & SMC &0.607 & 0.609 & 0.576 & 0.012 & 0.014 & 0.032 \\
     \hline
     \multirow{3}{*}{\shortstack{$RP$}} & MW &0.394 & 0.347 & 0.481 &  0.009 &0.012 & 0.016\\
     & LMC &0.293 & 0.279 & 0.272 & 0.005 & 0.005 & 0.016 \\
     & SMC &0.394 & 0.388 & 0.419 &0.012 & 0.012 & 0.028 \\
     \hline
     \multirow{3}{*}{\shortstack{$I$}} & MW &0.429 & 0.412 & 0.550 & 0.013 & 0.016 & 0.020 \\
     & LMC &0.411 & 0.386 & 0.512 & 0.004 & 0.004 & 0.016\\
     & SMC &0.456 & 0.451 & 0.569 & 0.010 & 0.011 & 0.027 \\
     \hline
     \multirow{3}{*}{\shortstack{$W_{VI}$}} & MW &0.158 & 0.164 & 0.386 & 0.006 & 0.010 & 0.015 \\
     & LMC &0.186 & 0.176 & 0.254 & 0.003 & 0.004 & 0.013 \\
     & SMC &0.196 & 0.194 &0.424 & 0.009 & 0.008 & 0.020 \\
     \hline
     \multirow{3}{*}{\shortstack{$W_{G}$}} & MW &0.143 & 0.098 & 0.184 & 0.006 & 0.006 & 0.010\\
     & LMC &0.148 & 0.129 & 0.342 & 0.004 & 0.004  & 0.013 \\
     & SMC &0.157 & 0.157 & 0.259 & 0.009 & 0.009  & 0.021 \\
     \hline 
     \hline
    \end{tabular}
    \begin{tablenotes}
        \item $^{\rm a}\Delta\beta_{\lambda} = \left|{\rm min(\beta_{\lambda}(\Phi))} - {\rm max(\beta_{\lambda}(\Phi))}\right|$,
    \end{tablenotes}
    \label{tab:4}
\end{table*}
The $\beta_{\lambda}$ values for Cepheids of all periods and short-periods are observed to get brighter in the phase range: $0.6 \leq \Phi < 0.8$. On the other hand, those of the longer periods get fainter in the same phase range. However, we find that the intercept values for long-period FU Cepheids are larger in $V$ and $I$ bands than in comparison to Gaia $G, ~G_{\rm BP}$, and $G_{\rm RP}$ bands near phase $\Phi \sim 0.8$. Furthermore, in the $W_{VI}$-band, we make the same observation in a different phase range ($0.8 \leq \Phi < 1$). We present the comparison of the multiphase PL intercepts of Cepheids in all the three galaxies with short and long pulsation periods separately in Fig. \ref{fig:7}, \ref{fig:a2} and \ref{fig:a3}, respectively. 

\subsection{Multiphase metallicity coefficients} \label{section:metl_result}
\begin{figure*}
    \centering
    \begin{tabular}{c|c}
     \resizebox{0.49\linewidth}{!}{
     \includegraphics[width=0.49\linewidth, keepaspectratio]{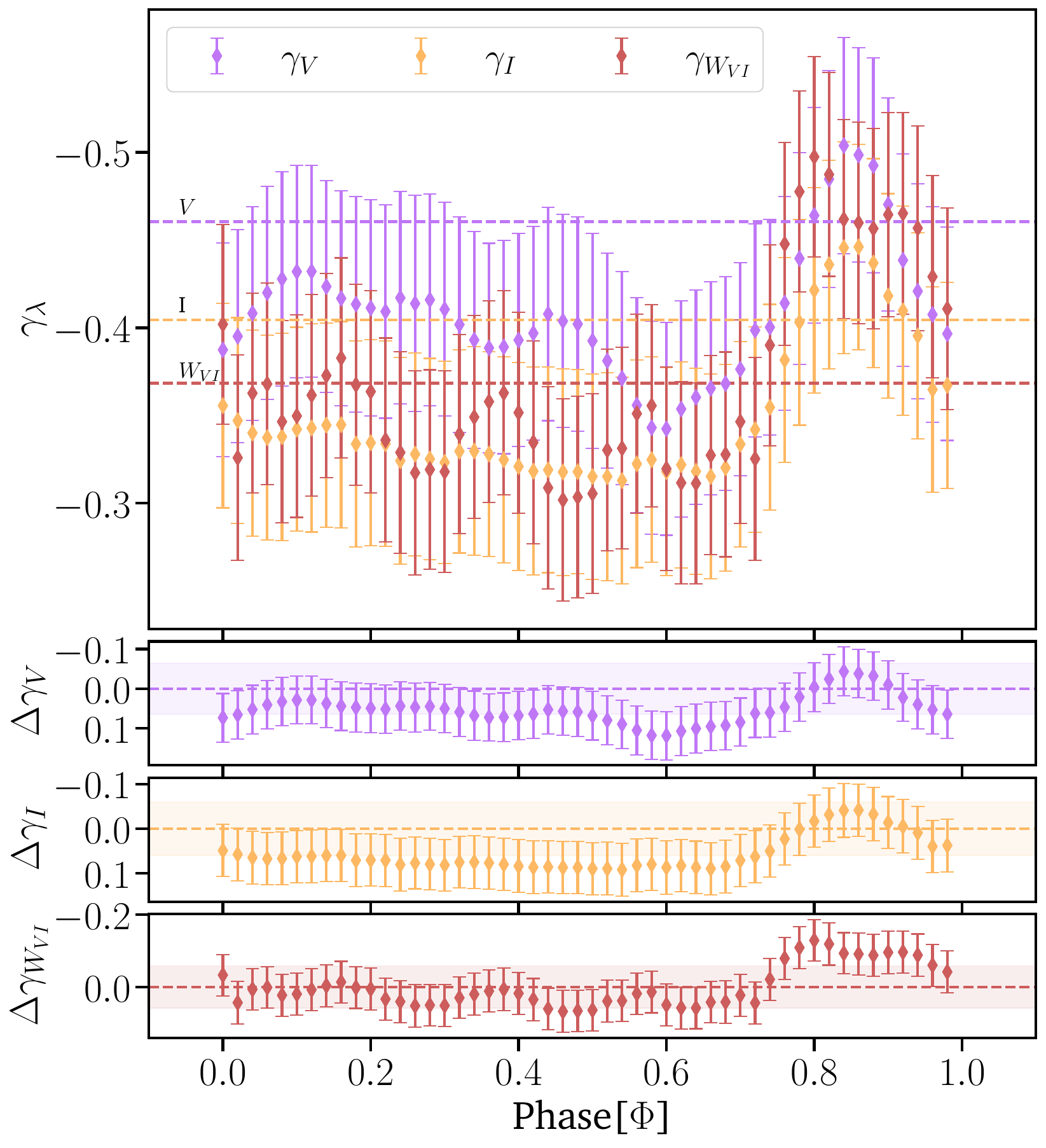}
     }
     \resizebox{0.5\linewidth}{!}{
     \includegraphics[width=0.5\linewidth, keepaspectratio]{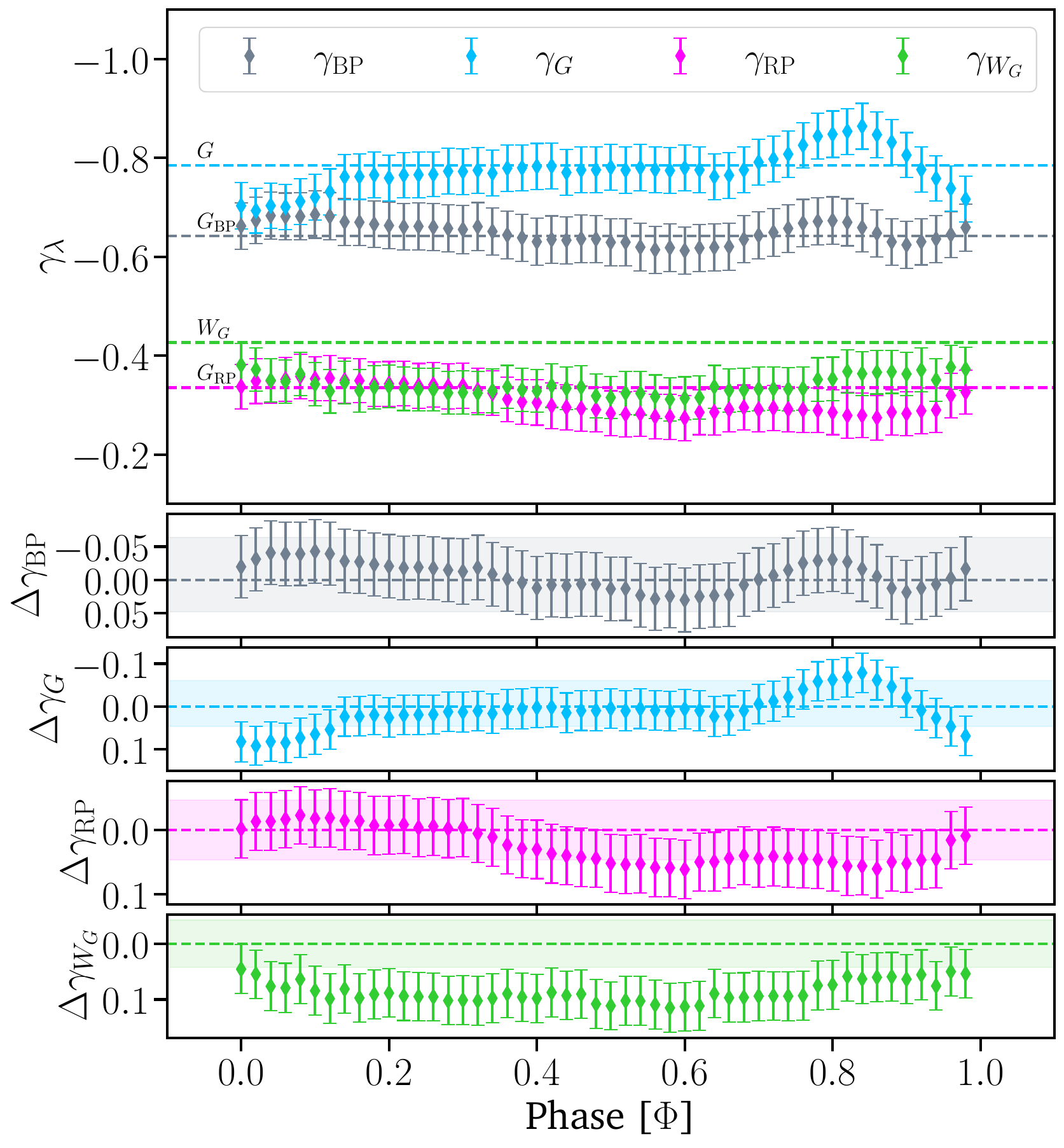}
     }\\ 
\end{tabular}    
    \caption{Coefficient of metallicity effect $(\gamma_{\lambda})$ as a function of the Cepheid pulsation phase $(\Phi)$. The \textit{left panel} shows the metallicity effects for $V$, $I$, and $W_{VI}$ Wesenheit band and the \textit{right panel} shows those in the $G_{\rm BP}$, $G$, $G_{\rm RP}$, and $W_{G}$ Wesenheit band, respectively. The $\gamma_{\lambda}$ values obtained using multiphase light PL relations of Cepheids of all periods $(0.4 \leq \log{P} < 2)$, assuming the PL slopes of MW, LMC, and SMC to be the same as those of LMC Cepheids. The dashed horizontal lines in the figure represent the $\gamma_{\lambda}$ values obtained using mean light PL relations in the respective bands. The smaller bottom panels on both sides show the difference between the $\gamma$ values obtained using mean-light and multiphase PL relations $(\Delta \gamma_{\lambda} = \gamma_{\lambda}^{(\rm mean ~PL) } - \gamma_{\lambda}(\Phi))$ in the respective bands. The shaded regions in the smaller bottom panels show the $1\sigma$ error bars of the $\gamma$ values obtained using the mean-light PL relations in the respective bands. }
    \label{fig:8}
\end{figure*}

Here we present the results of the metallicity effect on the Cepheid PL relations and their dependence on the pulsation phases over a complete pulsation cycle. The comparison of the multiphase metallicity coefficients ($\gamma_{\lambda,j}$) in all the photometric bands is shown in Fig. \ref{fig:8}. \textit{Left panel} shows the $\gamma$ values obtained in the $V$, $I$ and $W_{\rm VI}$ bands, while the \textit{right panel} shows the $\gamma$ values obtained in the $G_{\rm BP}$, $G$, $G_{\rm RP}$, and $W_{G}$ photometric bands. The horizontal lines overplotted on both panels of the figure represent the $\gamma$ values obtained using mean-light PL relations in the respective bands. We also plot the difference between the mean-light and multi-phase $\gamma_{\lambda}$ values along with their respective error bars as $\Delta \gamma_{\lambda}$ values as shown in the bottom panels on both sides of Fig. \ref{fig:8}. We observe that the phase-dependent variations in the $\gamma_{\lambda}$ values obtained from the multiphase PL relations are within the uncertainty limits of those obtained using mean-light PL relations. However, uncertainties in the $\gamma_{\lambda}$ values are dominated by the large systematic scatter in the $\langle$[Fe/H]$\rangle$ values of the three galaxies (Table \ref{tab:1}). Nevertheless, we observe that the $\gamma_{\lambda}$ values exhibit similar characteristic variations as seen in case of the multiphase PL slopes, specifically in the phase range: $0.6 \leq \Phi < 1$ in the $G_{\rm BP}$, $V$, $G$, $I$, and $W_{VI}$ bands. The amplitudes of variation of the $\gamma_{\lambda}$ values in all the photometric bands with different period ranges are provided in Table \ref{tab:5}. The largest amplitude of variation in the $\gamma_{\lambda}$ values for Cepheids of all-, short- and long-periods are observed in the $W_{VI}$- and $G_{\rm BP}$-band ($\sim 0.190, ~0.299,$ and $ 0.299$ mag/dex), respectively. We observe that the dynamical variation of $\gamma_{\lambda}$ values from the less negative towards more negative values in the phase range: $0.6 \leq \Phi < 1$ for the  $V$-, $I$-, $G$- and $W_{VI}$-band. On the other hand, no such significant dynamical variations are observed for the $G_{\rm BP}$-, $G_{\rm RP}$- and $W_{G}$-band for the same phase range. 

\subsection{Metallicity effect assuming a PL break} \label{section:metl_break_result}
\begin{figure*}
\begin{tabular}{c|c}
    \resizebox{0.45\linewidth}{!}{
    \includegraphics[width=0.45\linewidth, keepaspectratio]{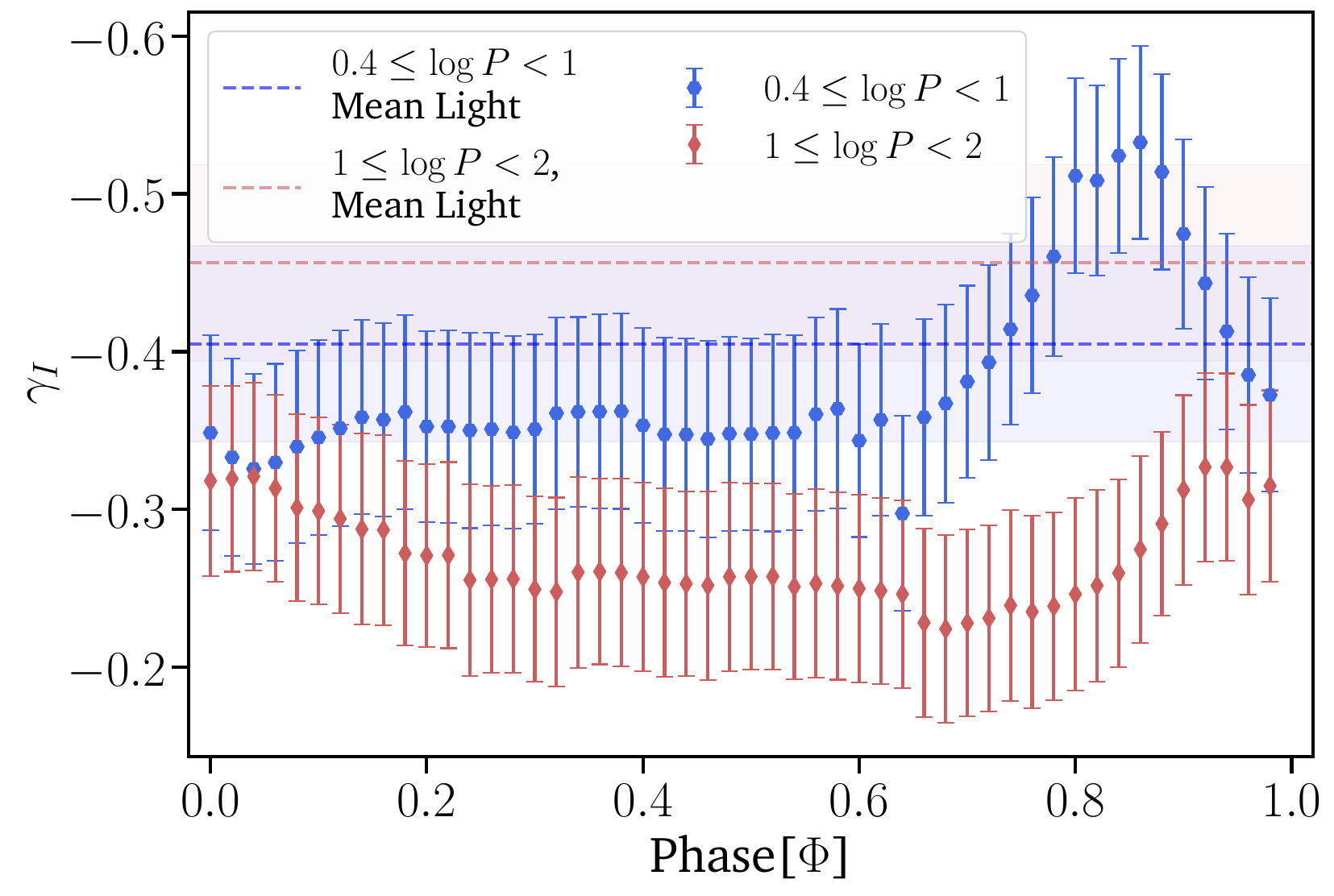}
     }
    \resizebox{0.45\linewidth}{!}{
    \includegraphics[width=0.45\linewidth, keepaspectratio]{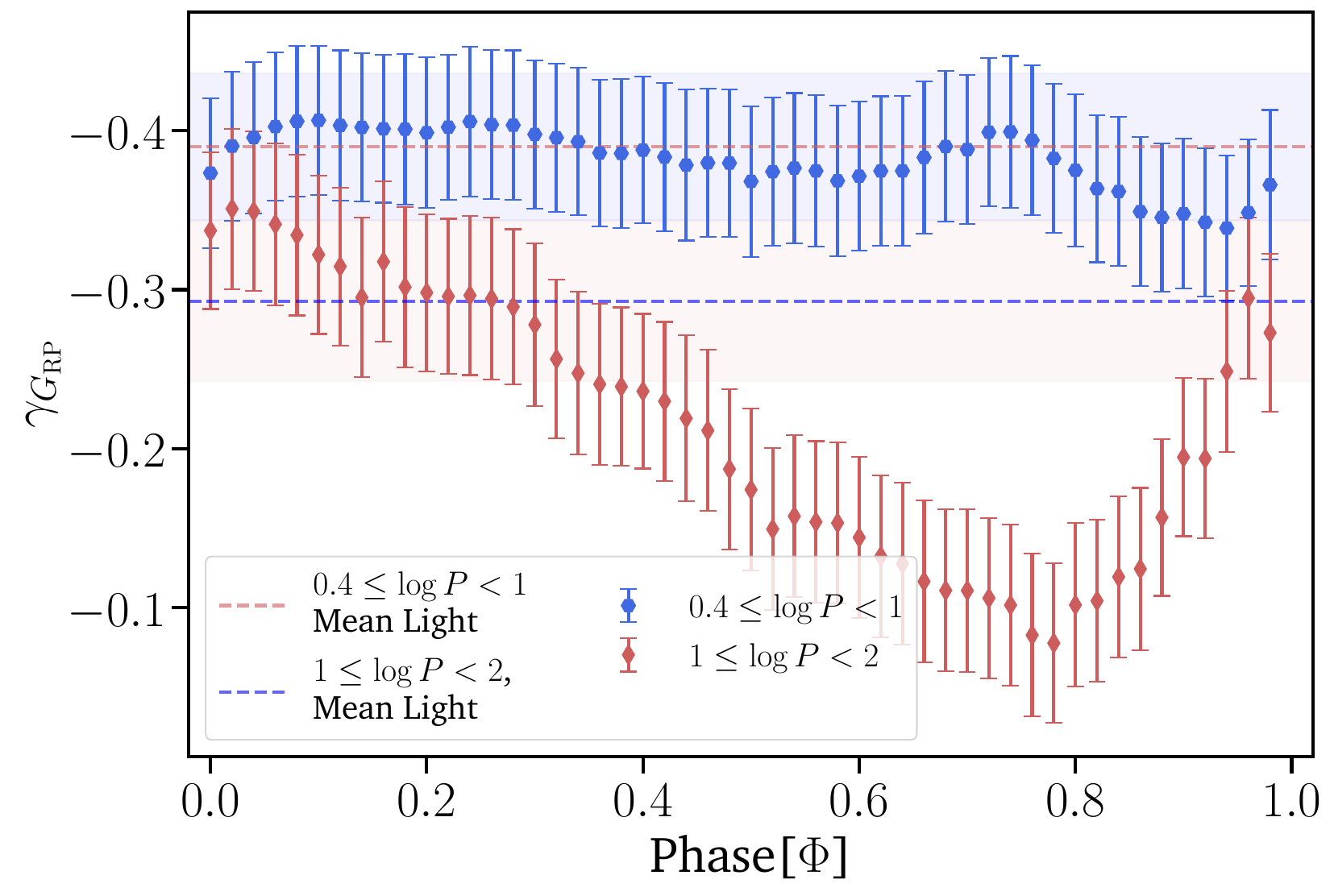}
     } \\
     \resizebox{0.45\linewidth}{!}{
     \includegraphics[width=0.45\linewidth, keepaspectratio]{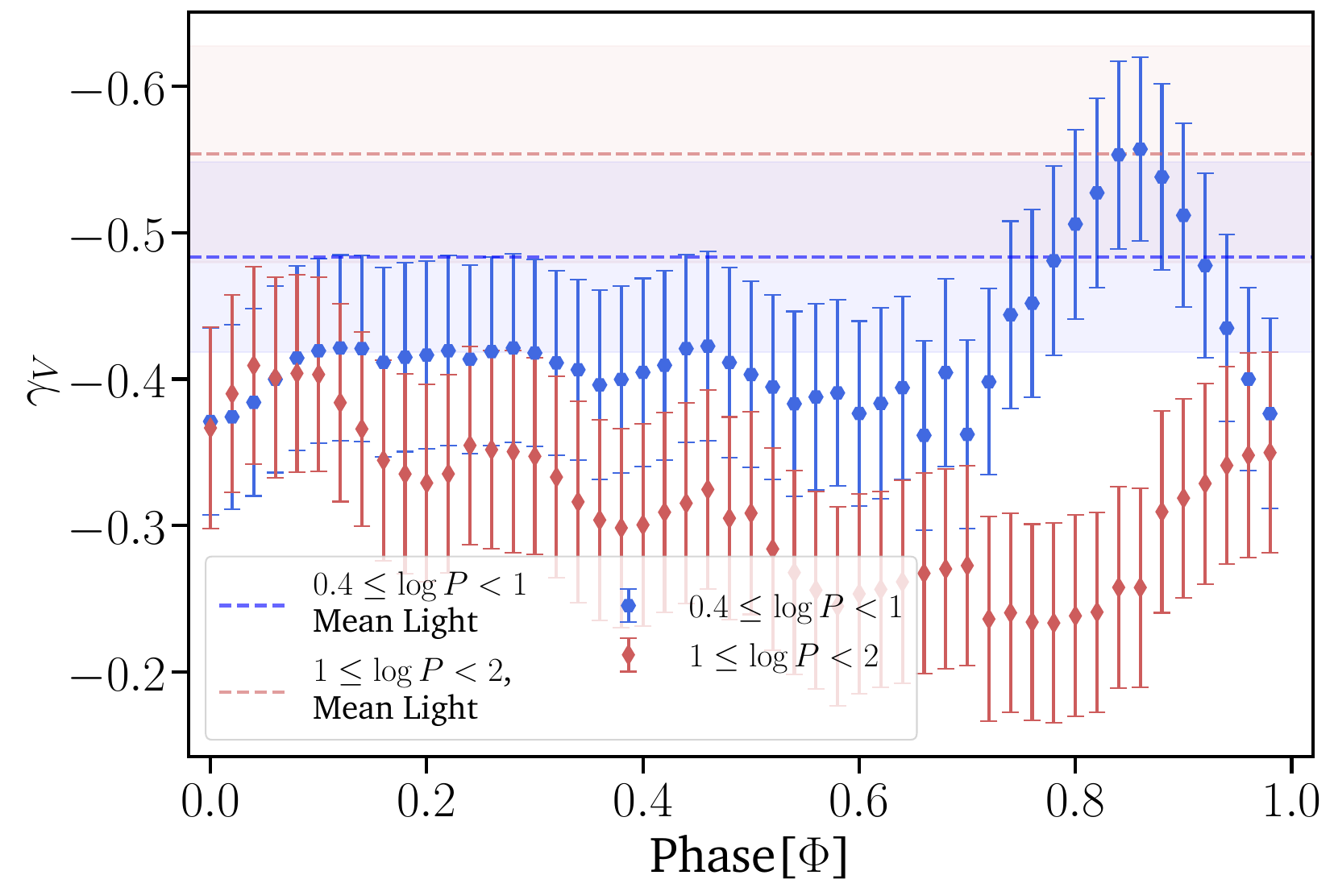}
     }
     \resizebox{0.45\linewidth}{!}{
     \includegraphics[width=0.45\linewidth, keepaspectratio]{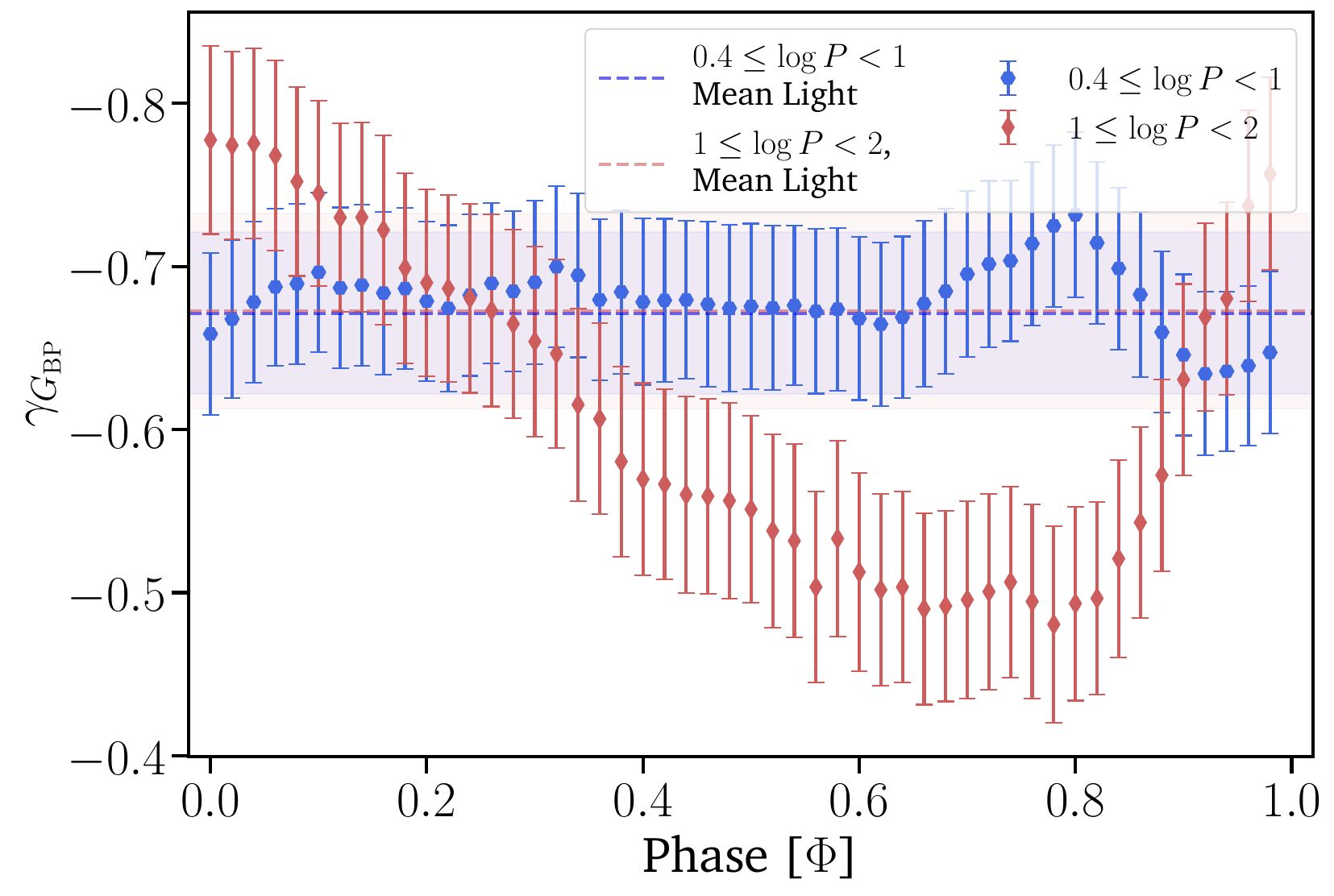}
     } \\

     \resizebox{0.45\linewidth}{!}{
     \includegraphics[width=0.45\linewidth, keepaspectratio]{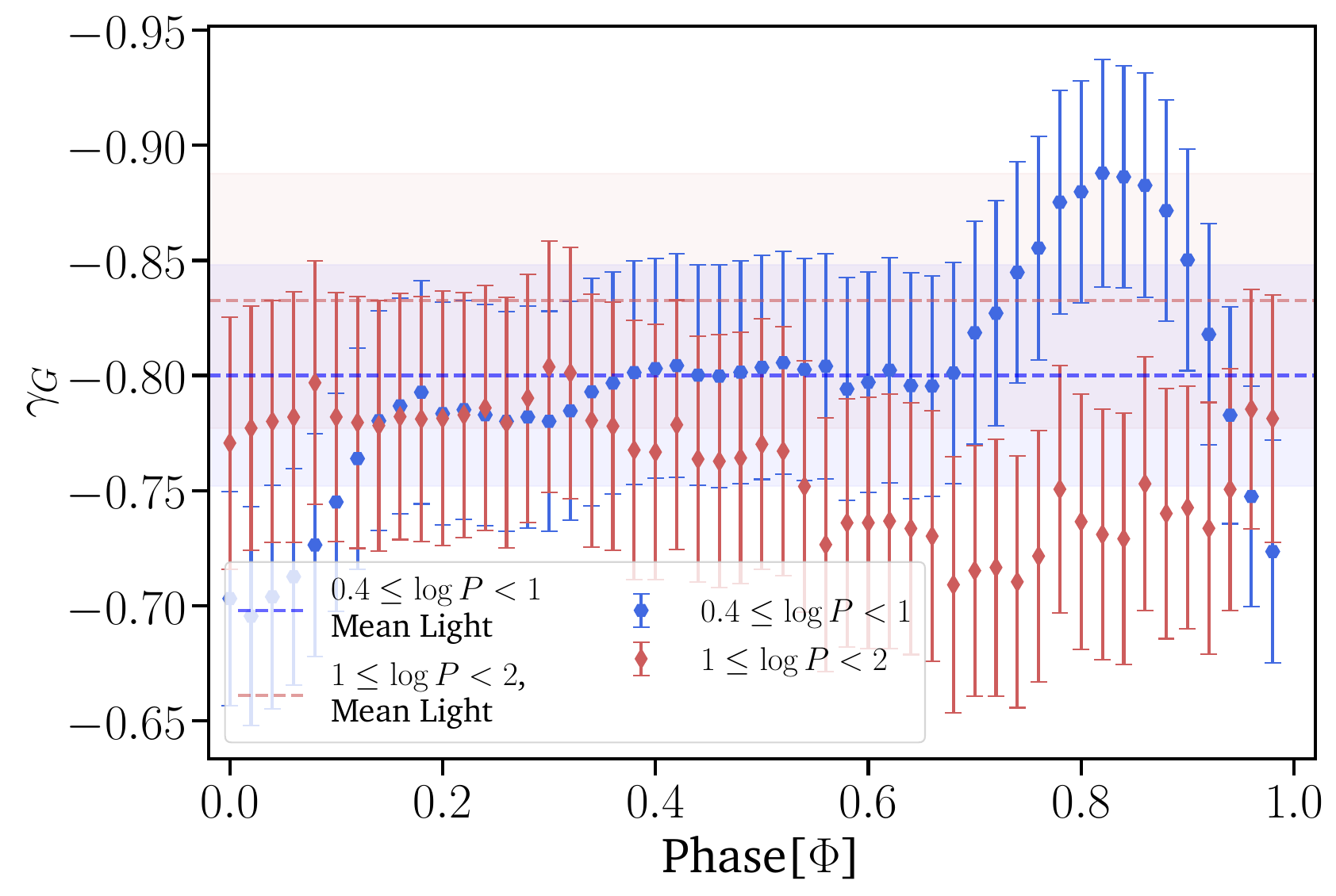}
     }
     \resizebox{0.45\linewidth}{!}{
     \includegraphics[width=0.45\linewidth, keepaspectratio]{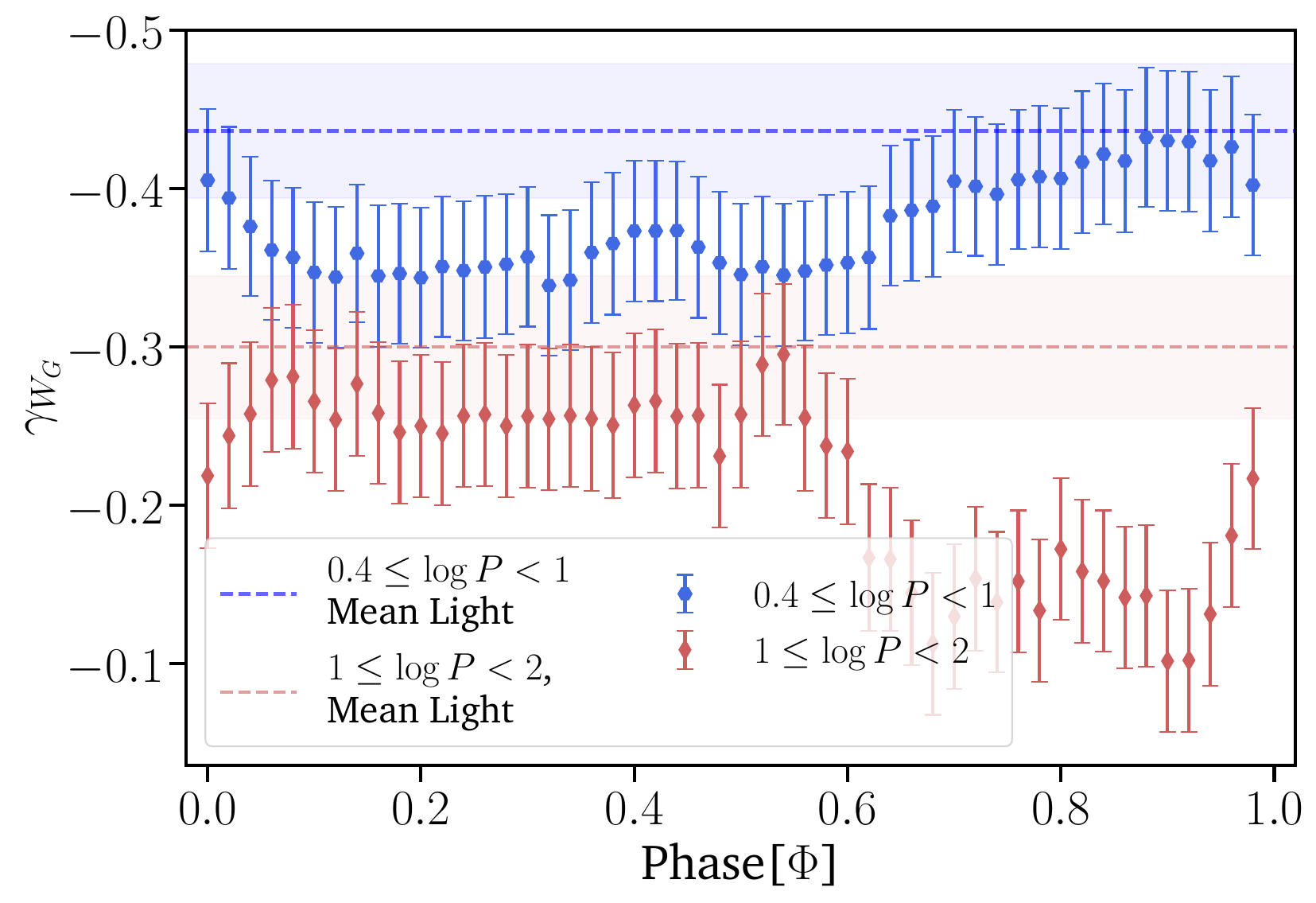}
     } \\

     \resizebox{0.45\linewidth}{!}{
     \includegraphics[width=0.45\linewidth, keepaspectratio]{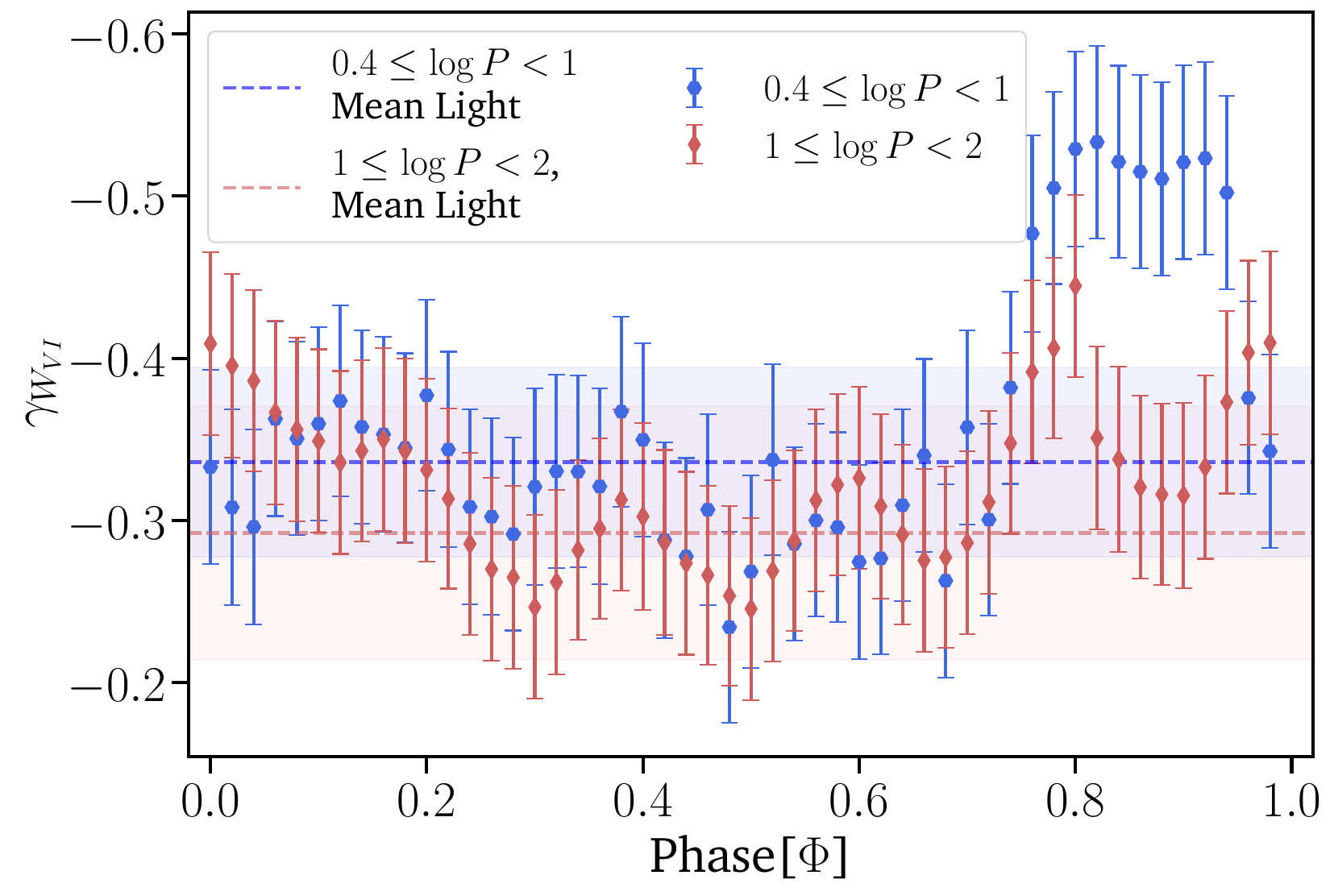}
     } \\     
\end{tabular}
\caption{Coefficient of metalliciy effect as a function of pulsation phases for short-period $(0.4 \leq \log{P} < 1)$ and long-period $(1 \leq \log{P} < 2)$ FU Cepheids in the MW, LMC and SMC in the photometric bands: $G_{\rm BP}$, $V$, $G$, $ G_{\rm RP}$, $I$, $W_{G}$ and $W_{VI}$ respectively. The dashed horizontal lines with the shaded regions in the figure represent the metallicity effect and their uncertainties obtained using mean light PL relations in respective bands for both short- and long-period Cepheids, respectively. The typical uncertainties are represented by the median uncertainties on $\gamma_{\lambda}$ values in each photometric band.}
\label{fig:9}
\end{figure*}

Several earlier studies have suggested the existence of a significant break in PL relations of FU Cepheids at $\log{P}=1$ or period, $P=10$ days \citep[and the references therein]{kanb04, bhar16, kurb23, bhuy24}. It is therefore important to investigate the non-linearity in the PLZ relations. Hence, we also separately determine the $\gamma_{\lambda}$ values of multiphase PLZ relations for short-period and long-period Cepheids. Fig. \ref{fig:9} shows the comparison of the $\gamma_{\lambda}$ values obtained based on the PL relations of short- and long-period Cepheids, respectively, in different photometric bands. We observe that the multiphase $\gamma_{\lambda}$ values obtained using short- and long-period Cepheids show a distinct nature of dynamical variations in all photometric bands, except for the $W_{VI}$ band. The $\gamma_{\lambda}$ values obtained using the short-period Cepheids are observed to move towards more negative values in the phase range: $0.6 \leq \Phi < 1$ in most photometric bands. On the other hand, the multiphase $\gamma_{\lambda}$ values derived using only the long-period Cepheids move towards less negative values in the same phase range in all the photometric bands. Such distinction in the variation of the $\gamma_{\lambda}$ values of short- and long-period Cepheids in the $0.6 \leq \Phi < 1$ phase range is similar to that observed for their respective PL slope and intercept values (Section \ref{section:pl_result}). However, we also find that in the $W_{VI}$-band the $\gamma_{\lambda}$ values of the long-period Cepheids either coincide with or vary similarly to those of the short-periods at most pulsation phases. We observe that dynamical variations of the $\gamma_{\lambda}$ values determined using Cepheids of all- and short-periods are similar. It suggests that the dynamical nature of multiphase PLZ relations of all-period Cepheids is dominated by the short-periods, due to their larger fraction in the total sample (Table \ref{tab:1}).

\begin{figure*}
    \begin{tabular}{c|c}
         \resizebox{0.45\linewidth}{!}{
         \includegraphics[width=0.45\linewidth, keepaspectratio]{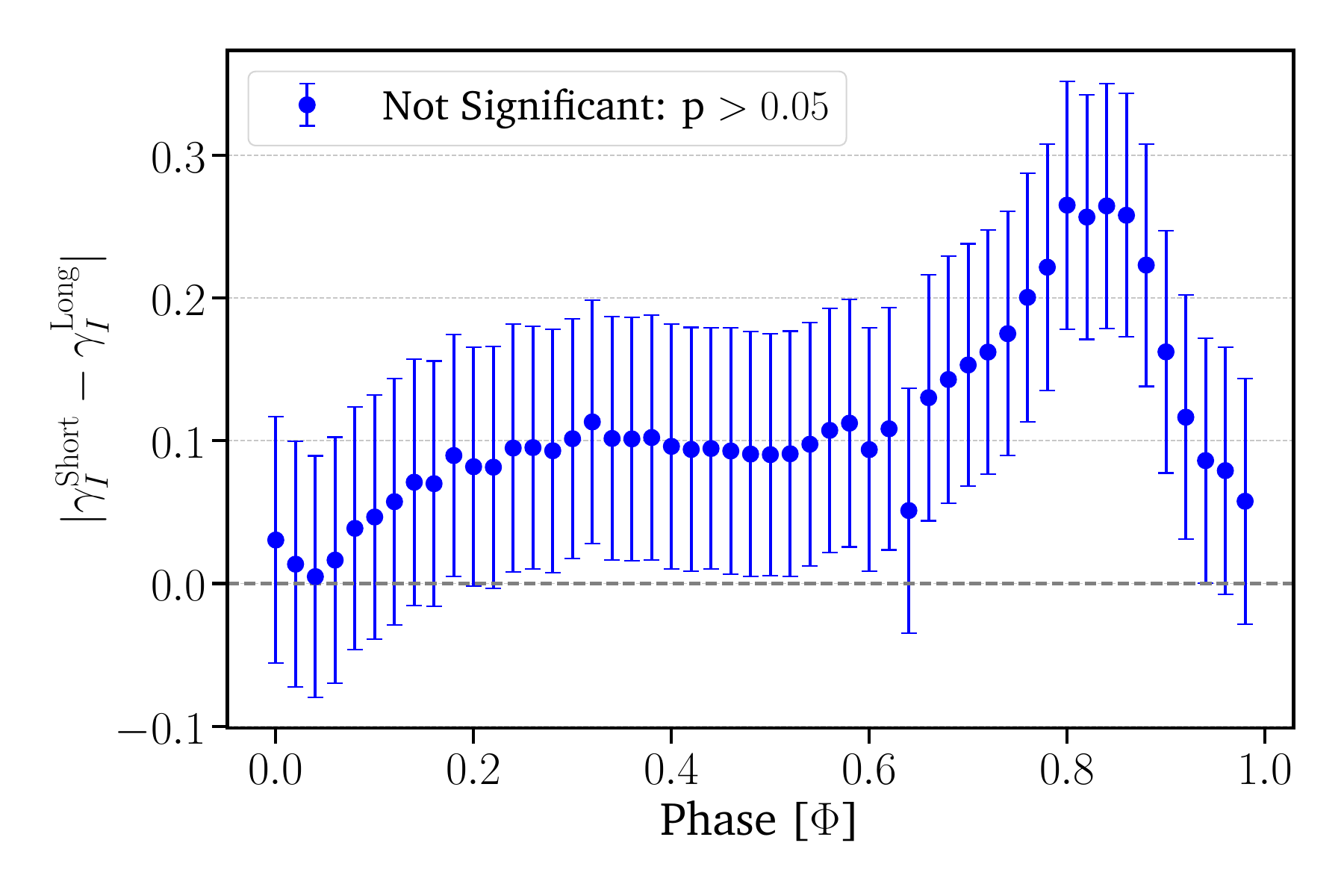}
         }
         \resizebox{0.45\linewidth}{!}{
         \includegraphics[width=0.45\linewidth, keepaspectratio]{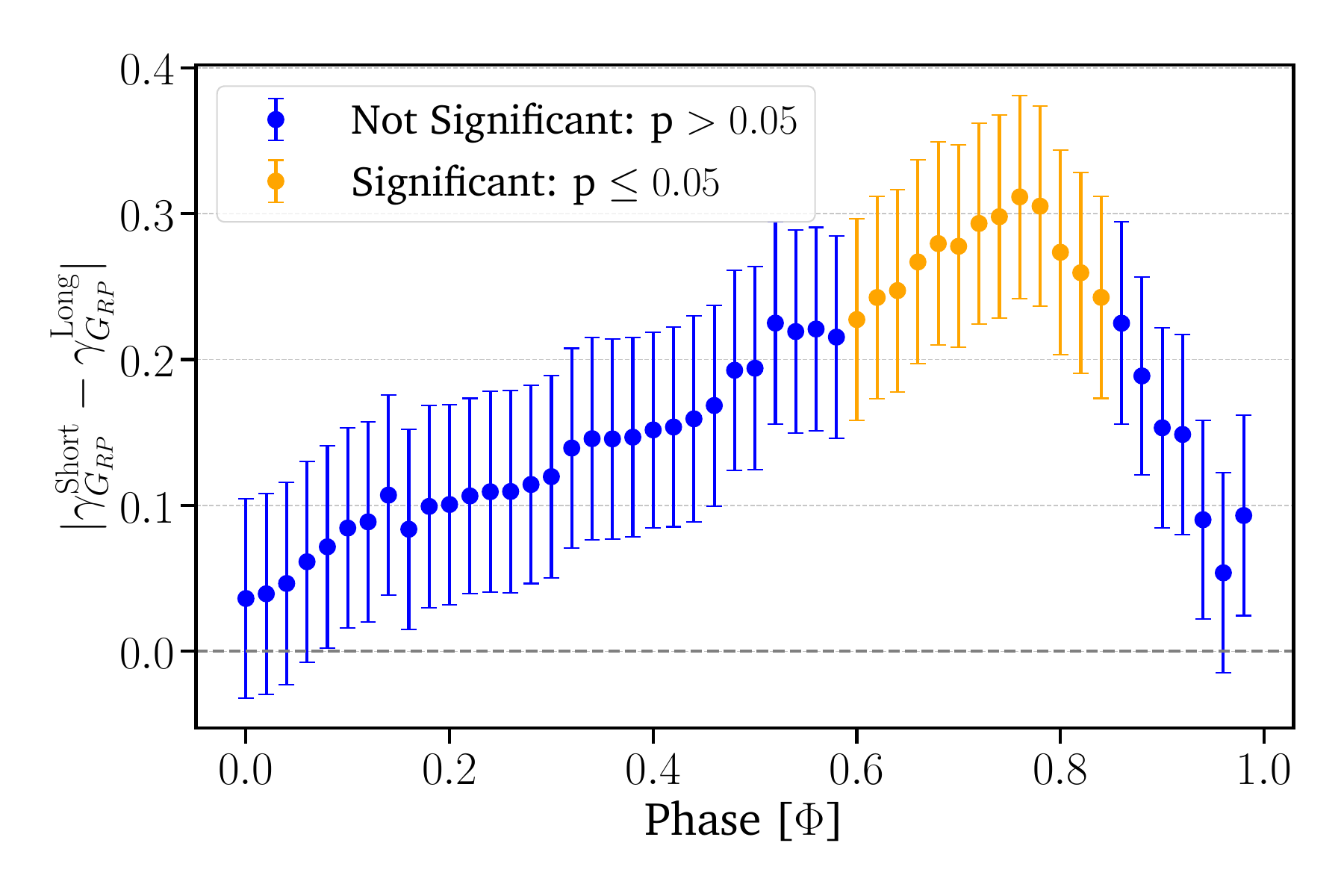}
         }\\
        \resizebox{0.45\linewidth}{!}{
         \includegraphics[width=0.45\linewidth, keepaspectratio]{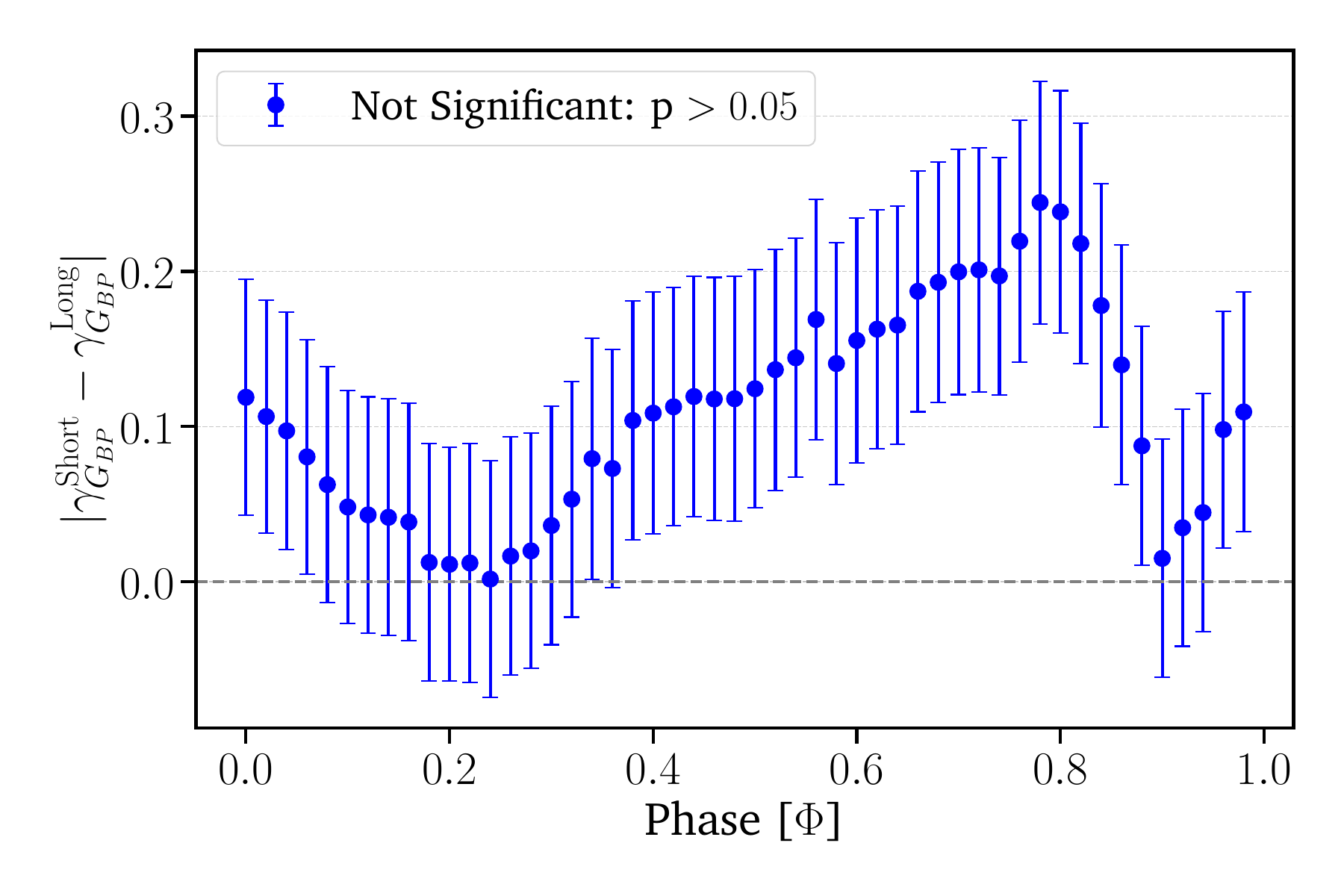}
         }
         \resizebox{0.45\linewidth}{!}{
         \includegraphics[width=0.45\linewidth, keepaspectratio]{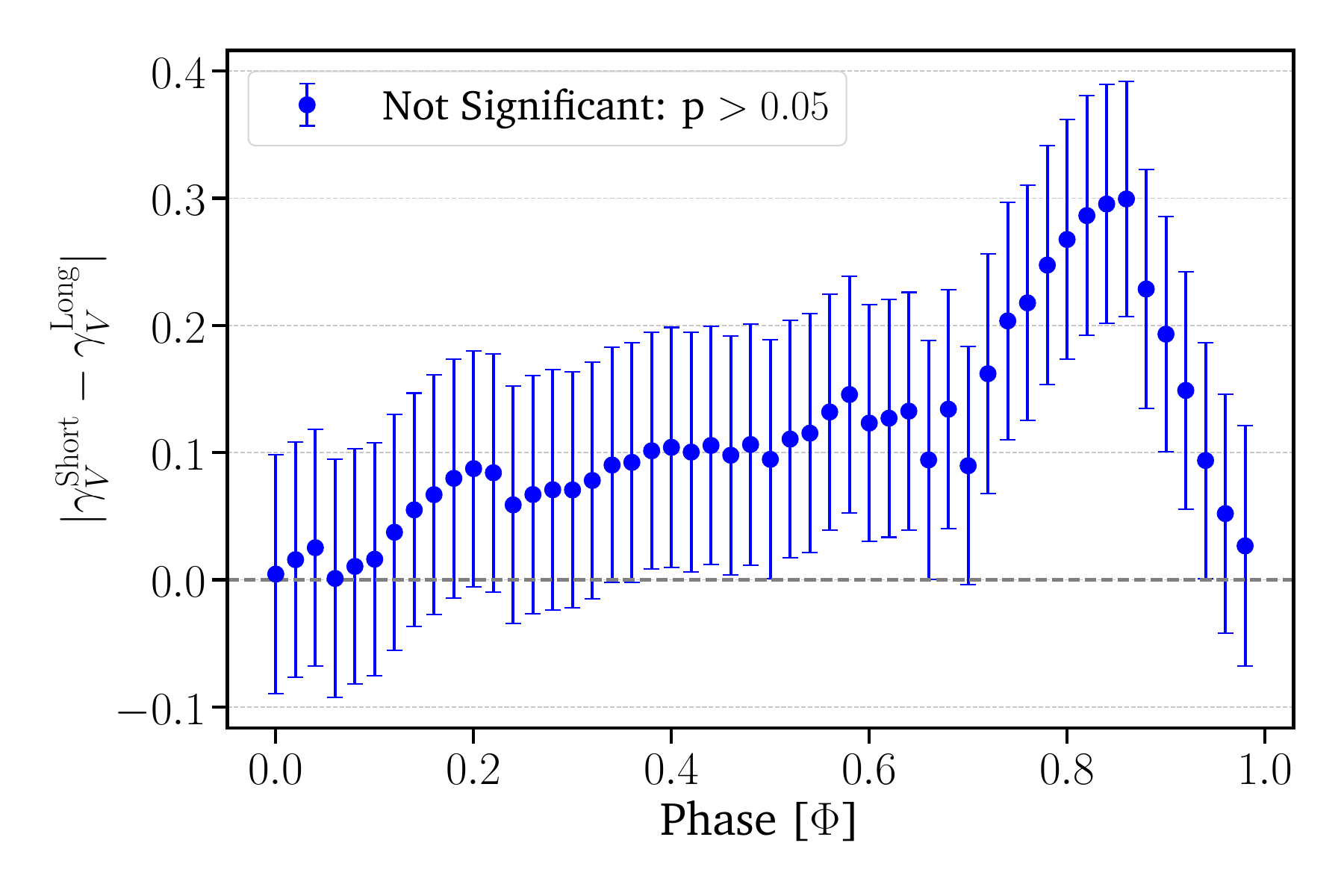}
         }\\

         \resizebox{0.45\linewidth}{!}{
         \includegraphics[width=0.45\linewidth, keepaspectratio]{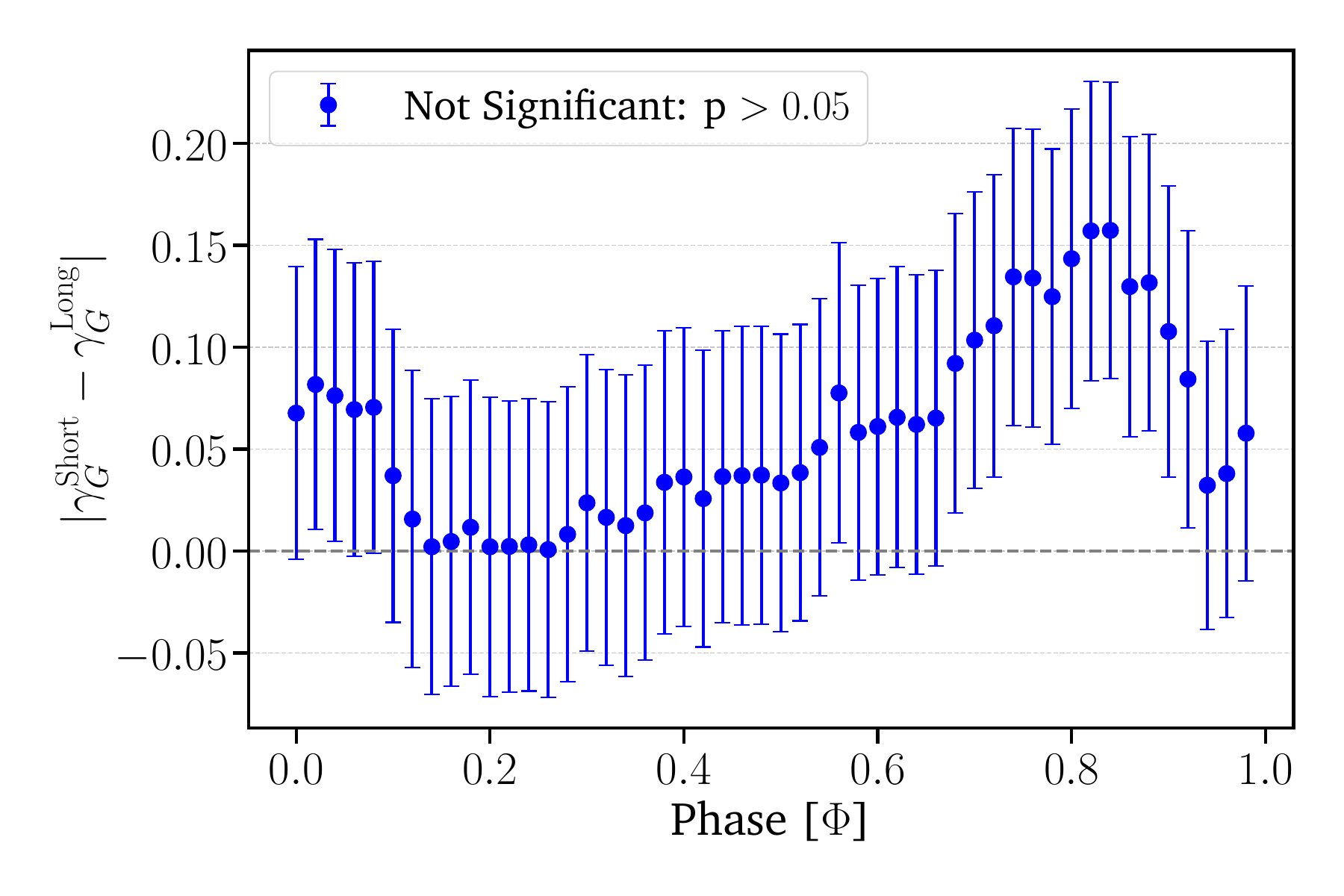}
         }
         \resizebox{0.45\linewidth}{!}{
         \includegraphics[width=0.45\linewidth, keepaspectratio]{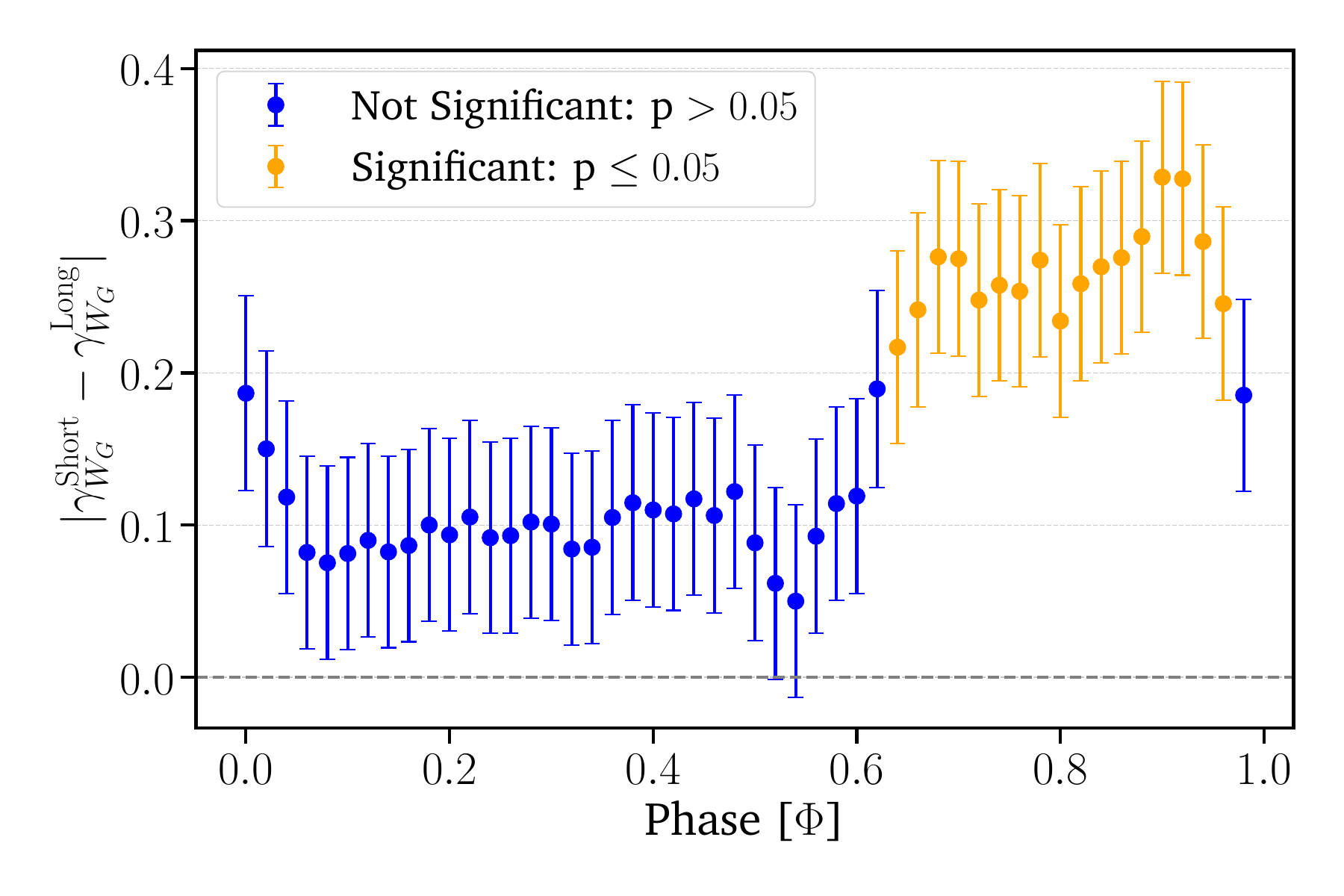}
         }\\
         \resizebox{0.45\linewidth}{!}{
         \includegraphics[width=0.45\linewidth, keepaspectratio]{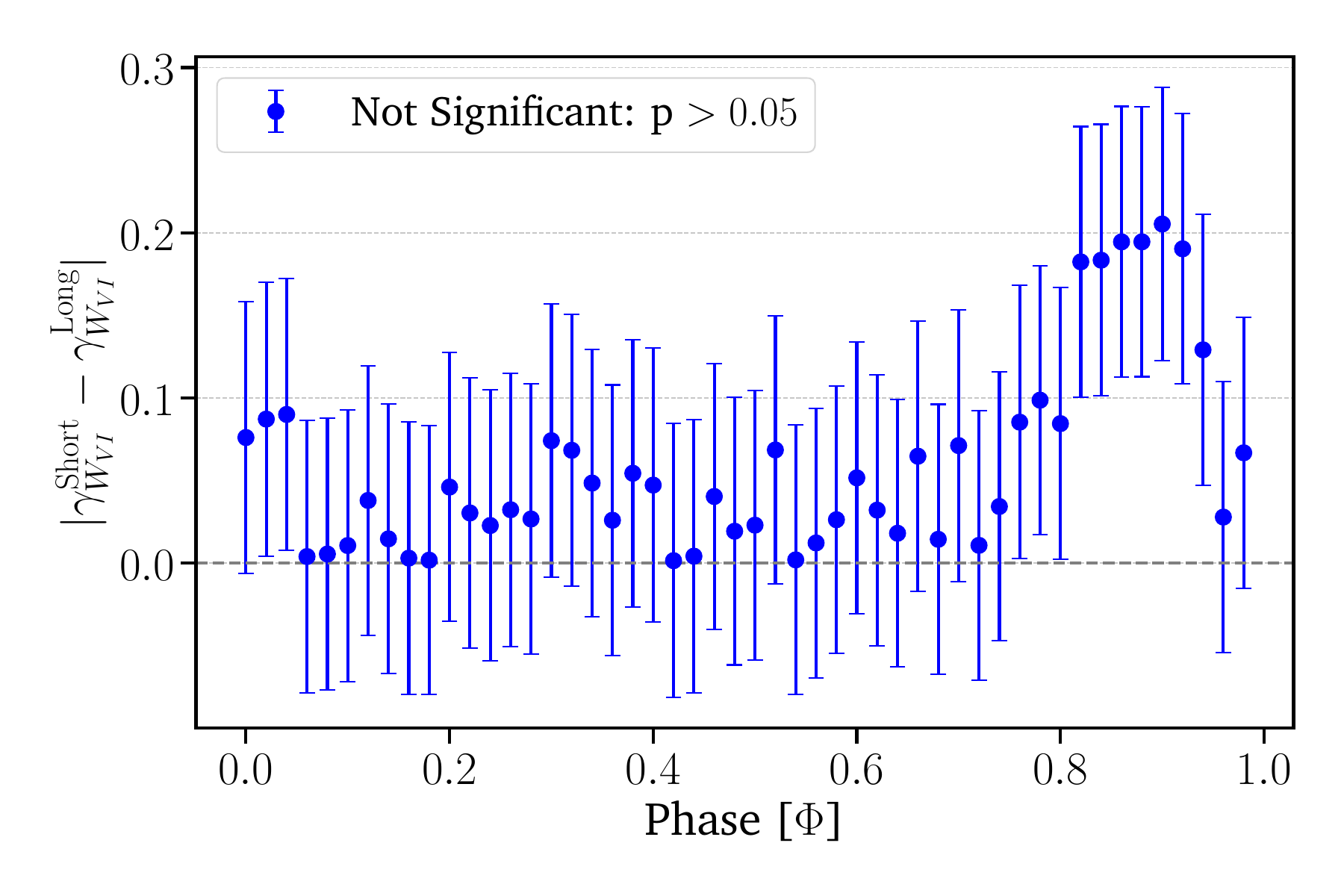}
         }
    \end{tabular}
    \caption{Results of the hypothesis testing based on the \textit{Bonferroni correction} method for the existence of different metallicity effects $(\gamma_{\lambda})$ on the multiphase PL relations of short-period $(0.4 \leq \log{P} < 1)$ and long-period $(1 \leq \log{P} < 2)$ Cepheids at $95\%$ confidence level $(\alpha = 0.05)$ in the photometric bands: $G_{\rm BP}$, $V$, $G$, $ G_{\rm RP}$, $I$, $W_{G}$ and $W_{VI}$ respectively.}
    \label{fig:10}
\end{figure*}

The $\gamma_{\lambda}$ values obtained using the two different period ranges of Cepheids appear distinct in most photometric bands. However, they might be well within $1\sigma-2\sigma$ levels of each other and hence be statistically similar. We have investigated this by utilizing statistical hypothesis testing based on the \textit{Bonferroni correction} \citep{dunn61} method. It is used to mitigate the risk of obtaining false positive results due to multiple comparisons, such as comparing $\gamma_{\lambda}$ values of short- and long-period Cepheids at multiple pulsation phases. In our analysis, the null hypothesis asserts that the metallicity dependence of multiphase PL relations of the short- and long-period Cepheids is statistically similar. If there are $N$ independent tests of the null hypothesis at the significance level $\alpha$, the probability that there will be no significant differences is $(1-\alpha)^{N}$ \citep{bland95}. Since we test the hypothesis at the significance level: $\alpha = 0.05$, which is much smaller than 1, we can write:
\begin{align}
    (1-\alpha)^{N} & \approx 1 - N\alpha,
\end{align}
using the binomial expansion of $(1-\alpha)^{N}$. For any $\alpha^{\prime} < \alpha/N$, the null hypothesis can be rejected at the significance level $\alpha$. Alternatively, it can be done by multiplying the $p-$values from multiple independent tests by the number of independent tests, $N$ and rejecting the null hypothesis for any $pN < \alpha$ \citep{bland95}. In our analysis, $p-$values are determined based on the $z-$score at each pulsation phase, which is defined as:
\begin{align}
    z & = \frac{\left|\gamma_{\lambda}^{\rm short} - \gamma_{\lambda}^{\rm long}\right| }{\sqrt{\sigma^2_{\rm short} + \sigma^2_{\rm long}}}.
\end{align}
For more details on the Bonferroni correction method, interested readers are redirected to \citet{dunn61} and \citet{bland95}. Results of the statistical hypothesis testing in all five photometric bands and two Wesenheit indices based on the Bonferroni method are presented in Fig. \ref{fig:10}. It is found that there are significant differences in the metallicity effect on PL relations between short- and long-period Cepheids in the  $G_{\rm RP}$- and $W_{G}$-band, in the $0.6 \leq \Phi < 1$ phase range. However, it is not observed in all the other photometric bands.

\begin{table*}
\centering
\caption{Amplitudes of variation and median uncertainties of the multiphase metallicity effect in five photometric bands: $G_{\rm BP}$, $V$, $G$, $G_{\rm RP}$, $I$, and two Wesenheit indices: $W_{VI}$ and $W_{G}$, respectively. These values are determined with Cepheids in different period ranges as shown in the table.}
\begin{threeparttable}
\begin{tabular}{cccccccc}
\hline
\hline
Bands & \multicolumn{3}{c}{$^{\rm a}\Delta\gamma_{\lambda}$ (mag/dex) } & \multicolumn{3}{c}{Median uncertainties $\left(\sigma^{\rm  median}_{\gamma_{\lambda}}\right)$ (mag/dex)} \\
\hline
&  $0.4 \leq \log{P} < 2$ & $0.4 \leq \log{P} < 1$ & $1 \leq \log{P} < 2$ & $0.4 \leq \log{P} < 2$ & $0.4 \leq \log{P} < 1$ & $1 \leq \log{P} < 2$ \\
\hline
 $BP$ & 0.073 & 0.098 &0.297 & 0.048 & 0.050 & 0.058 \\
 $V$ &0.162 & 0.195 & 0.176 & 0.061 & 0.064 & 0.068 \\
 $G$ &0.171 & 0.193 & 0.095 & 0.046 & 0.048 & 0.054 \\
 $RP$ &0.084 & 0.068 & 0.273 & 0.045 & 0.047 & 0.051 \\
 $I$ &0.133 & 0.236 & 0.103 & 0.059 & 0.062 & 0.059 \\
 $W_{VI}$ &0.190 & 0.299 & 0.199 & 0.057 & 0.060 & 0.056 \\
 $W_{G}$ &0.069 & 0.094 & 0.193 & 0.044 & 0.044 & 0.045 \\
\hline 
\hline
\end{tabular}
\begin{tablenotes}
    \item $^{\rm a}\Delta\gamma_{\lambda} = \left|{\rm min(\gamma_{\lambda}(\Phi))} - {\rm max(\gamma_{\lambda}(\Phi))}\right|$
\end{tablenotes}
\end{threeparttable}
\label{tab:5}
\end{table*}

\subsection{Comparison with literature}
Here we compare the metallicity effect results obtained in this study with those from earlier studies. However, the phase-dependent $\gamma_{\lambda}$ values cannot be directly compared to those obtained using mean-light PL relations. We calculate the weighted averages of multiphase $\gamma_{\lambda}$ values obtained in this work, which are also presented in Table \ref{tab:6}. We perform this exercise as a sanity check on the methods and the results of this study.

\begin{figure*}
    \centering
    \includegraphics[width=1.0\linewidth, keepaspectratio]{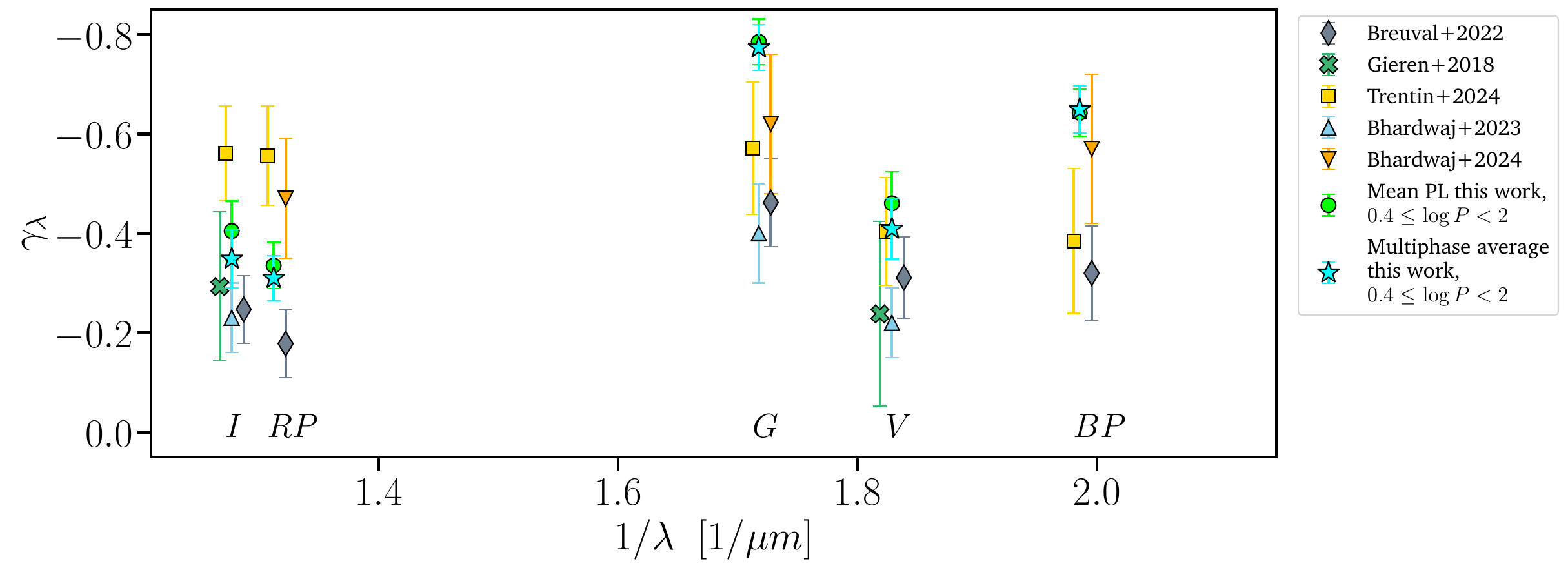}
    \caption{Comparison of the metallicity coefficients $(\gamma_{\lambda})$ of PLZ relations as a function of the inverse of wavelengths of five photometric bands $(V, I, G, G_{\rm BP} {\rm ~\&~} G_{\rm RP})$. The $\gamma_{\lambda}$ values are obtained using both mean light and multiphase PL relations of Cepheids with periods in the range: $(0.4 \leq \log{P} < 2)$. The weighted averages of metallicity coefficients ($\gamma_{\lambda}$) obtained from multiphase PL relations (cyan star symbols) are compared to those obtained based on mean-light PL relations by \citet{gier18}, \citet{breu22}, \citet{bhar23, bhar24}, \citet{tren24} as well as in this work (green circles). The error bars on coefficients obtained from mean light PL relations represent the $68\%$ confidence interval, and those on the coefficients obtained from multiphase PL relations represent the median uncertainties. The abscissa for the literature values in the corresponding photometric bands has been slightly offset to enhance clarity and to avoid overlapping of the $1\sigma$ error-bars.}
    \label{fig:metl_avg}
\end{figure*}

A comparison of $\gamma_{\lambda}$ values obtained in this study (both from multiphase average and mean light PL) and those from the literature is shown as a function of wavelengths in Fig. \ref{fig:metl_avg}. The typical uncertainties on the multiphase average of $\gamma_{\lambda}$ are denoted by their corresponding median uncertainties. Studies such as the C-MetaLL survey \citep{tren24, bhar24} leverage individual metal abundances ([Fe/H]) of MW Cepheids, yielding a larger metallicity effect (in the absolute sense) than those obtained by comparing the PL intercepts in different galaxies \citep{gier18, breu22, bhar23}. The multiphase averages of $\gamma_{\lambda}$ values obtained in this study vary between $(-0.310 \pm 0.045) $ mag/dex and $(-0.774 \pm 0.046)$ mag/dex. We find that they are in good agreement within $2\sigma$ error bars with those obtained by \citet{gier18, tren24, bhar24} in all photometric bands. However, the results obtained in this study differ by more than $3\sigma$ with those from \citet{breu22} and \citet{bhar23} in the $G$- and $G_{\rm RP}$-band while being consistent in the $V$-, $I$-, $W_{VI}$-,  $G_{\rm BP}$- and $W_{G}$-band. We note that the multiphase averaged $\gamma_{\lambda}$ values in the $W_{VI}$ obtained in this study are larger than those obtained by \citet{breu22} and \citet{bhar23}. It may be primarily due to the $R_{\lambda}(=1.387)$ values used in constructing the $W_{VI}$ index in these two studies are different from the one used in this study $(R_{\lambda} = 1.55)$.

We also note that \citet{mado25} report a metallicity dependence consistent with zero based on open cluster Cepheids in the Milky Way and field Cepheids in several nearby galaxies. The multiphase averages of $\gamma_{\lambda}$ values obtained in our analysis differ by more than $2\sigma$ from their results. However, there are caveats to their treatment of the Gaia EDR3 parallaxes for MW Cepheids: instead of adopting the parallax offset recommended by the Gaia Team \citep{lind21b}, they apply a magnitude offset of 0.26 mag to the MW Cepheid distance moduli to match the scatter of the LMC PL relation when combined with earlier HST FGS parallaxes \citep[see details in][]{breu2025}.

\begin{figure*}
    \begin{tabular}{c|c}
         \resizebox{\linewidth}{!}{
         \includegraphics[width=\linewidth, keepaspectratio]{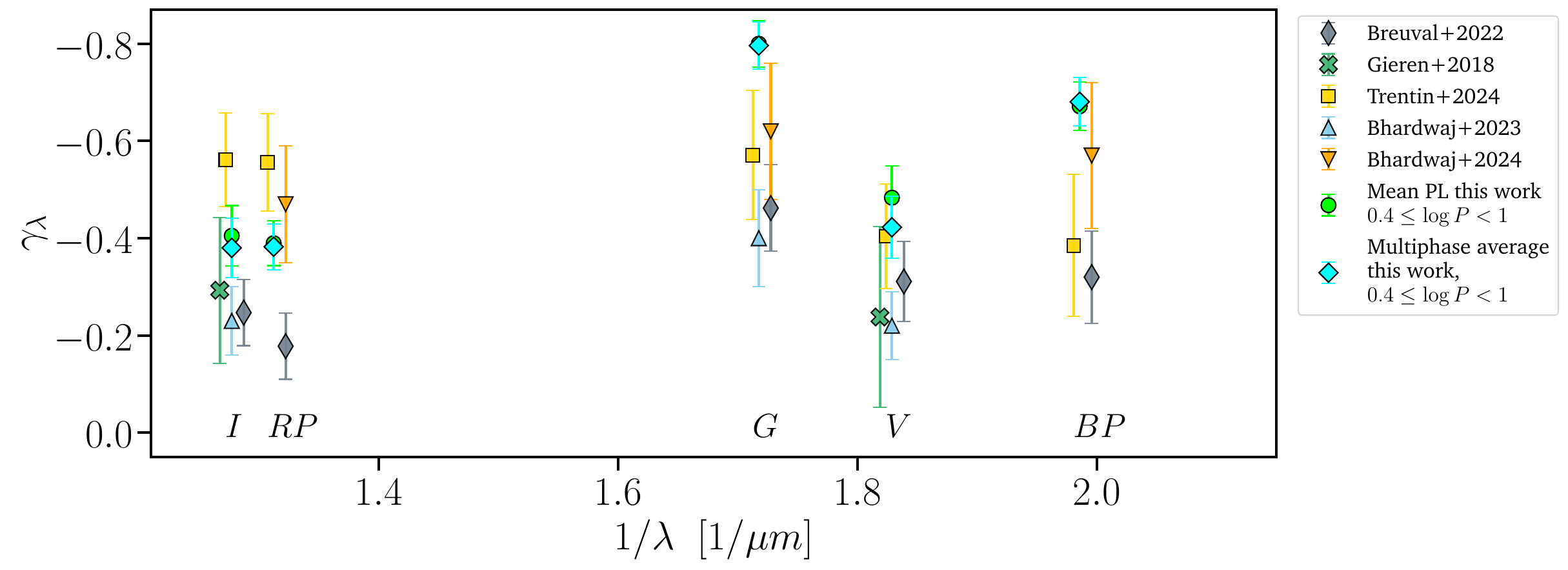}
         }\\
         \resizebox{\linewidth}{!}{
         \includegraphics[width=\linewidth, keepaspectratio]{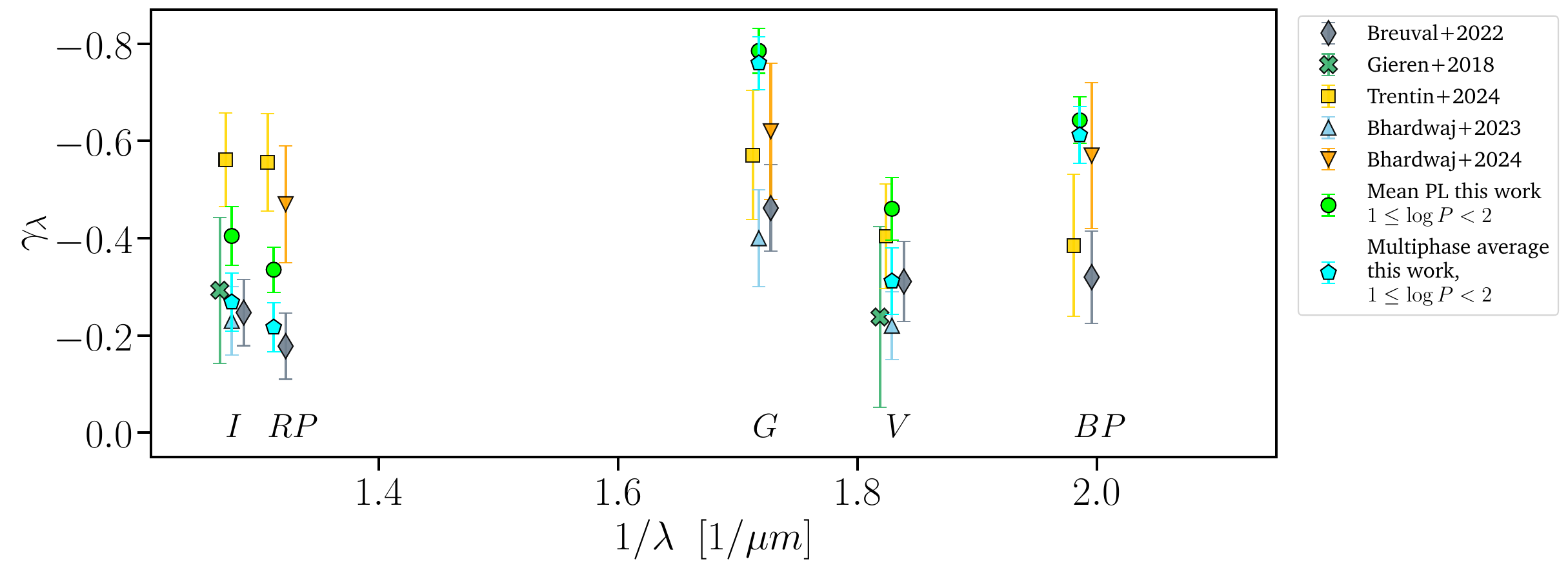}
         } 
    \end{tabular}
    \caption{Same as the comparison displayed in Fig. \ref{fig:8}. Multiphase averages of the metallicity coefficients of PLZ relations for short-period $(0.4 \leq \log{P} < 1)$ (cyan diamonds) and long-period $(1 \leq \log{P} < 2)$ (cyan pentagons) Cepheids, respectively, are compared to the metallicity coefficients determined in recent literature as well as in this work (green circles). \textit{Top panel} shows the metallicity coefficients determined using only the short-period Cepheids in this study, compared to the literature values. \textit{Bottom panel} shows the comparison of metallicity coefficients determined using only the long-period Cepheids in this study with the literature values. }
    \label{fig:11}
\end{figure*}

We also calculate the weighted averages of multiphase $\gamma_{\lambda}$ values for both short- and long-period Cepheids, which are compared to the results from other studies as shown in Fig. \ref{fig:11}. The typical uncertainties on the multiphase average of $\gamma_{\lambda}$ are denoted by their corresponding median uncertainties and are similar to those in the recent literature. These values are also presented in Table \ref{tab:5}. The multiphase averages of $\gamma_{\lambda}$ values of short-period Cepheids vary between $(-0.361 \pm 0.060)$ mag/dex and $(-0.796 \pm 0.048)$ mag/dex. Furthermore, those of the long-period Cepheids vary between $(-0.217 \pm 0.051)$ mag/dex and $(-0.760 \pm 0.054)$ mag/dex. Comparing the multiphase averaged $\gamma_{\lambda}$ values of the short-period Cepheids with those from other studies, we find that they compare similarly as the $\gamma_{\lambda}$ values of Cepheids of all periods. On the other hand, those of the long-period Cepheids are found to be consistent within $1-2\sigma$ error bars with the results from \citet{bhar24} and \citet{tren24} in $G$- and $G_{\rm BP}$-band; \citet{gier18}, \citet{breu22}, and \citet{bhar23} in $V$-, $I$-, and $W_{VI}$-band, respectively. However, the $\gamma_{\lambda}$ values of long-period Cepheids differ by more than $2.5\sigma$ level in $G$-, $W_{G}$- and $G_{\rm BP}$-band when compared to the $\gamma$ values obtained by \citet{breu22}, \citet{bhar23}; in $I$-band as compared to the results from \citet{tren24}. These differences may result from either using Cepheids only in a specific period range ($1 \leq \log{P} < 2$) or different reddening maps for the MW Cepheids or both.

Metallicity effect on the PL slope parameters have been explored in previous studies by \citet{ripe20}, and \citet{ripe21} based on individual metal abundances ([Fe/H]) of MW Cepheids. Their results indicate that the PL slope of fundamental-mode Cepheids shows only a marginal dependence on metallicity, while the intercept is more significantly affected. Given the small reported variation in the slope, we expect any impact on our results to be negligible.

\begin{table*}
    \centering
    \caption{Multiphase average of the coefficient of metallicity effect  $(\gamma_{\lambda})$ in five photometric bands: $G_{\rm BP}$, $V$, $G$, $G_{\rm RP}$, $I$, and two Wesenheit indices: $W_{VI}$ and $W_{G}$, respectively. The $\gamma_{\lambda}$ values are obtained using Cepheids of all periods $(0.4 \leq \log{P} < 2)$, short-periods $(0.4 \leq \log{P} < 1)$ and long-periods $(1 \leq \log{P} < 2)$, respectively. The quoted uncertainties represent the median uncertainties on the multiphase $\gamma_{\lambda}$ values. The $\gamma_{\lambda}$ values available from the literature are also presented for comparison. }
    \begin{tabular}{ccccc}
    \hline
    \hline
    Band & \multicolumn{4}{c}{ $\gamma_{\lambda}$ (mag/dex)} \\
    % & $\gamma_{\lambda}$ (mag/dex) from literature \\
    \hline 
    & & This work & & Literature$^{*}$ \\
    \hline
     & $0.4 \leq \log{P} < 2$ & $0.4 \leq \log{P} < 1$ & $1 \leq \log{P} < 2$ &  \\
    \hline
    \multirow{3}{*}{$BP$} & & & & $-0.320 \pm  0.095$ \citep{breu22} \\
    & $-0.649 \pm 0.048$ & $-0.681 \pm 0.050$ & $-0.612 \pm 0.059$ & $-0.385 \pm 0.146$ \citep{tren24} \\
    & & & & $-0.570 \pm 0.150$ \citep{bhar24} \\
    \hline
    \multirow{4}{*}{$V$} & & & & $-0.220 \pm 0.070$ \citep{bhar23} \\
    & $-0.409 \pm 0.061$ & $-0.422 \pm 0.064$ & $-0.311 \pm 0.068$ & $-0.238 \pm 0.186$ \citep{gier18} \\
    & & & &  $-0.311 \pm 0.082$ \citep{breu22} \\    
    & & & & $-0.404 \pm 0.108$ \citep{tren24}\\    
    \hline 
    \multirow{4}{*}{$G$} & & & & $-0.400 \pm 0.100$ \citep{bhar23}  \\
    & $-0.774 \pm 0.046$ & $-0.796 \pm 0.048$ & $-0.760 \pm 0.054$ & $-0.462 \pm 0.089$ \citep{breu22} \\
    & & & & $-0.571 \pm 0.133$ \citep{tren24} \\    
    & & & &  $-0.620 \pm 0.140$ \citep{bhar24} \\
    \hline 
    \multirow{3}{*}{$RP$} & & & & $-0.178 \pm 0.068$ \citep{breu22} \\
    & $-0.310 \pm 0.045$ & $-0.382 \pm 0.047$ & $-0.217 \pm 0.051$ & $-0.470 \pm 0.120$ \citep{bhar24}\\
    & & & & $-0.556 \pm 0.100$ \citep{tren24} \\
    \hline
    \multirow{4}{*}{$I$} & & & & $-0.230 \pm 0.070$ \citep{bhar23} \\
    & $-0.349 \pm 0.059$ & $-0.380 \pm 0.062$ & $-0.268 \pm 0.059$ & $-0.247 \pm 0.068$ \citep{breu22} \\
    & & & & $-0.293 \pm 0.150$ \citep{gier18} \\   
    & & & & $-0.561 \pm 0.096$ \citep{tren24}\\    
    \hline
    \multirow{2}{*}{$W_{VI}$} & & & & $-0.201 \pm 0.071$ \citep{breu22} \\
    & $-0.353 \pm  0.057$ & $-0.361 \pm 0.060$ & $-0.323 \pm 0.056$ & $-0.210 \pm 0.070$ \citep{bhar23} \\   
    \hline
    \multirow{3}{*}{$W_{G}$} & & & & $-0.384 \pm 0.051$ \citep{breu22} \\
    & $-0.340 \pm 0.044$ & $-0.376 \pm 0.044$ & $-0.215 \pm 0.045$ & $-0.430 \pm 0.060$ \citep{bhar23} \\   
    & & & & $-0.470 \pm 0.100$ \citep{bhar24} \\   
    \hline
    \hline
\end{tabular}
    \begin{tablenotes}
      \item  $^{*}$ The $\gamma_{\lambda}$ values available from the literature are based on Cepheids in different period ranges and employing various techniques.
    \end{tablenotes}
    \label{tab:6}
\end{table*}

\section{Summary \& Conclusions} \label{sec:conc}
The metallicity dependence of Cepheid PL relations is one of the least understood and most debated systematic effects of Cepheid distances. The HIF-photosphere interactions influenced by the different radiation hydrodynamical conditions in the stellar envelope lead to different PC relations and hence PL relations at the phases of minimum and maximum light. Furthermore, differences in metallicity may alter the envelope opacity and energy transport, which in turn impact the HIF-photosphere interaction and determine the morphology of the light curve. Earlier studies have shown that Cepheid PC relations vary along a pulsation cycle, especially around the phase of maximum and minimum light. The existence of multiphase PL relations has been demonstrated by several earlier studies \citep{ngeo12, kurb23, bhuy24}. A reasonable theoretical justification for the existence of such multiphase relations can also be provided by the existing theoretical framework based on period-mean density theorem, Stefan-Boltzmann law through the period-luminosity-color relation \citep{kanb96}. We extend it to investigate whether similar multiphase characteristics also exist in the metallicity effect on PL relations.

Using multi-epoch light curves of FU-mode Cepheids in the MW, LMC, and SMC from the OGLE-IV and Gaia DR3 databases, we quantify the metallicity effect on Cepheid PL relations in these three galaxies based on their phase-dependent properties over a pulsation cycle in five photometric bands: $G_{\rm BP}, V, G, I, $ and $G_{\rm RP}$ and two Wesenheit indices: $W_{VI}$ and $W_{G}$. The results obtained for the metallicity effect on multiphase PL relations at 50 different pulsation phases over a full cycle are summarized as follows:

\begin{enumerate}
\item The multiphase PL slopes in the LMC show dynamical variations over a full pulsation cycle. The largest amplitude of variation in PL slopes is observed in the $V$ band ($\sim 0.415 - 1.496$ mag/dex) across all period ranges. 

\item The multiphase PL intercepts of Cepheids in each galaxy vary dynamically over a full pulsation cycle. The largest amplitude of variation in PL intercepts is observed in the $V$ band ($\sim 0.645 -0.949$ mag) across all period ranges in all three galaxies. 

\item The dynamical variations in the PL slopes/intercepts of short-period Cepheids follow an opposite trend to that seen for the long-period Cepheids in all photometric bands in the phase range: $0.6 \leq \Phi \leq 1$. The multiphase PL slopes $(\alpha_{\lambda})$ and intercepts $(\beta_{\lambda})$ are found to be consistent with those in the recent literature.

\item The metallicity effect $(\gamma_{\lambda})$ shows dynamical variations over a complete pulsation cycle in all photometric bands. The largest amplitude of variation is observed in the $ \gamma_{W_{VI}}$ coefficients ($\sim 0.299, ~0.190$ mag/dex) for short-period and all-period Cepheids, and $\gamma_{G_{\rm BP}}$ coefficient ($\sim 0.297$ mag/dex) for the long-period Cepheids.

\item Comparing the multiphase $\gamma_{\lambda}$ values determined using short- and long-period Cepheids, we find that they vary oppositely in the phase range: $0.6 \leq \Phi < 1$ in all photometric bands except the $W_{VI}$-band.

\item We compare the weighted averages of multiphase $\gamma_{\lambda}$ values with those determined in several recent studies using mean-light PL relations. We find them to be consistent within $2\sigma$ level. This is used as a sanity check in this study.

\item Furthermore, our analysis is able to statistically quantify the distinctive nature of the metallicity effect between short- and long-period Cepheids within specific phase intervals using the Bonferroni correction method in two bands: $G_{\rm RP}$ and $W_{G}$. These results offer a better opportunity to constrain the metallicity effect on PL relations.  
\end{enumerate}

This study has effectively demonstrated the phase-dependent nature of the metallicity influence on Cepheid PL relations. Our study sheds light on the distinct nature of the $\gamma_{\lambda}$ values of short- and long-period Cepheids observed at phases around $0.6-1$. It will prove useful in understanding the effect of metallicity on the Cepheid PL relations as well as in tightly constraining Cepheid pulsation models. The underlying physical effects responsible for such distinct variations can be investigated in future works as the next step forward. One of the possibilities is the Hertzsprung progression, whereby a bump on the light curves of Cepheids moves from the ascending to the descending branches around a period $P=10$ days.

However, it is not clear at the moment if our phase-dependent method can be applicable or extended for investigating the PLZ relations of Cepheids for distant galaxies in the extragalactic distance ladder. The data for Cepheids in these host galaxies are sparse and the phase coverage is usually limited due to significant telescope time required at these distances, making it difficult to obtain complete light curves. Upcoming surveys such as the Legacy Survey of Space and Time (LSST) at the Vera C. Rubin Observatory are expected to provide well-sampled light curves of variable stars in distant galaxies, which would make it possible to explore the metallicity effect on the extragalactic distance scale based on the phase-dependent properties of Cepheids in future works.
 
\section*{Acknowledgements}
The authors acknowledge the use of highly valuable publicly accessible archival data from OGLE-IV and Gaia DR3. GB is grateful to the Department of Science and Technology (DST), Govt. of India, New Delhi, for providing the financial support of this study as a Senior Research Fellow (SRF) through the DST INSPIRE Fellowship research grant (DST/INSPIRE/Fellowship/2019/IF190616). SD thanks CSIR, Govt. of India, New Delhi, for the financial support received through a research grant ``03(1425)/18/EMR-II''. 
SMK thanks State University of New York, Oswego, NY 13126, USA and Cotton University, Guwahati, Assam for the support. 
A.B. thanks the funding from the Anusandhan National Research Foundation (ANRF) under the Prime Minister Early Career Research Grant scheme (ANRF/ECRG/2024/000675/PMS).
KK thanks the Council of Scientific and Industrial Research (CSIR), Govt. of India for the Senior Research Fellowship (SRF). This research was supported by the International Space Science Institute (ISSI) in Bern/Beijing through ISSI/ISSI-BJ International Team project ID $\#$24-603 – ``EXPANDING Universe'' (EXploiting Precision AstroNomical Distance INdicators in the Gaia Universe). The authors acknowledge IUCAA, Pune, for providing access to the Pegasus High Performance Computing facility. Finally, the authors are grateful to the anonymous reviewers for the useful comments and constructive suggestions, which have significantly improved the presentation of the manuscript.

\section*{Data Availability}
The OGLE-IV data is collected from \url{http://ftp.astrouw.edu.pl/ogle/ogle4/OCVS}. The Gaia DR3 data is downloaded from \url{http://cdn.gea.esac.esa.int/Gaia/gdr3} using Python astro-query. The \citet{skow21} reddening map is downloaded from \url{http://ogle.astrouw.edu.pl/}. 



\bibliographystyle{mnras}
\bibliography{example} % if your bibtex file is called example.bib

%\begin{thebibliography}{99}
%\end{thebibliography}

%%%%%%%%%%%%%%%%%%%%%%%%%%%%%%%%%%%%%%%%%%%%%%%%%%

%%%%%%%%%%%%%%%%% APPENDICES %%%%%%%%%%%%%%%%%%%%%

\appendix

\section{Additional Figures}

\begin{figure*}
\begin{tabular}{c|c}
    \resizebox{0.5\linewidth}{!}{
    \includegraphics[width=0.5\linewidth, keepaspectratio]{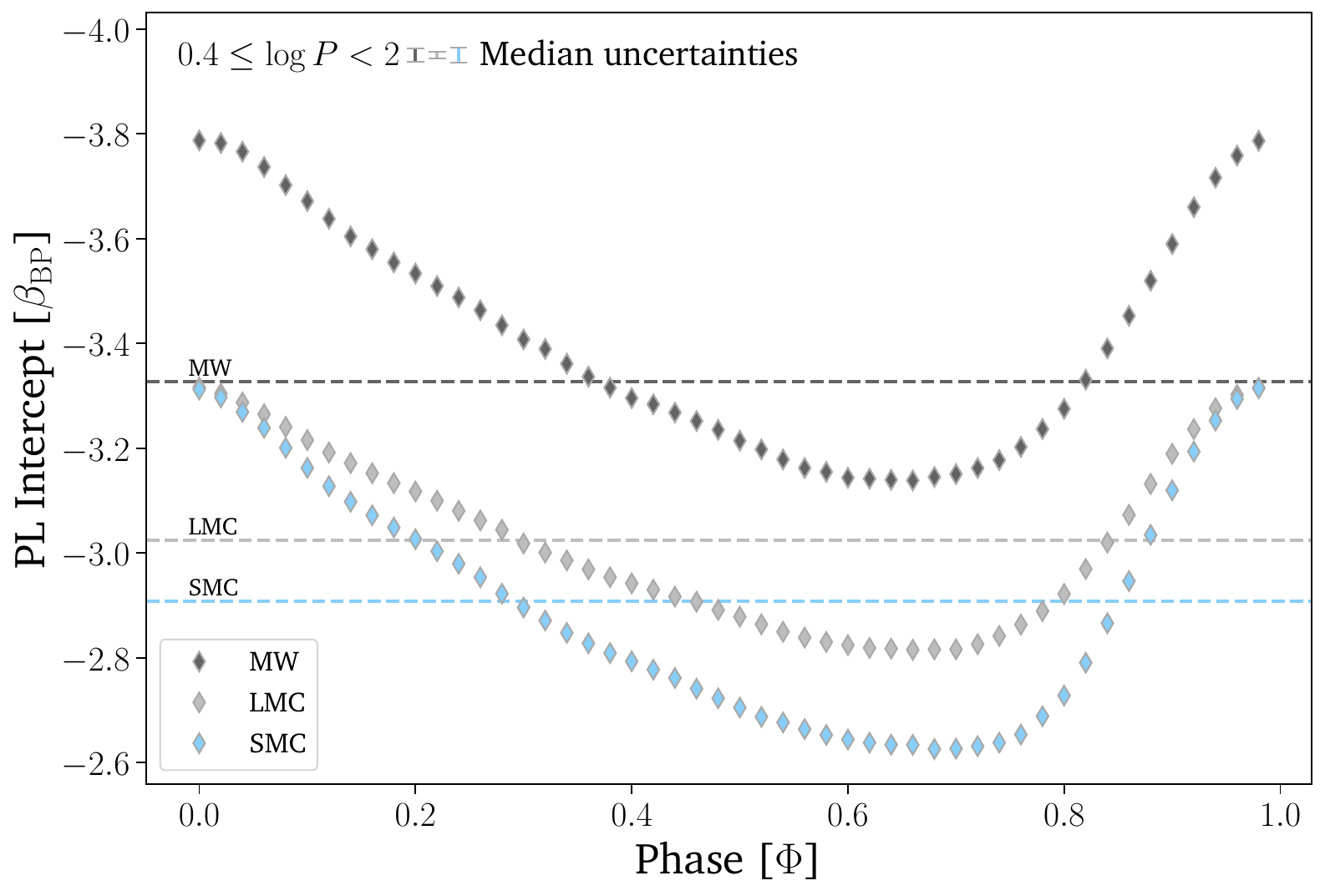}
     }
    \resizebox{0.5\linewidth}{!}{
    \includegraphics[width=0.5\linewidth, keepaspectratio]{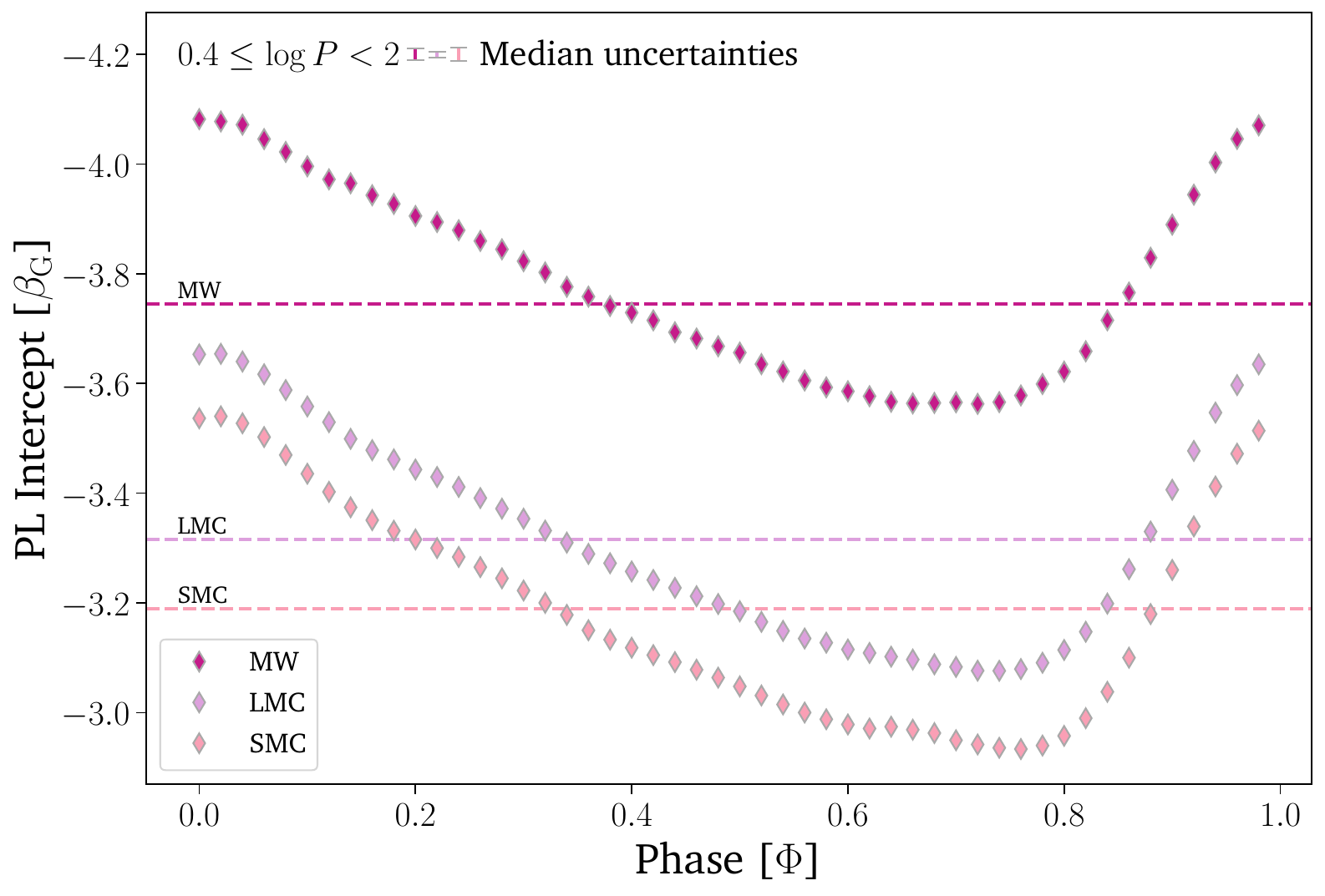}
     }\\
     % \hline 
    \resizebox{0.5\linewidth}{!}{
    \includegraphics[width=0.5\linewidth, keepaspectratio]{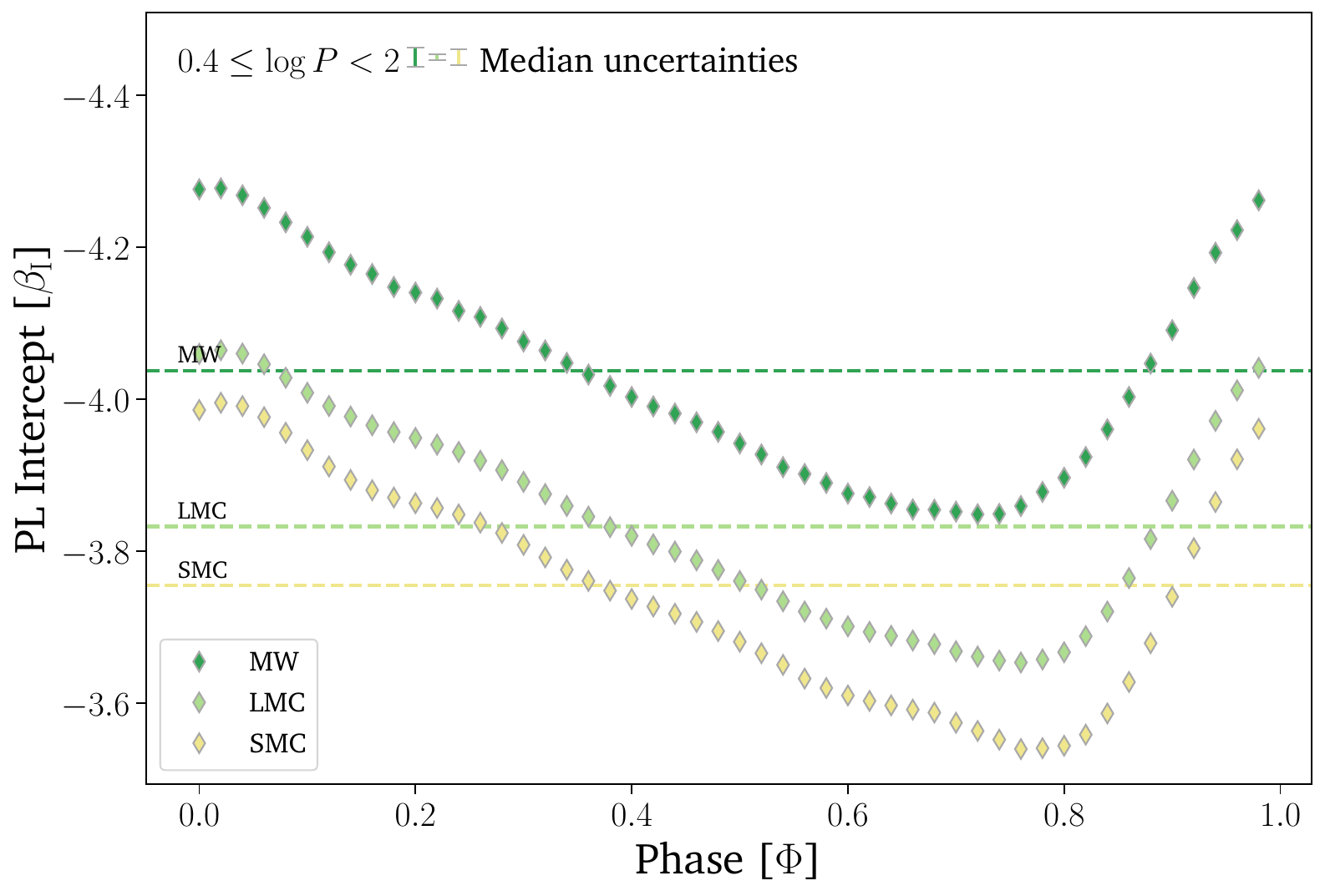}
     }
    \resizebox{0.5\linewidth}{!}{
    \includegraphics[width=0.5\linewidth, keepaspectratio]{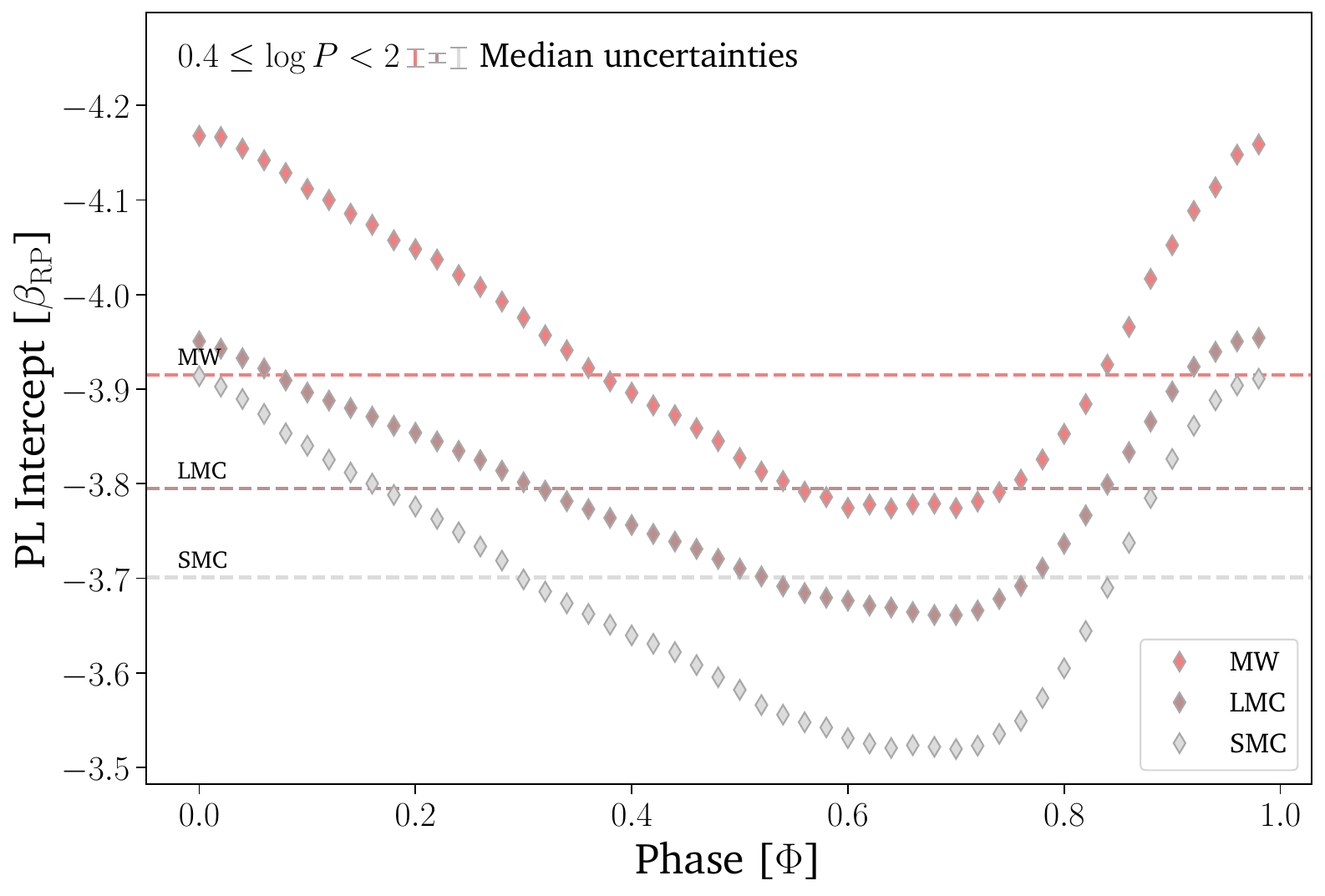}
     } \\
     % \hline
    \resizebox{0.5\linewidth}{!}{
    \includegraphics[width=0.5\linewidth, keepaspectratio]{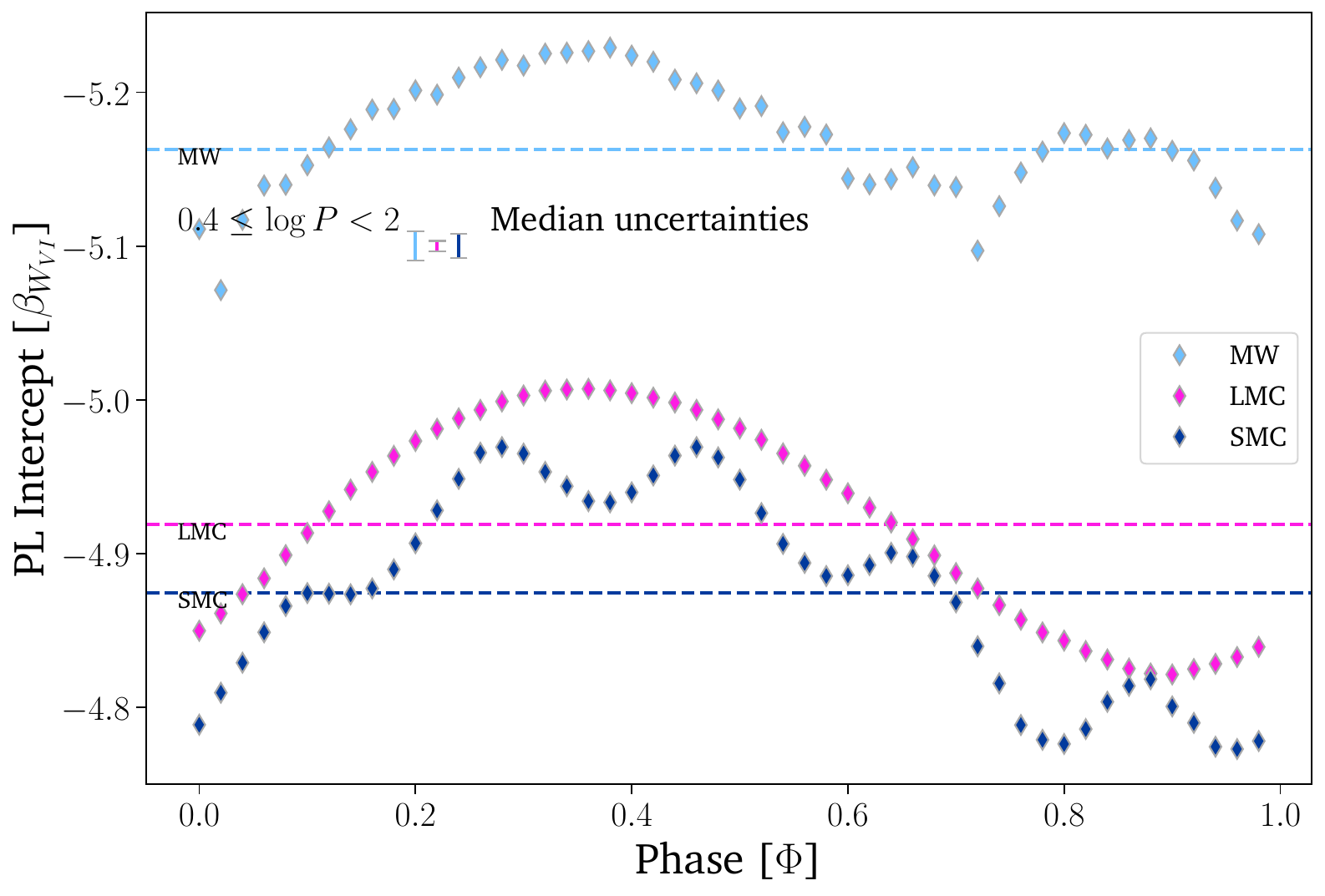}
     }
    \resizebox{0.5\linewidth}{!}{
    \includegraphics[width=0.5\linewidth, keepaspectratio]{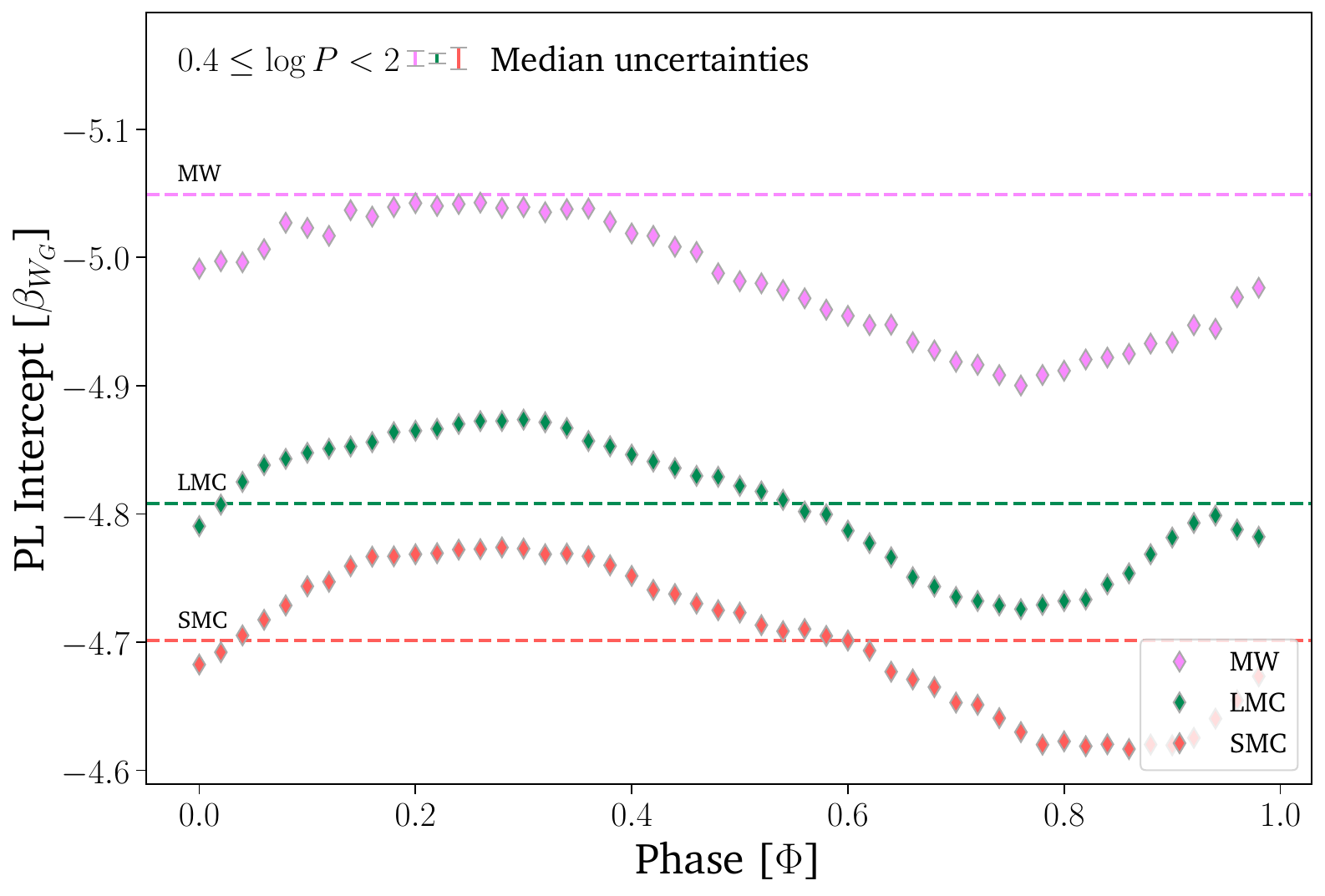}
     }
\end{tabular}
\caption{Same as Fig. \ref{fig:5} for Cepheids in the MW, LMC and SMC in the photometric bands: $G_{\rm BP}, G, I$ and $ G_{\rm RP}$ respectively. The dashed horizontal lines in the figure represent the PL intercepts obtained using mean magnitudes of Cepheids for different galaxies. Typical uncertainties are represented by the median uncertainties on $\beta_{\lambda}$ values in each photometric band.}
\label{fig:a1}
\end{figure*}

\begin{figure*}
    \begin{tabular}{c|c}
     \resizebox{0.5\linewidth}{!}{
     \includegraphics[width=0.5\linewidth, keepaspectratio]{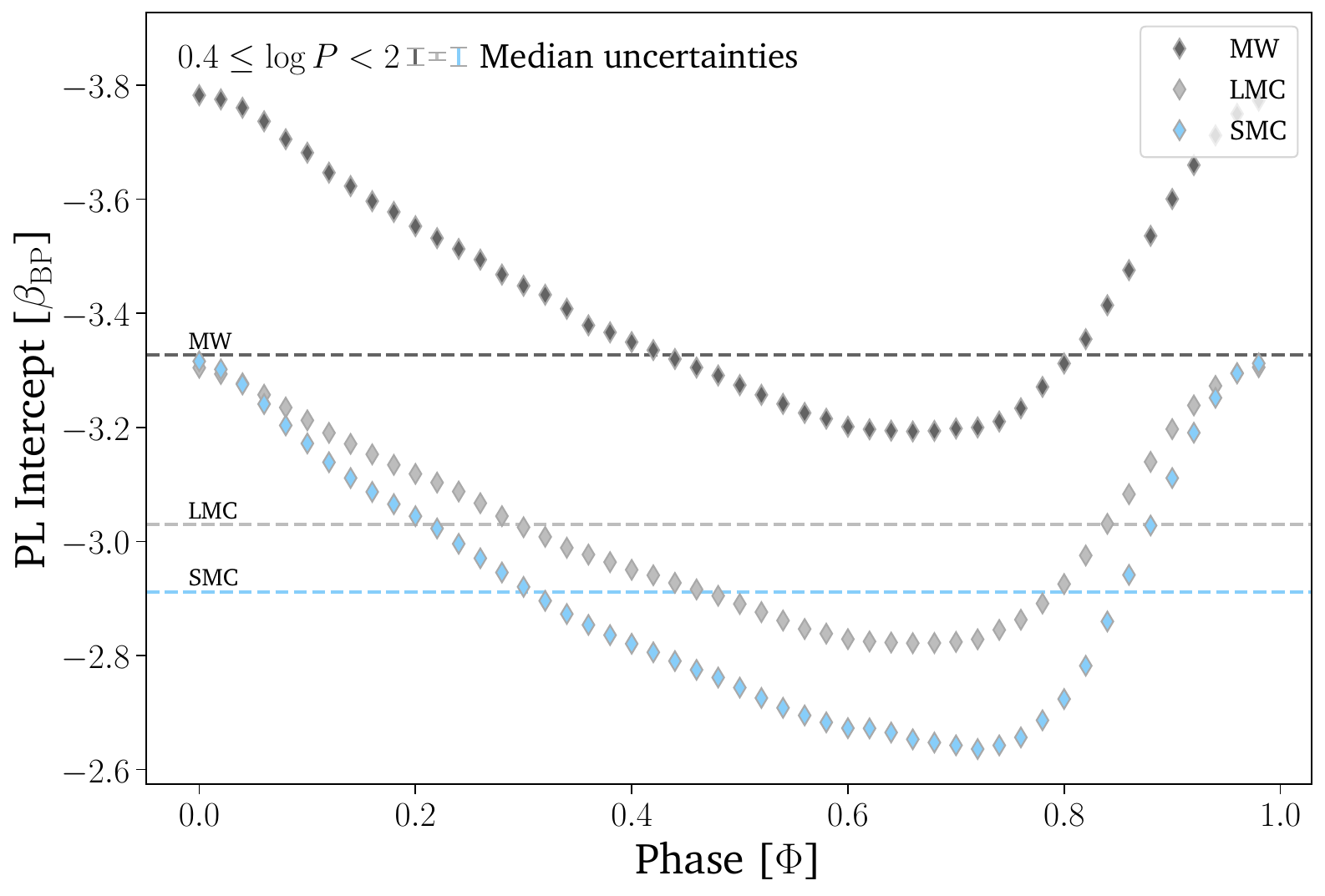}
     }
     \resizebox{0.5\linewidth}{!}{
     \includegraphics[width=0.5\linewidth, keepaspectratio]{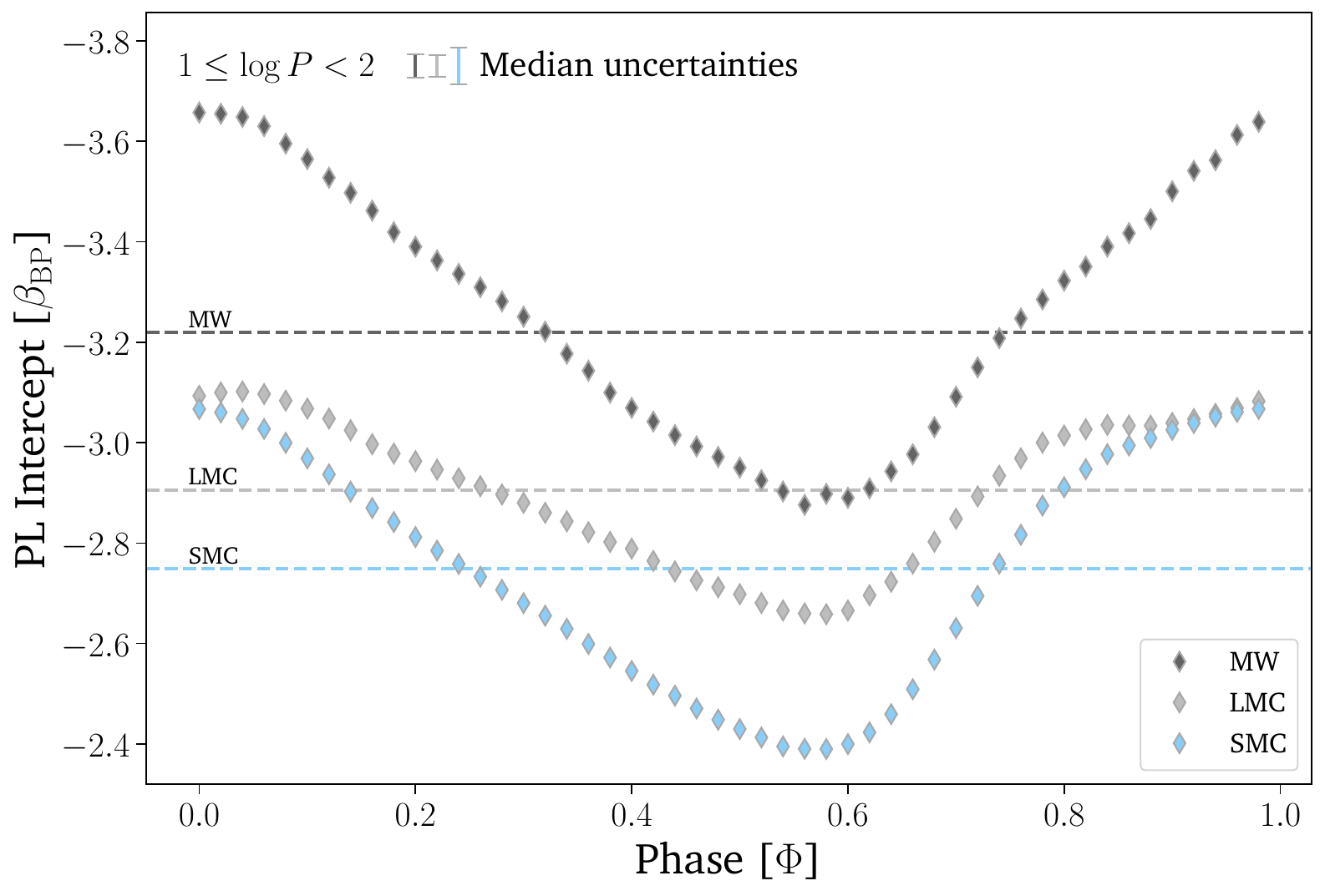}
     } \\

     \resizebox{0.5\linewidth}{!}{
     \includegraphics[width=0.5\linewidth, keepaspectratio]{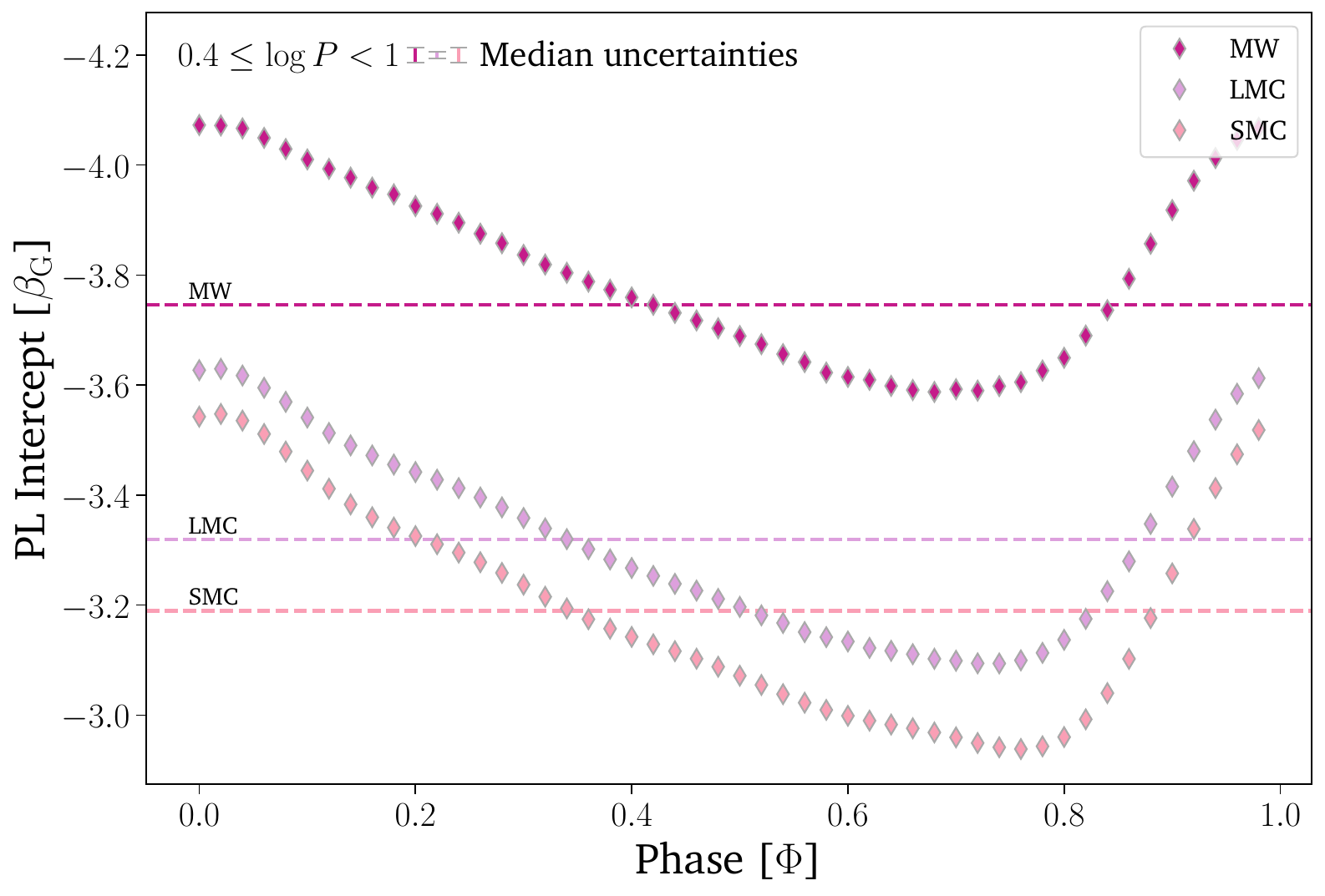}
     }
     \resizebox{0.5\linewidth}{!}{
     \includegraphics[width=0.5\linewidth, keepaspectratio]{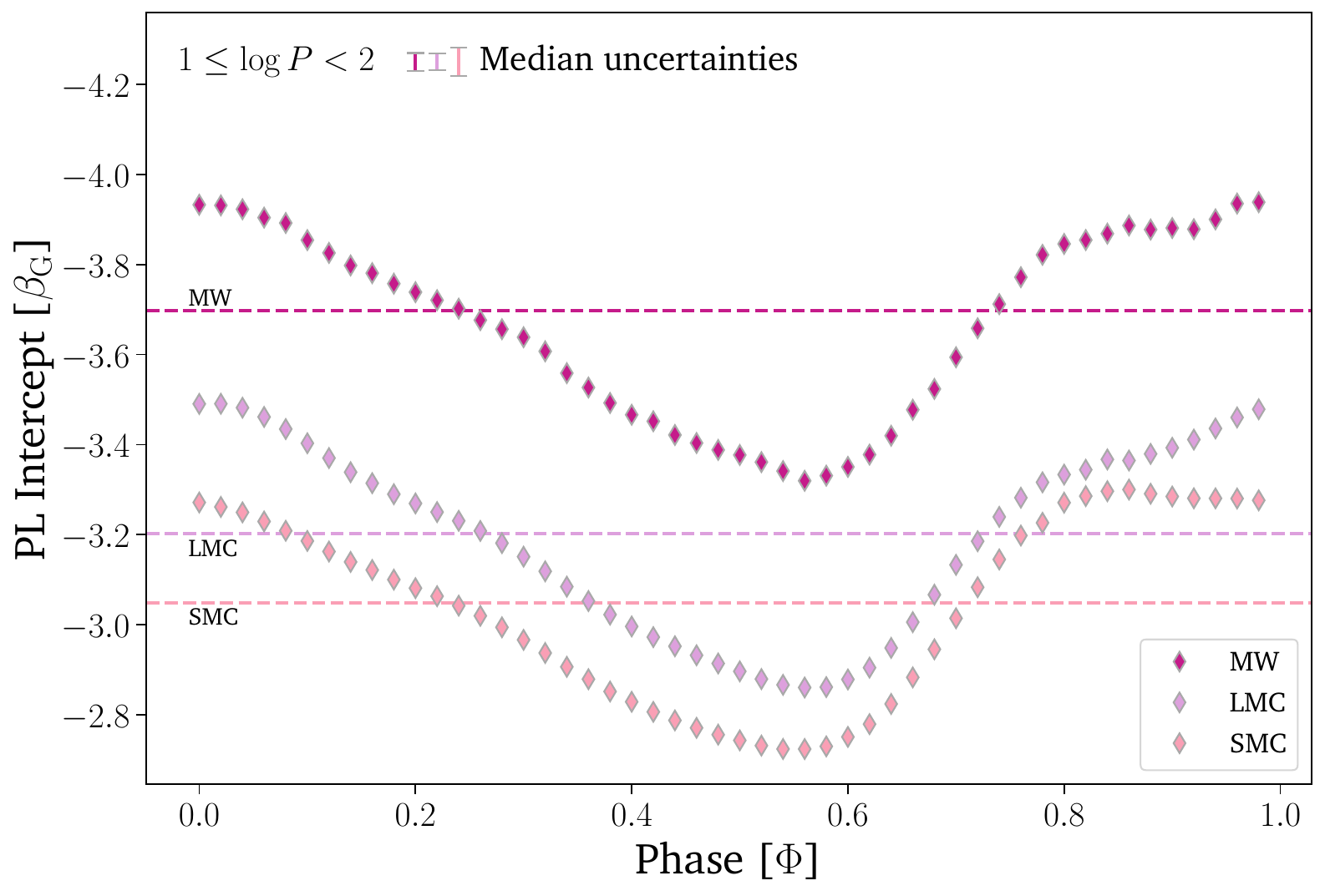}
     } \\

     \resizebox{0.5\linewidth}{!}{
     \includegraphics[width=0.5\linewidth, keepaspectratio]{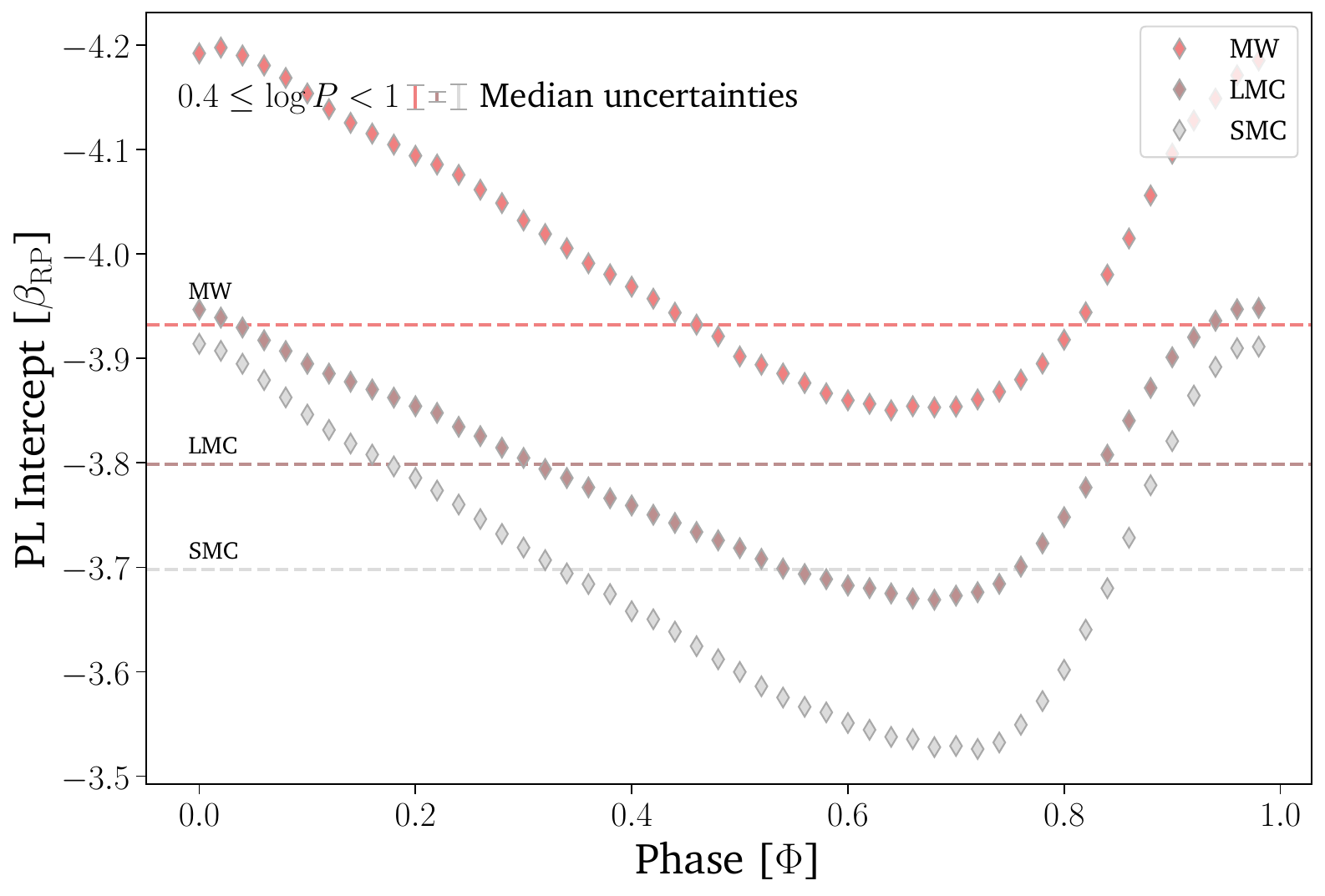}
     }
     \resizebox{0.5\linewidth}{!}{
     \includegraphics[width=0.5\linewidth, keepaspectratio]{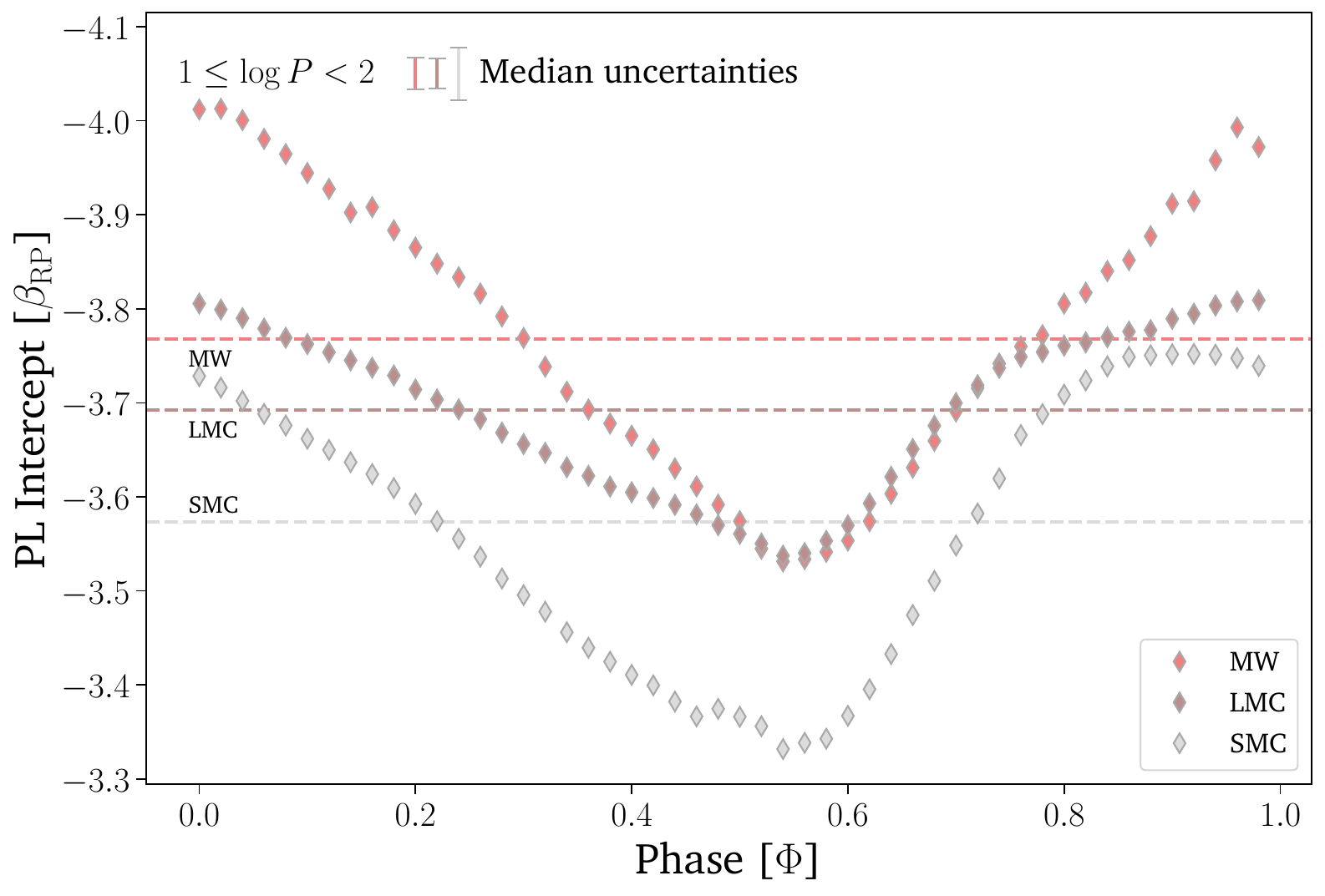}
     } \\     
    \end{tabular}
    \caption{Comparison of the multiphase PL intercepts of Cepheids in the MW, LMC, and the SMC in the photometric bands: $G_{\rm BP}$, $G$, and $G_{\rm RP}$. \textit{Left panels} compare the PL intercepts determined using only the short-period Cepheids. \textit{Right panels} compare the PL intercepts determined using only the long-period Cepheids. The dashed horizontal lines in the figure represent the PL intercepts obtained using mean magnitudes of the short- and long-period Cepheids, respectively, for different galaxies. The median uncertainties represent the typical uncertainties of the multiphase PL intercepts.}
    \label{fig:a2}
\end{figure*}

\begin{figure*}
    \begin{tabular}{c|c}
    
     \resizebox{0.5\linewidth}{!}{
     \includegraphics[width=0.5\linewidth, keepaspectratio]{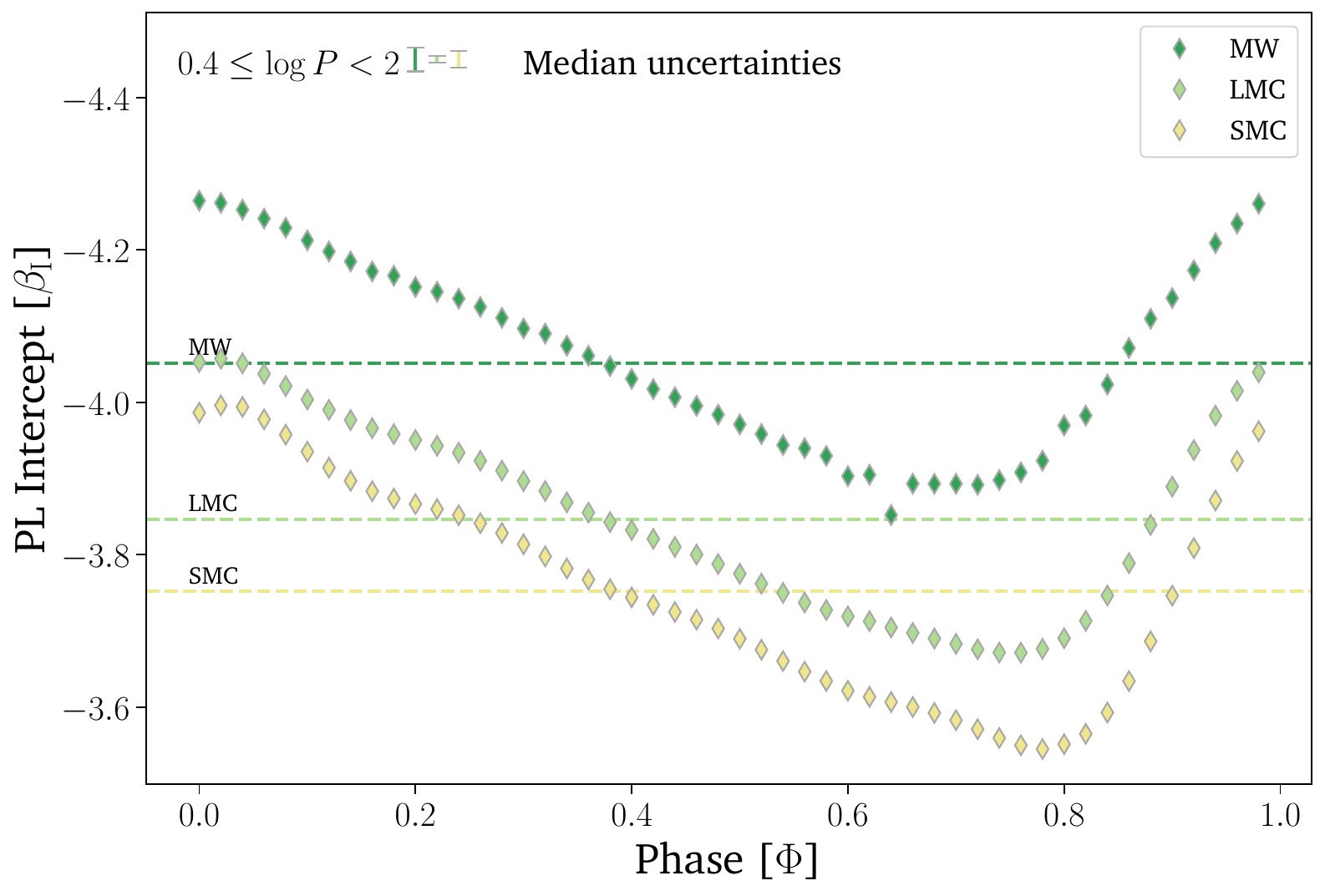}
     }
     \resizebox{0.5\linewidth}{!}{
     \includegraphics[width=0.5\linewidth, keepaspectratio]{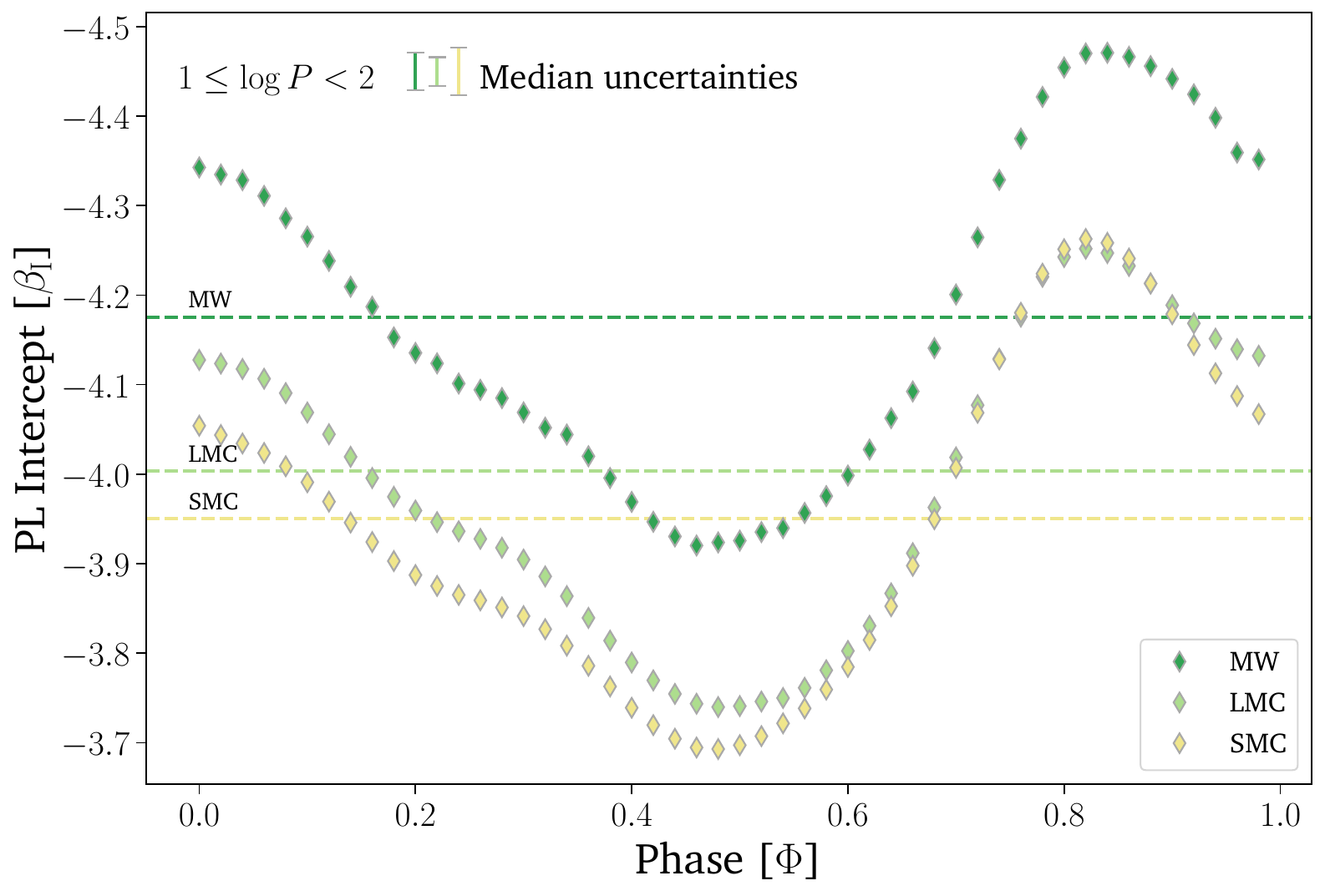}
     } \\

     \resizebox{0.5\linewidth}{!}{
     \includegraphics[width=0.5\linewidth, keepaspectratio]{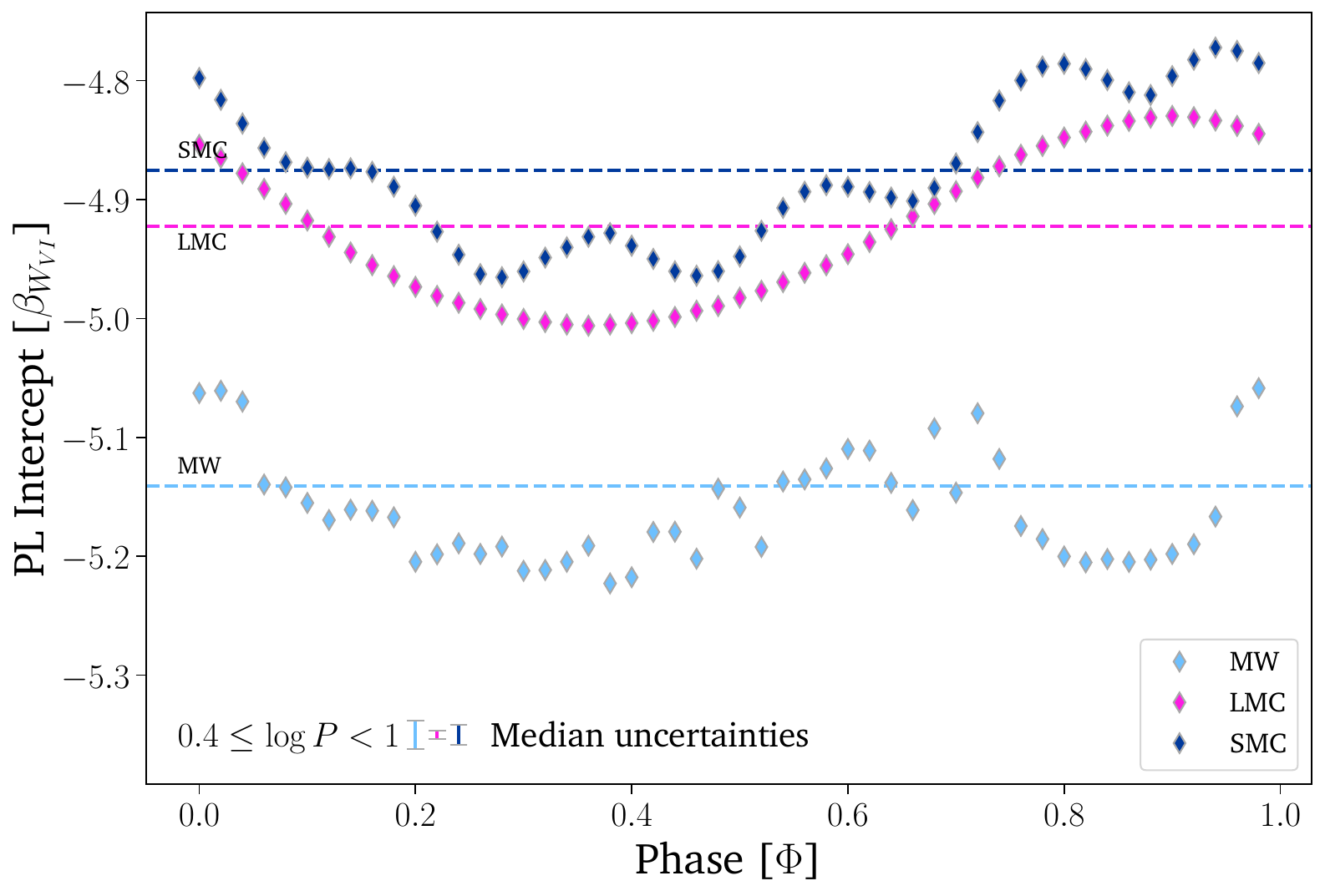}
     }
     \resizebox{0.5\linewidth}{!}{
     \includegraphics[width=0.5\linewidth, keepaspectratio]{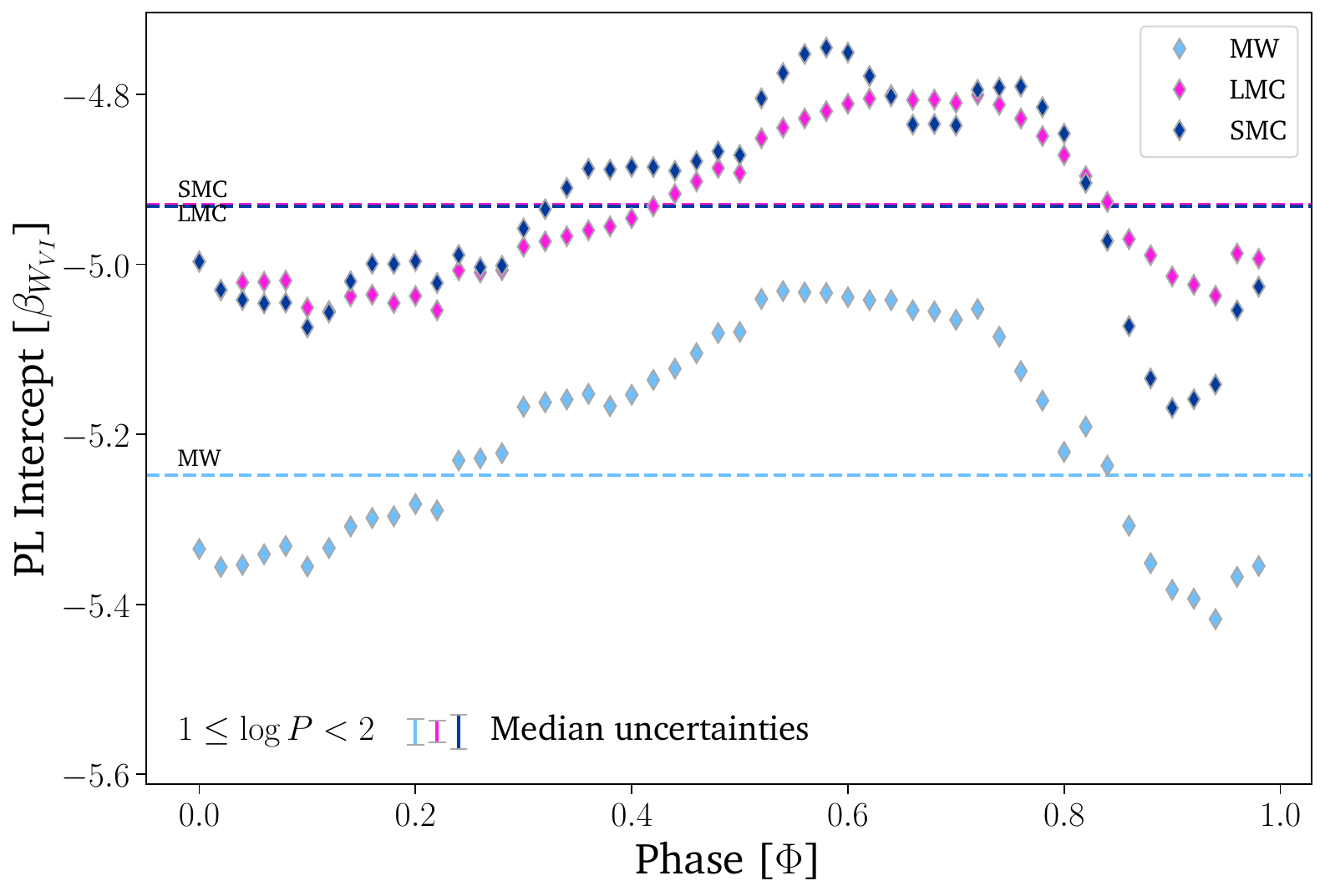}
     } \\

     \resizebox{0.5\linewidth}{!}{
     \includegraphics[width=0.5\linewidth, keepaspectratio]{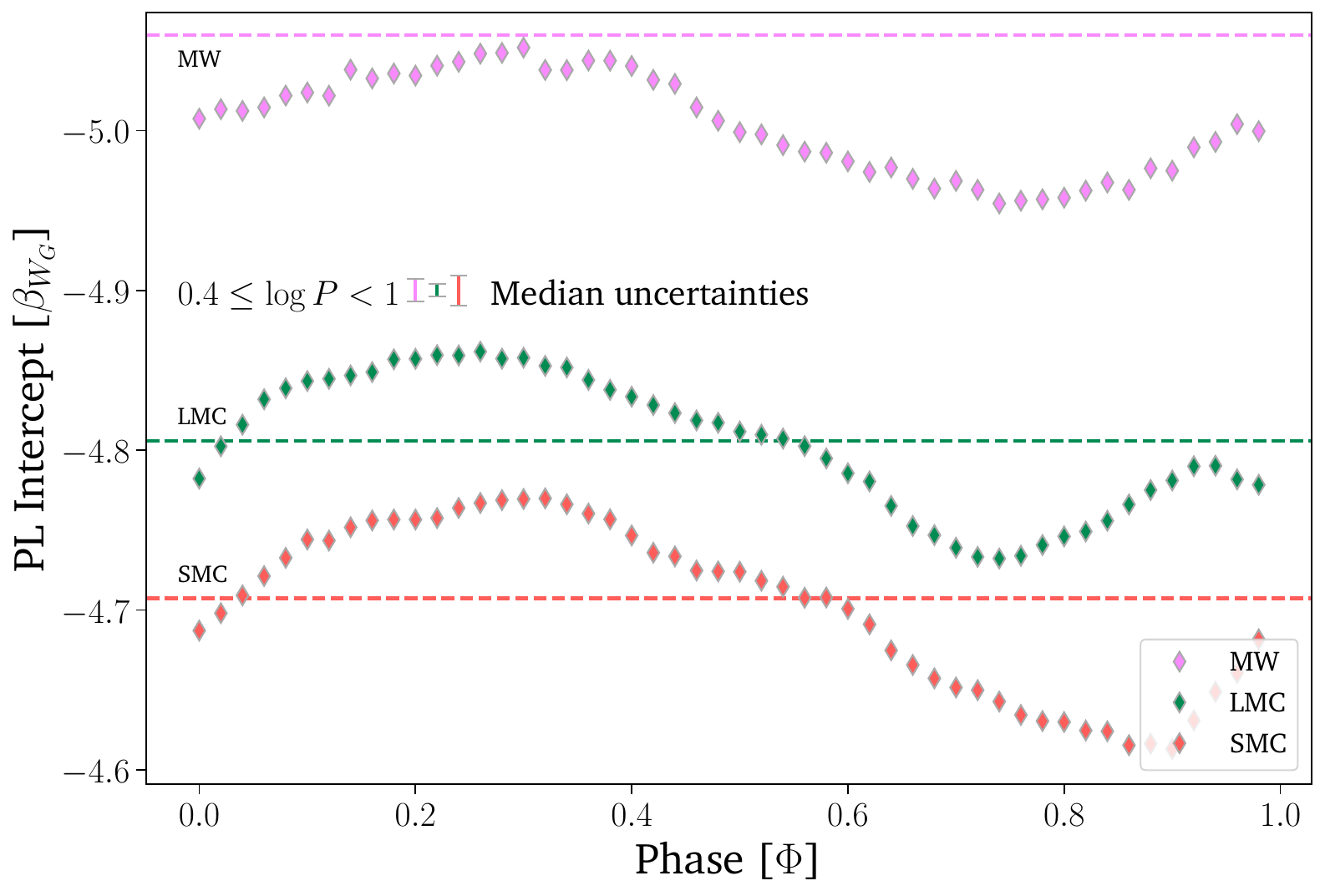}
     }
     \resizebox{0.5\linewidth}{!}{
     \includegraphics[width=0.5\linewidth, keepaspectratio]{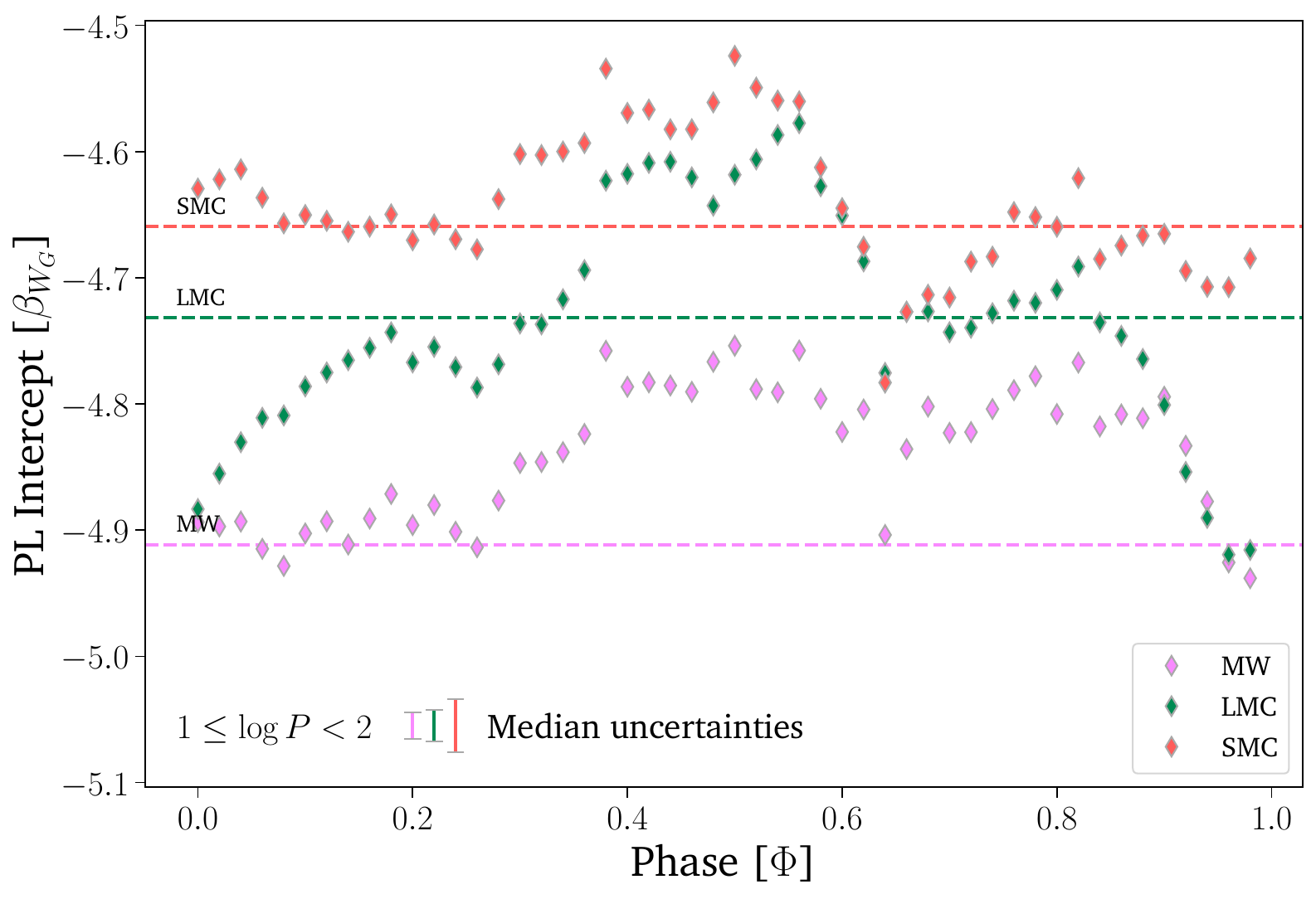}
     } \\     
    \end{tabular}
    \caption{Comparison of the multiphase PL intercepts of Cepheids in the MW, LMC, and the SMC in the photometric bands: $I$, $W_{VI}$, and $W_{G}$. \textit{Left panels} compare the PL intercepts determined using only the short-period Cepheids. \textit{Right panels} compare the PL intercepts determined using only the long-period Cepheids. The dashed horizontal lines in the figure represent the PL intercepts obtained using mean magnitudes of the short- and long-period Cepheids, respectively, for different galaxies. The median uncertainties represent the typical uncertainties of the multiphase PL intercepts.}
    \label{fig:a3}
\end{figure*}

\bsp	% typesetting comment
\label{lastpage}
\end{document}